\documentclass[pdflatex, sn-mathphys-num]{sn-jnl}

\usepackage{graphicx}%
\usepackage{multirow}%
\usepackage{amsmath,amssymb,amsfonts}%
\usepackage{amsthm}%
\usepackage{mathrsfs}%
\usepackage[title]{appendix}%
\usepackage{xcolor}%
\usepackage{textcomp}%
\usepackage{manyfoot}%
\usepackage{booktabs}%
\usepackage{algorithm}%
\usepackage{algorithmicx}%
\usepackage{algpseudocode}%
\usepackage{listings}%

\usepackage{comment}
\usepackage{caption}
\usepackage{subcaption}
\usepackage{enumitem}
\usepackage{makecell}
\usepackage{pdflscape}
\usepackage{soul}
\usepackage{longtable}

\usepackage{graphicx}
\usepackage{float}

\theoremstyle{thmstyleone}%
\theoremstyle{thmstyletwo}%

\theoremstyle{thmstylethree}%

\begin{document}


\title[Article Title]{\centering 
\fontsize{12.5pt}{15pt}\selectfont
Imputation-free transformer learning\\ enables robust Alzheimer's disease prediction and calibrated uncertainty quantification across heterogeneous clinical cohorts}


\author{
    \centering
    \vspace{2mm}
    {\fnm{Christelle} \sur{Schneuwly Diaz}}$^{1,2}$\footnote{christelle.schneuwly@chuv.ch}, 
    {\fnm{Narmina} \sur{Baghirova}}$^{1,2}$, 
    {\fnm{Duy-Thanh} \sur{V\~u}}$^{1,2}$, 
    {\fnm{Duy-Cat} \sur{Can}}$^{1,2}$, 
    {\fnm{Gilles} \sur{Allali}}$^{3}$, 
    {\fnm{Philippe} \sur{Ryvlin}}$^{4}$, 
    and {\fnm{Oliver Y.} \sur{Chén}}$^{1,2}$\footnote{olivery.chen@chuv.ch} \\[4mm]
      \vspace{2mm}
    {for the Alzheimer's Disease Neuroimaging Initiative (ADNI)}\footnote{Part of the data used in this article was from the Alzheimer's Disease Neuroimaging Initiative (ADNI) database (adni.loni.usc.edu). As such, the investigators within the ADNI contributed to the design and implementation of ADNI and/or provided data but did not participate in the analysis or writing of this paper. A complete list of ADNI investigators is available at \url{http://adni.loni.usc.edu/wp-content/uploads/how_to_apply/ADNI_Acknowledgement_List.pdf}.} , \\
    {for the Australian Imaging, Biomarkers and Lifestyle (AIBL)}\footnote{Part of the data used in the preparation of this article was obtained from the Australian Imaging Biomarkers and Lifestyle flagship study of ageing (AIBL) funded by the Commonwealth Scientific and Industrial Research Organisation (CSIRO) which was made available at the ADNI database (\url{www.loni.usc.edu/ADNI}). The AIBL researchers contributed data but did not participate in analysis or writing of this report. AIBL researchers are listed at \url{https://data.aibl.org.au/adni/}.}\hfill  study and \\
    {for the Open Access Series of Imaging Studies (OASIS)}\footnote{Part of the data used in this article was from the Open Access Series of Imaging Studies (OASIS) (\url{https://sites.wustl.edu/oasisbrains/}).}
    \vspace{1.5mm}
\vspace{5mm}
}

\affil[1]{
\orgdiv{Platform of Bioinformatics}, \orgname{Lausanne University Hospital}}

\affil[2]{\orgdiv{Faculty of Biology and Medicine}, \orgname{University of Lausanne}}

\affil[3]{\orgdiv{Leenaards Memory Centre}, \orgname{Lausanne University Hospital}}

\affil[4]{\orgdiv{Department of Clinical Neurosciences}, \orgname{Lausanne University Hospital}}



\abstract{Accurate diagnostic classification and disease-severity prediction for Alzheimer's disease are hampered by the pervasive incompleteness and heterogeneity of real-world clinical data. Left unaddressed, these barriers prevent reliable disease modelling and hinder effective clinical evaluation. Conventional imputation strategies introduce systematic bias, distort inter-feature relationships, and yield overconfident predictions, limitations that are especially consequential in diagnostic settings. 
Here, we propose NITROGEN\footnote{NITROGEN: No-imputation Inter-sample-attention TRansformer Oriented for GENeral datasets.} \hspace{.05mm}, an imputation-free transformer that jointly models within-patient feature dependencies and between-patient relational structure through masked and intersample attention, enabling robust multimodal learning directly from partially observed records. 
We trained NITROGEN on the Alzheimer’s Disease Neuroimaging Initiative (ADNI; N=7858 scans), and evaluated it, without further retraining, on two independent cohorts: the Open Access Series of Imaging Studies (OASIS-3; N=2675 scans) and the Australian Imaging, Biomarkers and Lifestyle (AIBL; N=1286 scans) study.
Across all cohorts and various tasks including binary and multi-class diagnostic classification as well as continuous cognitive score prediction, NITROGEN showed robust probability calibration and uncertainty quantification advantages over tree-based ensemble methods, while maintaining competitive discriminative and continuous cognitive score prediction performance. Moreover, cross-cohort and cross-method analyses identified cortical thickness in the temporal pole, age, and APOE genotype as important, though not individually sufficient, features for robust AD status classification. To address prediction reliability under incomplete data, we further introduced a modality-aware uncertainty adjustment that augments predictive uncertainty in proportion to the estimated importance of absent modalities, enabling models to express calibrated confidence when critical diagnostic information is unavailable.
Taken together, our results show that imputation-free attention learning preserved meaningful discrimination under external cohort shift, while revealing the expected performance degradation on more distributionally different cohorts. These findings further demonstrate that evaluating machine learning models for neurological disease along dimensions of calibration, interpretability, and cross-cohort reliability---not predictive accuracy alone---is essential for better clinical deployment. Code to reproduce our experiments is publicly available at \url{https://github.com/cschneuw/nitrogen}.
}




\maketitle

\begin{figure}[t!]
    \centering
\includegraphics[width=\linewidth]{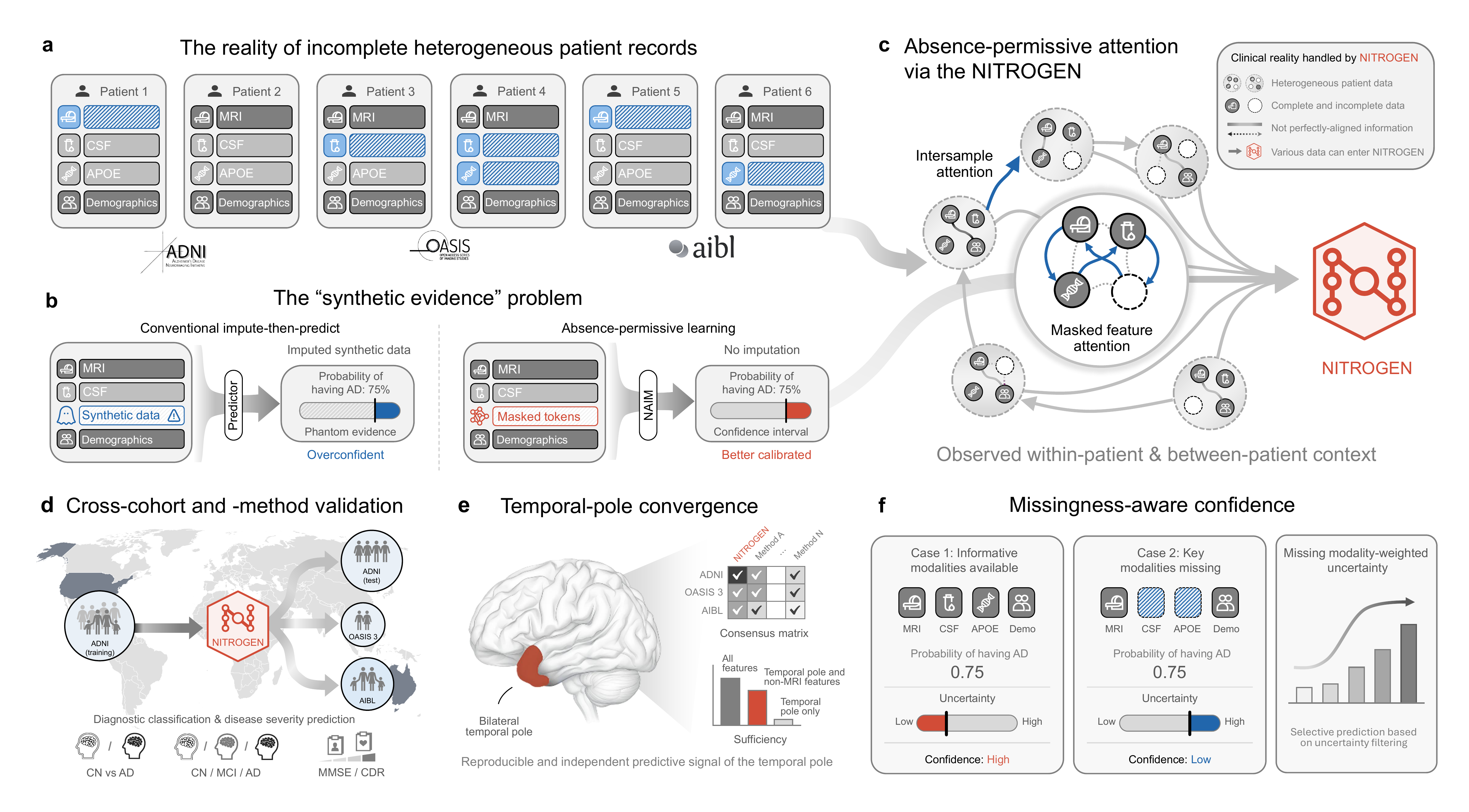}
    \caption{ 
    \textbf{Overview of the NITROGEN framework for imputation-free multimodal learning from incomplete Alzheimer's disease cohorts.}
    \textbf{(a)} Alzheimer’s disease cohorts integrate complementary sources of information, including demographic, genetic, neuroimaging, cognitive, and molecular measurements, but these modalities are unevenly available across individuals and studies. This patient-level heterogeneity creates a major challenge for multimodal learning and cross-cohort generalisation.
    \textbf{(b)} Conventional impute-then-predict strategies replace missing measurements with synthetic values, potentially introducing artificial evidence, altering data structure, and affecting the reliability of model confidence estimates.
    \textbf{(c)} NITROGEN learns directly from partially observed patient records by combining masked feature-level attention within individuals with intersample attention across heterogeneous patients, allowing integration of available evidence without explicit imputation.
    \textbf{(d)} The NITROGEN framework is trained on ADNI and evaluated across held-out ADNI participants and independent OASIS-3 and AIBL cohorts, covering diagnostic classification, disease-state discrimination, and continuous cognitive score prediction.
    \textbf{(e)} Cross-cohort and cross-method attribution analyses identify temporal pole cortical thickness (right hemisphere: TempPole~1; left hemisphere: TempPole~3) as important MRI-derived features associated with Alzheimer's disease classification, showing consistent importance rankings across datasets and models.
    \textbf{(f)} Modality-aware uncertainty adjustment incorporates the importance of missing information into individual predictions, increasing uncertainty when highly informative modalities are unavailable and providing confidence estimates that reflect patient-specific evidence availability.
    }
    \label{fig:pipeline}
\end{figure}

\section{Introduction}\label{sec:introduction}

Alzheimer’s disease (AD) is a progressive neurodegenerative disorder characterised by a long preclinical phase, heterogeneous clinical presentation, and complex interactions between molecular pathology, brain structure, and cognitive decline~\cite{safiri2024alzheimer}. Globally, more than 55 million people live with dementia, with AD accounting for the majority of cases, a number projected to nearly triple by 2050 as populations age~\cite{Nichols2022dementia}. The personal cost is severe: patients lose independence, memory, and identity over years, while caregivers shoulder an enormous unpaid burden, estimated at approximately 82 billion hours of informal care annually provided to people with dementia living at home worldwide~\cite{Wimo2018informalcare}. Health systems face mounting pressure, with dementia-related expenditure already exceeding \$1313.4 billion per year worldwide and rising~\cite{Wimo2023costs}. 

Despite this burden, therapeutic progress in AD has long been limited. However, the recent approval of several disease-modifying therapies marks an important transition from decades of largely unsuccessful drug development toward clinically available interventions~\cite{chengInnovationBreakthroughAlzheimers2026}. These advances are expected to stimulate the development of additional therapeutic targets and treatment strategies. Nevertheless, the clinical benefits observed to date remain modest, a limitation that has been attributed in part to the late stage at which treatment is initiated, after substantial neurodegeneration has already occurred~\cite{hunterAlzheimersDiseaseBiomarkers2025}. Consequently, improving disease detection, staging, and the prediction of disease severity may facilitate the identification of therapeutic windows during which interventions are most effective.

In parallel to methods development, central to earlier and more reliable diagnosis are multimodal data. Accurate characterisation of AD progression increasingly relies on the integration of diverse biological measurements that capture the multifaceted and heterogeneous nature of the disease. Key variables include demographic information such as age, sex, and years of education, genetic risk factors such as \textit{APOE}$\varepsilon$4 genotype, structural neuroimaging features derived from MRI, and molecular biomarkers of amyloid and tau pathology measurable via cerebrospinal fluid assays or positron emission tomography. 

Demographic and genetic factors shape baseline AD risk. Age remains an important non-genetic risk factor for late-onset AD, whereas \textit{APOE}$\varepsilon$4 genotype substantially increases lifetime risk. AD is also more prevalent in women, although the basis of this difference remains debated and may involve both longevity and interactions between sex and \textit{APOE}-associated risk~\cite{dessyDisentanglingEffectsSex2025}. Brain imaging data such as magnetic resonance imaging (MRI) data capture  structural neurodegeneration; they provide a key observable marker of disease progression~\cite{hunterAlzheimersDiseaseBiomarkers2025}. In particular, medial temporal lobe atrophy is among the earliest detectable changes, followed by involvement of parietal, frontal and cingulate cortices; the hippocampus, in particular, exhibits accelerated and non-linear morphological decline~\cite{frisoniClinicalUseStructural2010a, hunterAlzheimersDiseaseBiomarkers2025}. These structural alterations correlate with symptom severity and are widely used to track disease progression and treatment response. 

Although amyloid and tau pathology represent defining molecular hallmarks of AD, their direct quantification typically requires invasive or high-cost procedures, such as cerebrospinal fluid (CSF) sampling or positron emission tomography (PET) imaging. Consequently, disease characterisation often relies on heterogeneous and partially observed combinations of molecular, neuroimaging, genetic, demographic, and cognitive measurements. Because no single modality fully captures disease progression, integrating complementary sources of information has become increasingly important for studying AD heterogeneity and progression, particularly at early and prodromal stages~\cite{hunterAlzheimersDiseaseBiomarkers2025}. In practice, however, multimodal datasets remain inherently incomplete due to differences in acquisition protocols, study design, patient compliance, and follow-up schedules, resulting in substantial variability in modality availability across individuals and cohorts~\cite{vavekanandMultimodalMachineLearning2026}.

\begin{figure}[t!]
\includegraphics[clip, trim=0cm 0cm 0cm 0cm, width=\textwidth]{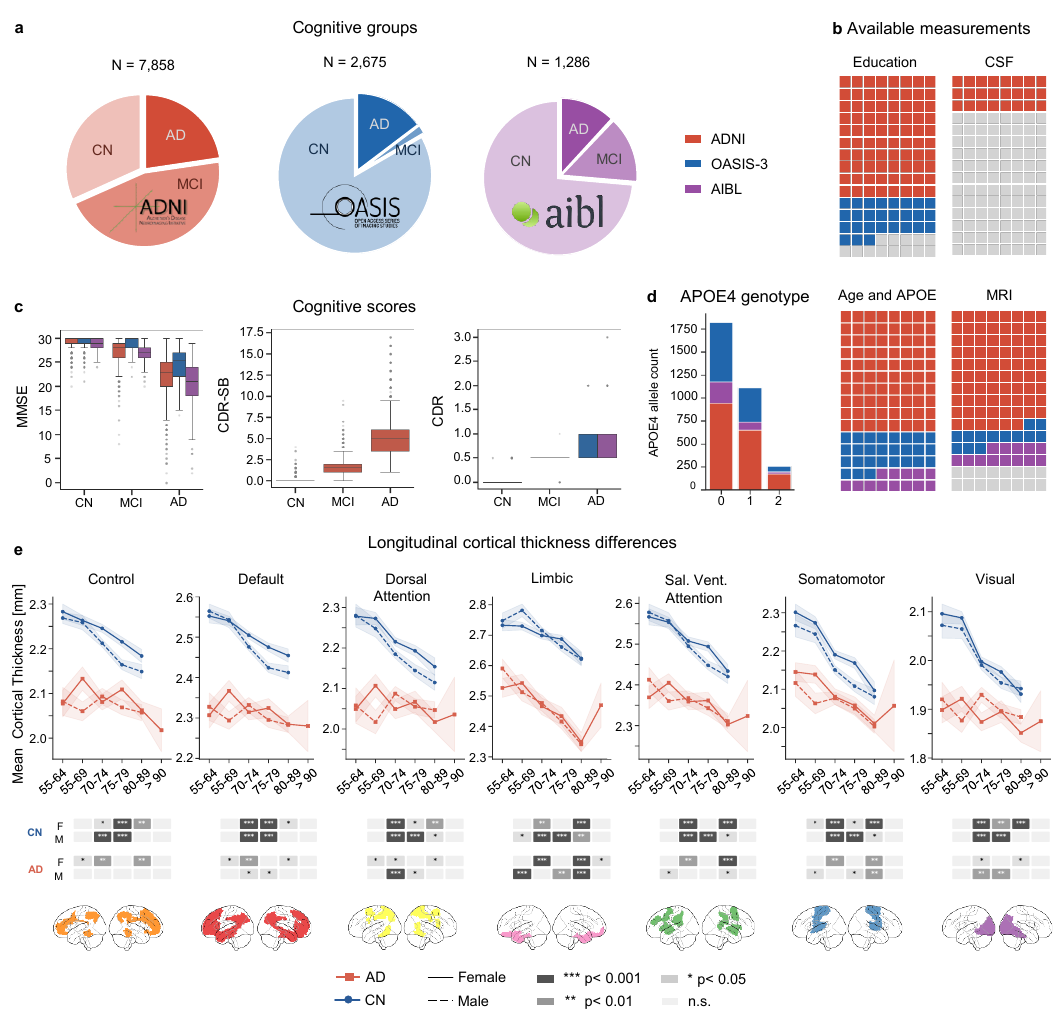}
\caption{\textbf{Dataset characteristics and cortical thickness analyses.}
Panels (a--d) are colour-coded by cohort: Alzheimer's Disease Neuroimaging Initiative (ADNI, red), Open Access Series of Imaging Studies-3 (OASIS-3, blue), and Australian Imaging, Biomarkers and Lifestyle Flagship Study of Ageing (AIBL, purple).
\textbf{(a)} Distribution of cognitively normal (CN), mild cognitive impairment (MCI), and Alzheimer's disease (AD).
\textbf{(b)} Availability of multimodal data across cohorts (60 participants per cell; gray: missing).
\textbf{(c)} Box-plots of cognitive scores: Mini-Mental State Examination (MMSE), Clinical Dementia Rating (CDR), and CDR Sum of Boxes (CDR-SB).
\textbf{(d)} Distribution of \textit{APOE}$\varepsilon$4 allele counts across cohorts.
\textbf{(e)} Mean cortical thickness trajectories as a function of age for CN (blue) and AD (red) participants, by sex and functional brain network. Adjacent age bins were compared using Welch's \textit{t}-tests with Benjamini--Hochberg false discovery rate correction. Shaded bands indicate 95\% confidence intervals; horizontal strips denote significant comparisons.
}
\label{fig:dataset_characteristics}
\end{figure}

This is not a peripheral inconvenience; rather, it is the crucial reality of AD research. The use of multimodal data for AD prediction, therefore, necessitates addressing the challenge of missing data. Traditionally, missing data in clinical machine learning has been predominately handled through explicit data imputation prior to model training, a strategy commonly referred to as \textit{impute-then-classify} for classification tasks and, broadly, \textit{impute-then-predict}. While this approach is compatible with standard algorithms, imputation  inevitably introduces synthetic measurements that can distort feature distributions, disrupt inter-feature dependencies, and compromise uncertainty estimation~\cite{laroseImpactMissingValues2019}. In diagnostic applications, these distortions are not merely technical inconveniences: overconfident or biased predictions can directly influence clinical decision-making, with potential downstream consequences for patients, including misdiagnosis or inappropriate treatment recommendations.

To mitigate these shortcomings, a growing body of work has begun to explore missing-aware models that operate directly on partially observed data, including modified decision trees and ensemble methods with missingness-aware splitting criteria~\cite{twalaGoodMethodsCoping2008}, linear models augmented with explicit missingness indicator variables~\cite{vannessMissingIndicatorMethod2023}, and neural network architectures that either incorporate masking mechanisms or explicitly modify the forward pass to exclude missing inputs contributions during computation~\cite{smiejaProcessingMissingData2018a, zengNeuralNetworksBased2023}. Broadly, these approaches fall into two categories: those that augment the input space with missingness indicators, and those that directly adapt model computation to exclude or down-weight missing contributions. 

Recently, attention-based deep learning models, particularly transformer architectures, have demonstrated considerable flexibility in handling tabular data, extending well beyond natural language processing~\cite{gorishniyRevisitingDeepLearning2021, huangTabTransformerTabularData2020a}. Leveraging self-attention and masked attention mechanisms, these models have been adapted to generate predictions directly from tabular datasets with missing values~\cite{carusoNotAnotherImputation2025}, removing the need for explicit imputation at the input stage. However, eliminating imputation addresses only how missing data are handled during representation learning; it does not by itself guarantee that the resulting predictions are reliable. Two further challenges therefore remain unresolved.

The first concerns the gap between discrimination and calibration. Most existing approaches are optimised and evaluated primarily on discriminative accuracy, while the reliability of individual-level probability estimates, a critical requirement for clinical decision support, is rarely assessed. 
Although the effect of imputation on uncertainty has been examined relative to complete-case analyses~\cite{laroseImpactMissingValues2019}, this question remains essentially unexplored in the harder setting when entire modalities, rather than scattered values, are missing. The second concerns the lack of modality-aware uncertainty. Existing uncertainty estimation methods typically treat all missing information as interchangeable, without accounting for how much a given absent modality would have contributed to the prediction. 

Here, we address these challenges within a unified framework called NITROGEN (No-imputation Inter-sample-attention TRansformer Oriented for GENeral datasets), which integrates Not ANother Imputation Method (NAIM)~\cite{carusoNotAnotherImputation2025}, a imputation-free method introducing masked self-attention, and SAINT, an inter-sample-attention transformer~\cite{somepalliSAINTImprovedNeural2021b}. 
NITROGEN is designed to handle incomplete multimodal clinical data by jointly capturing within-patient feature dependencies and inter-patient relationships, while explicitly accommodating partially missing observations through masked attention (Fig.~\ref{fig:pipeline} and~\ref{fig:model_architecture}). Unlike imputation-based pipelines, NITROGEN does not require missing values to be filled before training or inference. Instead, masked attention explicitly limits the contribution of missing observations during representation learning. At the same time, intersample attention enables the model to incorporate relational information across patients, which helps stabilise predictions in heterogeneous clinical cohorts with modality-level data loss. 

To evaluate the efficacy of the model, we first applied it to three predictive tasks: a binary disease classification, multi-class disease classification and continuous cognitive score prediction. We then assessed generalisation and the reproducibility of the framework via cross-cohort validation as well as cross-cohort and cross-method analyses across three independent datasets: the ADNI dataset ($N_{subjects}=1,776$), the OASIS-3 data ($N_{subjects}=1,078$) and AIBL ($N_{subjects}=1,286$), first applying the model trained on a part of the ADNI data on the held-out ADNI data, and subsequently to the latter two cohorts without further training or fine-tuning. Furthermore, we identified brain regions consistently selected across all three datasets, as well as those robustly implicated across all analytical methods and three cohorts. 

To address the reliability of individual-level probability estimates, we further examine the robustness  of individual-level confidence estimates under incomplete multimodal inputs. Specifically, we proposed an uncertainty adjustment, through which global estimates of modality importance are combined with subject-specific missingness patterns to increase uncertainty when informative modalities are absent. We designed the uncertainty adjustment because, intuitively, missing MRI, CSF, demographic, genetic, or cognitive information should not affect confidence equally: the absence of a modality that contributes strongly to prediction should lead to greater uncertainty regarding the final prediction than the absence of a less informative modality. Practically, this adjustment accounts not only for how much information is missing, but also for which type of information is missing. 

Collectively, by learning directly from incomplete, heterogeneous records rather than imputed ones, NITROGEN performs classification tasks and disease severity estimation across independently acquired, diverse cohorts while remaining calibrated and transparent about which information a given prediction is missing. This demonstrates that AD prediction models can remain trustworthy under missing values, missing modalities, and cohort heterogeneity---conditions under which conventional approaches typically degrade.

\section{Results}\label{sec:results}

In this section, we first present the results regarding three tasks: binary disease status classification, multi-class disease classification, and continuous cognitive score prediction. We then demonstrate the reliability and reproducibility of NITROGEN and derived neuroimaging features in the binary classification task as well as the impact of uncertainty adjustment techniques. 

To do that, we perform cross-cohort analysis using three independent datasets from the Alzheimer's Disease Neuroimaging Initiative (ADNI), the Open Access Series of Imaging Studies (OASIS-3) and the Australian Imaging, Biomarkers and Lifestyle (AIBL) study. We report predictive performance across binary classification, multi-class diagnostic classification, and continuous cognitive score regression tasks. 

To understand the potential drivers of model predictions, we perform permutation-based feature attribution analysis, identifying the top 30 most influential MRI regions on each dataset independently. We then examine the consistency of these rankings across all seven models and all three cohorts, retaining only regions that appear among the top attributed features across all methods and datasets. 

The cross-model and cross-cohort analysis identifies temporal pole cortical thickness as the most consistently attributed MRI feature, suggesting robustness of this finding to modelling choices and cohort characteristics. Furthermore, to assess whether this feature retains predictive value in isolation, we conduct two complementary sufficiency analyses in which feature attributions are recomputed on a held-out data partition, distinct from that used for the original attribution analysis, to prevent data leakage between attribution computation and sufficiency evaluation. Non-selected features are then progressively replaced with missing values, and model performance is evaluated on the remaining held-out subjects across all three cohorts. This allows us to quantify the extent to which the identified temporal pole feature, alone or in combination with a minimal set of non-imaging variables, supports diagnostic classification, independently of the remaining feature set.

Finally, we evaluate the proposed uncertainty adjustment by measuring the improvement in predictive accuracy obtained after excluding predictions with high estimated uncertainty, together with the corresponding reduction in the number of evaluated samples on the validation set.

\subsection{Dataset Characteristics and Cortical Thickness Analyses}

We first characterised the three cohorts included in this study. Figure~\ref{fig:dataset_characteristics} summarises cohort composition (Fig.~\ref{fig:dataset_characteristics}a), modality-specific missingness patterns (Fig.~\ref{fig:dataset_characteristics}b), cognitive score distributions (Fig.~\ref{fig:dataset_characteristics}c), and \textit{APOE}$\varepsilon$4 allele frequencies (Fig.~\ref{fig:dataset_characteristics}d), revealing substantial cohort heterogeneity across data modalities and datasets.

Because cortical thickness varies with age and sex across brain regions~\cite{vu2025residual}, we examined how these effects differ between diagnostic groups (cognitively normal (CN) vs.\ AD) and between sexes across age strata using Welch t-tests (Fig.~\ref{fig:dataset_characteristics}e). We found that AD participants exhibit significantly reduced cortical thickness relative to CN individuals across all functional brain networks and age groups, with this difference already present in the youngest stratum examined (55--64 years), suggesting that structural changes are present at the time of clinical diagnosis across the full age range studied. These findings support the use of MRI-derived cortical thickness as a discriminative feature for classification and regression tasks.

\subsection{Binary Disease Group Classification}

We evaluated binary diagnostic classification in two settings. First, we considered a classification task distinguishing AD from CN subjects. Second, we considered a broader impairment-detection task distinguishing CN subjects from subjects with either mild cognitive impairment (MCI) or AD.

For each task, the NITROGEN model was trained and evaluated on the ADNI cohort using a 60\%/20\%/20\% train-validation-test split. To assess generalisation under cohort shift, the ADNI-trained model was then applied, without further training or fine-tuning, to the independent OASIS-3 and AIBL cohorts. The same evaluation procedure was repeated for baseline models, including the original NAIM model, XGBoost, LightGBM, Random Forest, MA-Lasso, and MA-GBT. Classification metrics and model comparison results are reported in Table~\ref{tab:cn_ad_results_seeds}.

\begin{figure}[!htbp]
\centering
\includegraphics[width=.9\textwidth]{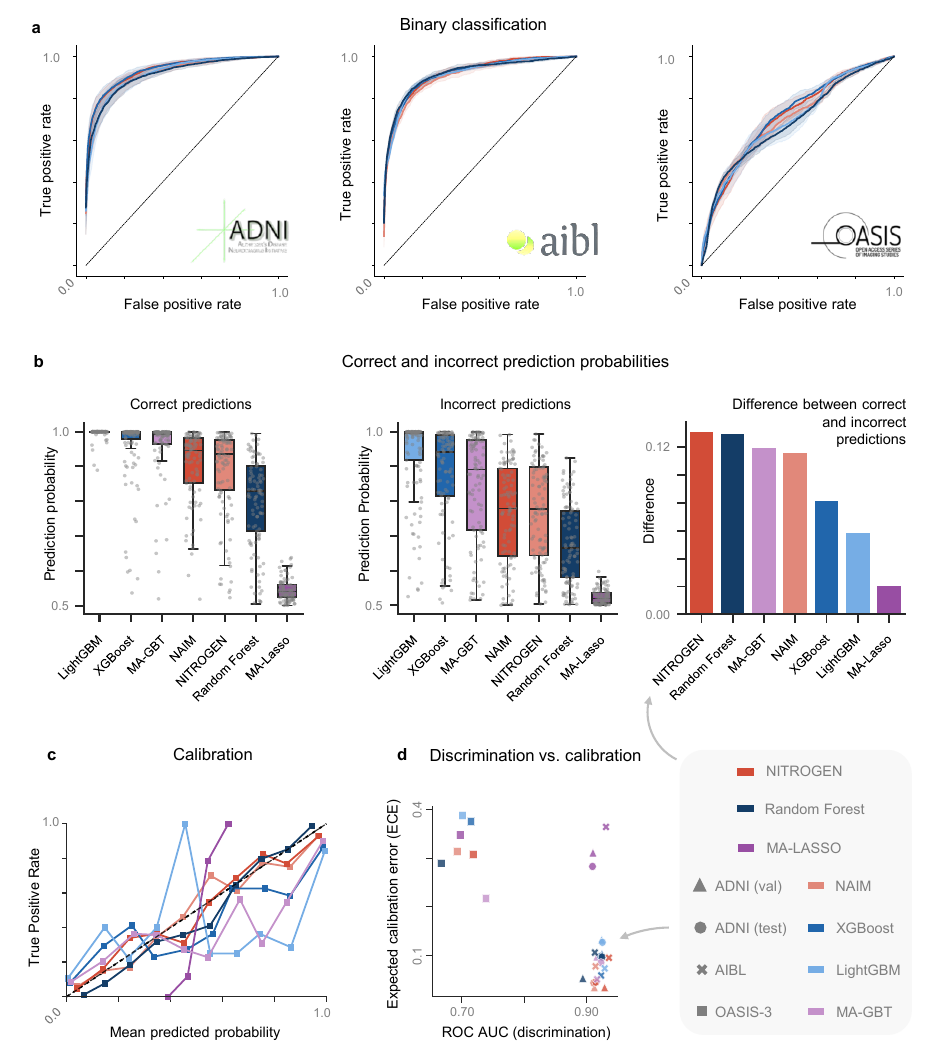}
\caption{\textbf{Classification performance, calibration, and discrimination-calibration 
trade-off across models and cohorts.}
\textbf{(a)} ROC curves for binary CN vs.\ AD classification on ADNI test set, AIBL, and OASIS-3. All models trained on ADNI. Curves represent the mean curve and shaded area the standard deviation across 10 random seeds
(seeds = [123, 42, 0, 7, 13, 21, 37, 55, 77, 99]).
\textbf{(b)} Prediction efficacy analysis on ADNI validation set (seed=123). Box-plots of predicted probabilities and 100 randomly drawn samples for correctly classified, incorrectly classified, and mean prediction probability gap bar plot. Prediction probability gap is the difference between mean probabilities for correct versus incorrect predictions.
\textbf{(c)} Reliability diagrams showing model calibration on ADNI validation set (seed=123). Mean predicted probability per bin ($x$-axis) versus observed fraction of AD cases ($y$-axis). Dashed diagonal indicates perfect calibration.
\textbf{(d)} Discrimination-calibration trade-off. $x$-axis: ROC-AUC; $y$-axis: Expected Calibration Error (ECE; lower is better). Evaluated on ADNI validation set, ADNI test set (seed=123), AIBL  data, and OASIS-3 data . Lower-right region: strong discrimination and good calibration.
}
\label{fig:classification_performance}
\end{figure}

Our results reveal several insights. For the CN vs.\ AD classification task, all models achieved strong discriminative performance on the ADNI test set, with AUC values ranging from 0.913 to 0.935 (Table~\ref{tab:cn_ad_results_seeds}, Fig.~\ref{fig:classification_performance}a). Performance transferred well to AIBL data (AUC: 0.901--0.918), likely reflecting the similar recruitment strategies and clinical protocols shared between ADNI and AIBL. Performance degraded on OASIS-3 data (AUC: 0.689--0.719), consistent with the marked distributional shift between these cohorts (Fig.~\ref{fig:dataset_characteristics}), as OASIS-3 specifically targets preclinical populations resulting in a higher proportion of CN participants and fewer AD cases. 
Across cohorts, attention-based models (NITROGEN, NAIM) and MA-Lasso consistently achieved higher sensitivity than tree-based ensemble methods, despite the latter achieving the highest AUC. On ADNI test data and AIBL data, MA-Lasso achieved the highest sensitivity among all models (0.833 and 0.761 respectively), followed closely by NITROGEN (0.832 and 0.697) and NAIM (0.808 and 0.695), while tree-based models showed comparatively reduced sensitivity, particularly on AIBL data (XGBoost: 0.610; LightGBM: 0.645). On OASIS-3 data, this pattern partially reversed, with XGBoost (0.770) and NAIM (0.763) achieving the highest sensitivity, followed closely by NITROGEN (0.751), while 
MA-Lasso (0.650) and Random Forest (0.636) performed comparatively worse. This pattern suggests that, with the exception of OASIS-3 data, tree-based ensemble methods are more conservative in predicting AD status, favouring specificity over sensitivity, whereas attention-based architectures maintain a more balanced operating point without explicit threshold tuning. Notably, Random Forest exhibited markedly elevated variance on OASIS-3 data across all metrics (accuracy standard deviation: 0.168; specificity standard deviation: 0.222), suggesting that its point estimates on this cohort are highly seed-dependent and should be interpreted with caution.

Second, we  evaluated all models on a more challenging binary classification task distinguishing cognitively normal participants from those with any level of cognitive impairment (MCI or AD). This setting reflects a relevant screening scenario but is inherently more difficult than CN vs.\ AD classification due to the heterogeneous and transitional nature of MCI, which exhibits overlapping characteristics with both CN and AD along the disease continuum~\cite{weiMappingHeterogeneousBrain2026, knopmanAlzheimerDisease2021}. Our results show that it is possible to generally distinguish cognitively normal participants from those with any level of cognitive impairment. As expected, performance decreased across all models and cohorts relative to the CN vs.\ AD task, with ADNI test data AUC ranging from 0.748 to 0.802 and OASIS-3 data AUC ranging from 0.639 to 0.700 (Table~\ref{tab:dementia_results}). This decline likely reflects both cross-cohort differences in acquisition protocols and the increased diagnostic ambiguity introduced by MCI, which represents an intermediate stage of cognitive decline with variable clinical presentation.

Third, beyond discrimination, we assessed model calibration, meaning the alignment between predicted class probabilities and observed outcome frequencies, as calibration reveals differences in model reliability not captured by AUC alone (Fig.~\ref{fig:classification_performance}b, c, d). Among tree-based models, LightGBM, MA-GBT and XGBoost achieved the best competitive discriminative performance but exhibited some overconfidence as shown by the poor alignment with identity line on the calibration plot (Fig.~\ref{fig:classification_performance}c) and on OASIS-3 and ADNI test sets with slightly higher Expected Calibration Error (ECE) values (Fig.~\ref{fig:classification_performance}d). In contrast, NAIM and NITROGEN produced better-calibrated probabilities across datasets, with NITROGEN achieving the largest prediction probability gap between correct and incorrect predictions on the validation set, followed by Random Forest, indicating superior uncertainty awareness (Fig.~\ref{fig:classification_performance}d). Most models maintained reasonable calibration on the ADNI and AIBL sets; however, calibration degraded on OASIS-3 for all models. These results suggest that whilst NITROGEN's sensitivity limitations at the default decision threshold merit consideration, its superior calibration, argue for its utility as a candidate for reliable probabilistic prediction in heterogeneous clinical settings. To further improve uncertainty estimates in settings with incomplete data, in Section~\ref{sec:uncertainty}, we explored entropy-based uncertainty quantification adjusted for modality-level missingness.

\subsection{Multi-class Diagnostic Classification}

Following the binary classification analyses, we evaluated model performance in a three-class diagnostic task distinguishing CN, MCI, and AD participants. This setting provides a relevant yet more challenging problem, as models must simultaneously differentiate normal cognition, mild cognitive impairment, and AD. We note that these categories correspond to clinician-assigned diagnoses rather than biologically defined disease stages. Global and per-class performance metrics are reported in Tables~\ref{tab:multiclass_global_metrics} and~\ref{tab:multiclass_perclass_metrics}.

Per-class analysis revealed interpretable heterogeneity in class-wise performance (Table~\ref{tab:multiclass_perclass_metrics}). Across all models and cohorts, AD discrimination was consistently the strongest task: per-class AUC on ADNI test ranged from 0.830–0.863, and NITROGEN achieved the highest AD AUC of any model on both ADNI test (0.863) and the fully held-out AIBL cohort (0.907), indicating that the captured signal associated with established disease generalises robustly even under domain shift. CN discrimination was likewise stable (0.744–0.799 on ADNI test), reflecting a consistent separation between cognitively normal individuals and clinical syndromes.

MCI was noticeably harder to discriminate (AUC 0.547–0.654 on ADNI test, narrowing further on OASIS-3 to 0.361–0.441 across models), a pattern consistent across every method we evaluated, including the tree-based ensembles. This is in line with MCI's established clinical status as an intermediate, heterogeneous category rather than a discrete diagnostic entity, and the reduced MCI sample size in OASIS-3 likely compounds this further. The convergence of all five models on this same difficulty pattern suggests it reflects a property of the diagnostic task itself rather than a limitation specific to any one architecture.

At the global level (Table~\ref{tab:multiclass_global_metrics}), tree-based ensembles achieved the highest macro AUC on ADNI test (0.762--0.771), with NITROGEN at 0.752 and  NAIM at 0.707. Under genuine domain shift to AIBL, NAIM improved slightly (0.707 to 0.729), while NITROGEN's macro AUC dropped modestly (0.752 to 0.719), smaller than XGBoost's decline (0.046). On OASIS-3, LightGBM and NAIM performed strongest (0.594 and 0.596), while NITROGEN and XGBoost ranked lower (0.583 and 0.574). These results demonstrate that in-distribution accuracy does not predict cross-cohort robustness. 

Taken together, these results indicate that all models capture AD-related signal, and that NITROGEN in particular combines competitive in-distribution discrimination with stable transfer to external cohorts, while MCI detection remains the principal opportunity for improvement across the field, consistent with its recognised status as the most clinically ambiguous diagnostic category in this three-class setting.

\subsection{Cognitive Score Prediction}

\begin{figure}[tp]
\centering
\includegraphics[clip, trim=0cm 0cm 0cm 0cm, width=1\textwidth]{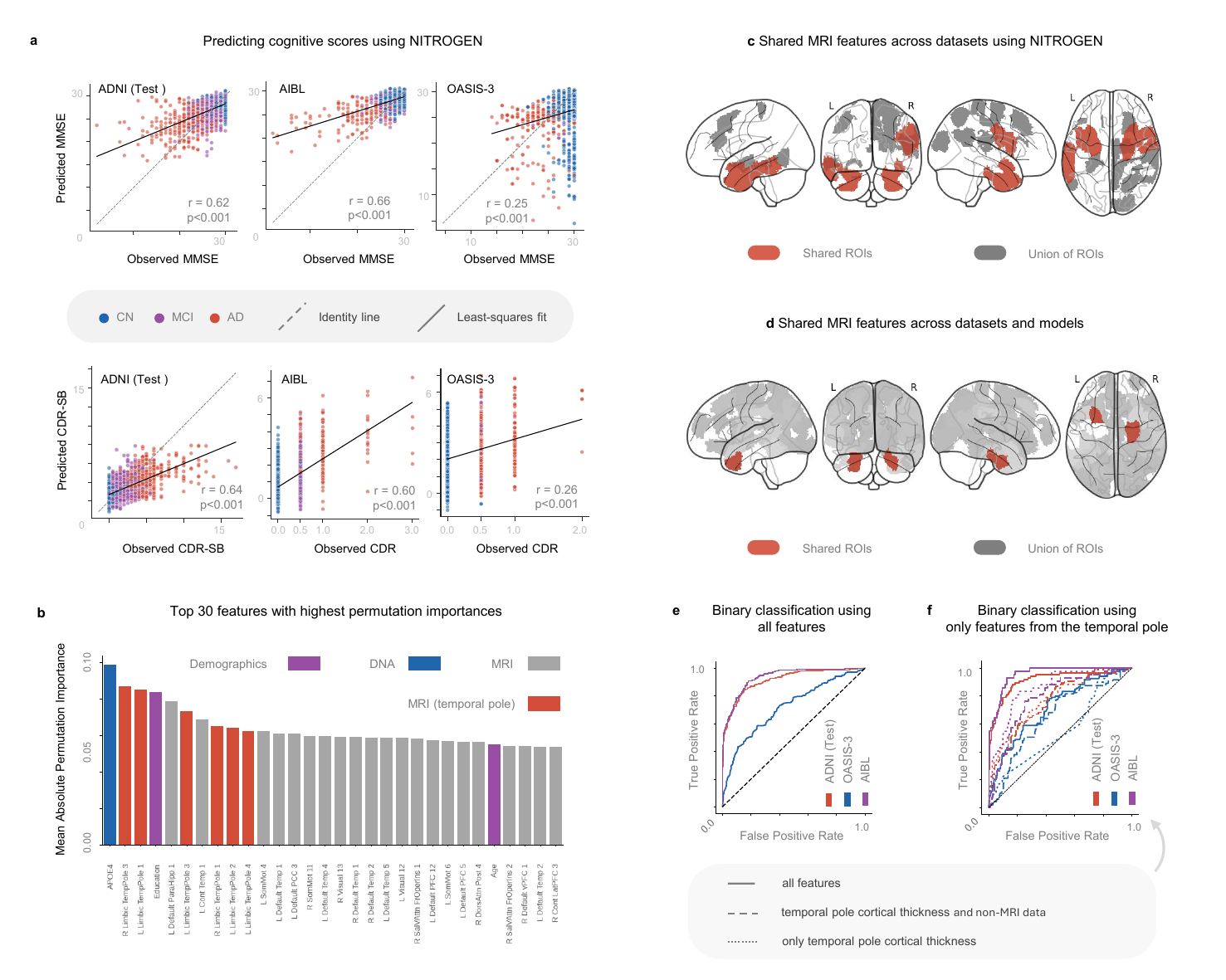}
\caption{\textbf{NITROGEN performance and feature attribution analysis.}
\textbf{(a)} Predicted vs. clinician-rated cognitive scores on ADNI, AIBL, and OASIS-3. Top: MMSE. Bottom: CDR-SB (ADNI test set) or CDR (OASIS-3 and AIBL data). Pearson $r$ (R) and $p$-values (p) shown.
\textbf{(b)} Top 30 features by mean absolute permutation importance on ADNI test set. 
\textbf{(c)} Top 30 MRI ROI features across datasets (ADNI test, AIBL and OASIS-3).
\textbf{(d)} Consensus MRI ROIs identified in bilateral temporal pole across cohorts and models.
\textbf{(e)} Classification accuracy on ADNI test set, AIBL and OASIS-3 data.
\textbf{(f)} Test of sufficiency using data from temporal pole only. ROC curves compare all features, consensus features from temporal pole plus non-imaging features, and features from temporal pole only.
}
\label{fig:feature_attributions}
\end{figure}

Next, we sought to predict continuous cognitive scores obtained during neuropsychological assessments that measure a more fine-grained characterisation of cognitive decline. Here, we consider the prediction of the Mini Mental State Examination (MMSE) scores and Clinical Dementia Rating (CDR) score (or their Sum of Boxes (CDR-SB) version). We aim to predict these two scores simultaneously in an end-to-end fashion using a single model. Full results are reported in Supplementary Table~\ref{tab:regression_results_supp} and NITROGEN's regression results are plotted in Figure~\ref{fig:feature_attributions}b.

On the ADNI test set, all models showed moderate-to-strong Pearson correlation with clinician-rated MMSE scores, with NAIM achieving the strongest correlation ($r$ = 0.627, $p = 1.58 \times 10^{-174}$), closely followed by NITROGEN ($r$ = 0.621, $p = 1.58 \times 10^{-170}$) and Random Forest ($r$ = 0.610, $p = 1.58 \times 10^{-163}$). On the fully external AIBL cohort, correlations for MMSE prediction improved for every model relative to ADNI test, with Random Forest and NITROGEN performing best ($r$ = 0.666 and 0.664, respectively), indicating that the ordinal relationship between predicted and true MMSE scores generalise to this cohort. Performance on OASIS-3 was weaker for all models ($r$ between 0.245 and 0.303), consistent with the markedly different case mix in this cohort (only 49 MCI and 398 AD cases, versus 184 and 155 in AIBL) rather than a cohort-specific failure of any one method.

For CDR-SB prediction, NITROGEN and NAIM were again the best-performing models on the ADNI test set ($r$ = 0.641 and 0.635, respectively), followed by Random Forest ($r$ = 0.633) and XGBoost ($r$ = 0.604). Due to the unavailability of CDR-SB in AIBL and OASIS-3, performance was assessed against the global CDR score as a surrogate target. $R^2$ values for this surrogate comparison were negative for all models in both external cohorts, reflecting the low variance of the global CDR scale in samples that are predominantly cognitively normal; we therefore focus on Pearson correlation, which remained informative, as the primary cross-cohort metric. On AIBL, Random Forest achieved the highest correlation ($r$ = 0.656), followed closely by NITROGEN ($r$ = 0.603). On OASIS-3, correlations were more attenuated but remained significant for all models ($r$ = 0.247--0.284), likely reflecting the coarse, low-variance nature of the global CDR scale in a cohort with very few non-CN cases.

Notably, NITROGEN's correlations on the ADNI test set and the two fully external cohorts were consistently close to one another for both MMSE and CDR-SB (e.g., $r$ = 0.621 on ADNI test versus $r$ = 0.664 on AIBL for MMSE), in contrast to the tree-based ensembles, whose strong in-distribution fit did not always translate into a comparable advantage once evaluated externally. This pattern echoes the classification results above and suggests that NITROGEN's representation of disease severity, while not always the single best performer on any one held-out set, transfers consistently across cohorts.

Taken together, these results indicate that agreement between predicted and observed clinical severity scores, as measured by Pearson correlation, was largely preserved across cohorts despite the pronounced scale and case-mix differences in OASIS-3 and AIBL. NITROGEN achieved correlations that were competitive with those of the other methods and, on AIBL, ranked among the best-performing models. Moreover, its performance changed comparatively little between in-distribution and external evaluation, suggesting greater robustness to cohort differences that may be beneficial for deployment in populations that differ from the training data.

\subsection{Model Explanation via Feature Attribution Analysis}

To better understand the potential biological underpinnings of the model predictions, we analysed feature attributions and modality contributions using permutation importance. We focus on feature attributions in the binary CN versus AD classification setting, as this contrast provided the clearest discrimination across cohorts. We quantify feature importance by measuring the drop in model performance induced by randomly permuting input feature values. 

The top 30 features with the largest impact on NITROGEN predictions on the ADNI test set are shown in Figure~\ref{fig:feature_attributions}b, with APOE$\varepsilon$4 identified as the most important feature, followed by MRI-derived cortical thickness features from bilateral temporal poles, other temporal and limbic regions (parahippocampal cortex, default mode network areas), and demographic variables including education and age.

When the same permutation-based attribution analysis was applied independently to held-out test sets from OASIS-3 and AIBL, the top-ranked features showed consistencies despite variations. Features that appeared in the top 30 across all three cohorts for NITROGEN are displayed in Figure~\ref{fig:feature_attributions}c and include: bilateral temporal poles (Left (L)/Right (R) Limbic Temporal Pole 1/3 and L Temporal Pole 2), parahippocampal cortex (L Default ParaHipp 1), additional temporal and default mode regions (L/R Default Temporal 1/2), somatomotor cortex (L SomMot 4), visual cortex (L Vis 12), a control network temporal region (L Cont Temp 1), APOE$\varepsilon$4, and age.


The prominence of temporal cortical regions is consistent with the well-established spatial progression of AD pathology. Amyloid-$\beta$ deposition begins years to decades before symptom onset and preferentially accumulates within regions of the default mode network, including lateral temporal, posterior cingulate, and medial prefrontal cortices~\cite{knopmanAlzheimerDisease2021}. In contrast, tau pathology is thought to originate within medial temporal structures and subsequently propagate to temporal and parietal association cortices before involving the broader neocortex, a process that more closely parallels the emergence of cognitive symptoms~\cite{frisoniClinicalUseStructural2010a,knopmanAlzheimerDisease2021}.

Our results extend these observations by showing that temporal and limbic features were consistently prioritised across independent cohorts and diverse machine learning models, suggesting that their predictive value is not specific to a single dataset or modelling framework. Although their relative rankings varied across cohorts, likely reflecting differences in participant characteristics and acquisition protocols, the cross-cohort and cross-method analyses (Section~\ref{Subsec:cross-cohort-and-method}) identified a subset of features that remained robustly selected despite these sources of variability.

\subsection{Consistently Attributed MRI Regions Across Cohorts and Methods}
\label{Subsec:cross-cohort-and-method}

To assess whether feature attributions yield consistent patterns despite cohort heterogeneity and methodological differences, we computed the intersection of top 30 features across all seven models and all three cohorts. This analysis identified a minimal core set of four features consistently selected across all 21 model–cohort combinations: bilateral temporal pole cortical thickness (L Limbic Temporal Pole 1, R Limbic Temporal Pole 3), APOE$\varepsilon$4, and age (Fig.~\ref{fig:feature_attributions}e). The persistence of such a small set across all combinations suggests these features consistently carry important signal for AD status, with limited sensitivity to methodological choice or to distributional differences between cohorts. The association between temporal pole volume loss and cognitive decline is relatively well established~\cite{frisoniClinicalUseStructural2010a}, as is its correlation with plasma tau in AD/MCI participants in ADNI~\cite{advancingresearchandtreatmentforfrontotemporallobardegenerationartflinvestigatorsDiagnosticValuePlasma2020}.  That the temporal poles carry consistent signal despite differences across datasets and methodological frameworks suggests that they may serve as consistently informative structural features in the context of AD.

\subsection{Sufficiency Analysis of Consistently Attributed Regions}

To evaluate whether the consistently attributed temporal pole region retained predictive value in isolation, we performed a sufficiency analysis using a two-stage evaluation protocol. Note that the feature selection performed here differs from the previous section to further guard against data leakage between attribution computation and sufficiency evaluation, attributions were recomputed on a restricted, non-overlapping data partition and extended to the full multimodal feature space rather than cortical-thickness features alone, yielding a distinct, stricter feature set. First, feature attributions were re-calculated with 70\% of the held-out test sets (ADNI test, OASIS-3, and AIBL). Second, performance was evaluated on the remaining 30\% of each test set using previously trained models, thereby preventing data leakage. Applying this stricter, multimodal feature selection identified two robustly attributed features: APOE$\varepsilon$4 and left limbic temporal pole cortical thickness (L Limbic Temporal Pole 3). We evaluated performance under two feature-restricted scenarios: (i) MRI, where the identified temporal pole MRI feature was retained but all other MRI features were replaced with missing values, while demographics, CSF biomarkers, and genotyping data remained available, and (ii) feature-masked (ALL), where only the temporal pole measure was retained and all other features were masked. Results are summarised in Table~\ref{tab:sufficiency_analysis}.

Retaining the temporal pole measure together with non-imaging variables (MRI setting) resulted in respectable discriminative performance. Across models, AUC values ranged from 0.76--0.80 on the ADNI test set (versus 0.90--0.92 under the full feature configuration), 0.59--0.69 on OASIS-3, and 0.57--0.81 on AIBL, with several models (e.g., NAIM, MA-Lasso) approaching or exceeding 0.80 on AIBL. That a single cortical region, combined only with routinely collected demographic and genetic variables, recovers the large majority of full-model discriminative performance underscores the value of the temporal pole as a parsimonious structural biomarker for AD.

Further, we investigated whether it is possible to classify patients using only the temporal pole feature (Table~\ref{tab:sufficiency_analysis}). Although AUC values suggested ranking-level discrimination (0.76–-0.77 on ADNI test, 0.83–-0.87 on AIBL), this did not translate into clinically usable classification . When required to make binary decisions, most models defaulted to constant predictions. NAIM Default, NITROGEN, and Random Forest converged to constant predictions of class AD, while LightGBM and MA-GBT converged to constant CN predictions. XGBoost produced random balanced accuracy (0.53 on ADNI test, 0.40 on AIBL, 0.28 on OASIS-3) without exhibiting constant predictions. Random Forest was the clear exception to the AUC pattern above, with AUC collapsing to 0.41-–0.57, consistent with its tree-splitting behaviour degenerating under a near-constant single-feature input. MA-Lasso was unique: it produced probability distributions tightly clustered around 0.50 (reflecting model uncertainty), yet still achieved non-random accuracy, most notably a balanced accuracy of 0.81 on AIBL. However, this performance must be contextualised: AIBL exhibits severe class imbalance (87.6\% CN, 12.4\% AD), and inspection of MA-Lasso's predictions reveals that its 0.81 balanced accuracy reflects assignment of different labels despite genuine probability uncertainty (probs~0.50), rather than confidence-driven discrimination. Constant-prediction models on AIBL appear deceptively high in AUC (0.80+) due to class imbalance, but achieve balanced accuracy of 0.50 and are unusable. In other words, these results suggest that most classifiers, lacking complementary features to anchor a calibrated decision boundary, default to a single constant label at the standard 0.5 threshold rather than exploiting the genuine ranking ability the temporal pole feature provides on its own---underscoring that AUC and balanced accuracy must be interpreted jointly when characterising single-feature sufficiency.

Taken together, these analyses indicate a feature-richness gradient with two distinct findings. First, temporal pole cortical thickness alone seems to carry a reproducible ranking signal for AD status. Second, this signal is not, by itself, sufficient to support stable decision-level classification, which requires the additional context provided by demographics, genotype, and complementary MRI regions. Comprehensive multimodal assessment remains preferable when all modalities are available, providing the highest diagnostic confidence. Nevertheless, the robustness of the temporal pole signal and its contribution alongside minimal non-imaging information indicate that relevant predictions can be maintained under reduced modality availability.

\subsection{Uncertainty adjustment}
\label{sec:uncertainty}

\begin{figure}[tp]
\includegraphics[clip, trim=0cm 0cm 0cm 0cm, width=1\textwidth]{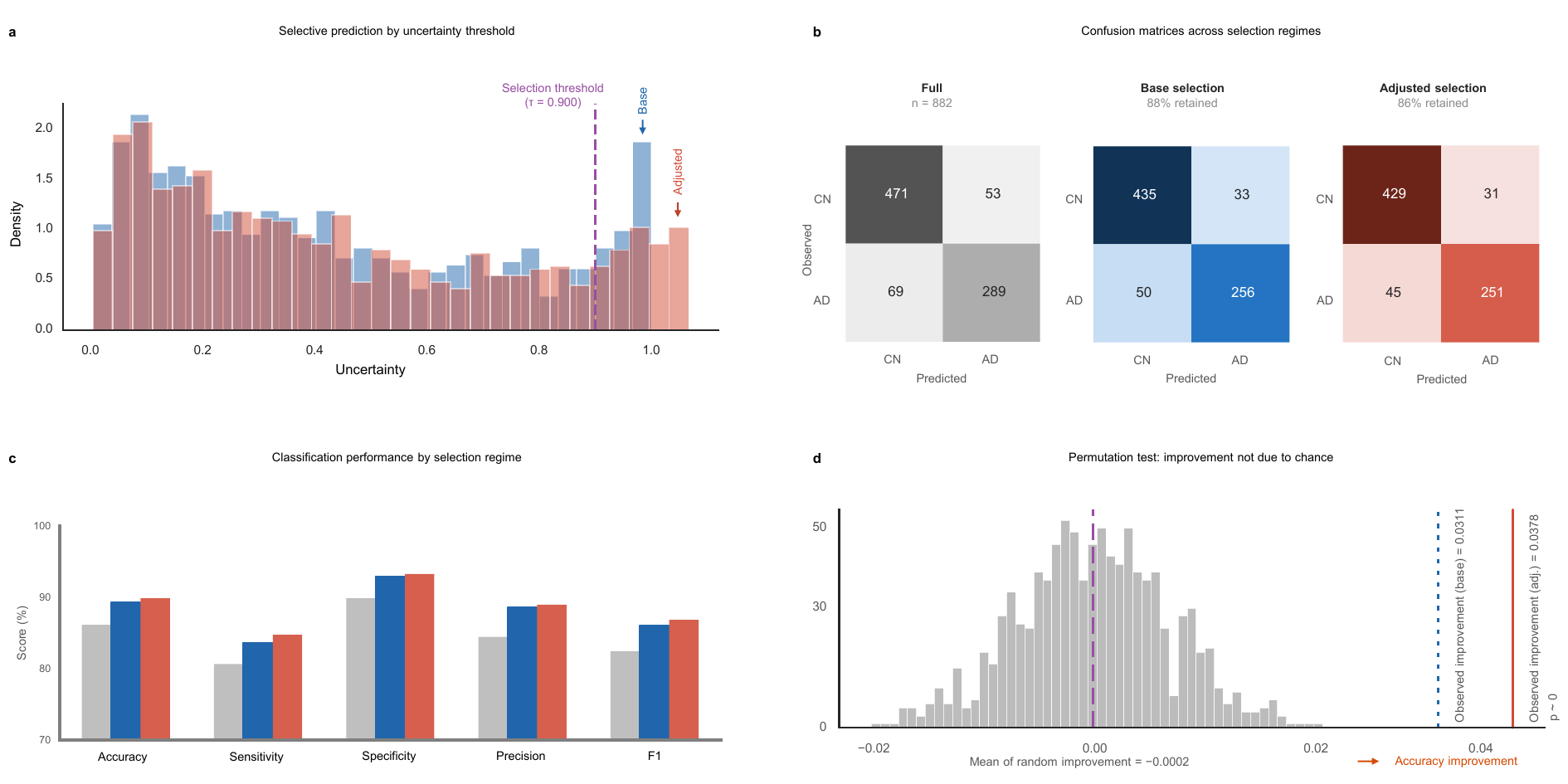}
\caption{\textbf{Missingness-aware uncertainty adjustment improves selective prediction reliability.}
\textbf{(a)} Distribution of base (blue) and missingness-adjusted (red) uncertainty scores on the ADNI validation set ($N=882$). The selection threshold ($\tau=0.900$) is shown as a solid vertical line; samples with uncertainty above this threshold are excluded from selective prediction.
\textbf{(b)} Confusion matrices for CN vs. AD classification under the full validation set (Full, $N=882$), the subset retained under base-uncertainty thresholding (Base selection, 88\% retained), and the subset retained under missingness-adjusted uncertainty thresholding (Adjusted selection, 86\% retained).
\textbf{(c)} Classification performance (accuracy, sensitivity, specificity, precision, and F1) for the full validation set and for the base- and adjusted-selection subsets.
\textbf{(d)} Permutation test assessing whether the accuracy improvement achieved through adjusted-uncertainty selection exceeds chance. Null distribution (gray bars, 1,000 permutations) of accuracy improvement under random sample exclusion at matched coverage; dashed line indicates the mean of the null distribution, dash-dotted line indicates the observed improvement under base-uncertainty selection (+0.0311), and solid line indicates the observed improvement under adjusted-uncertainty selection (+0.0378, $p<0.001$).
}
\label{fig:uncertainty}
\end{figure}

Although softmax-derived probabilities provide a default measure of confidence, they conflate two distinct sources of uncertainty: uncertainty arising from the inherent difficulty of a case, and uncertainty arising from missing diagnostic information. A model evaluated on a patient with all modalities present and a model evaluated on a patient missing a highly informative modality (e.g., CSF biomarkers) may report similar softmax confidence despite having access to substantially different amounts of evidence. This is undesirable in a clinical setting, where the reason for low confidence---case difficulty versus missing data---has direct implications for what action should follow (e.g., obtaining the missing test versus seeking a specialist opinion). Motivated by this gap, we sought to adjust predictive uncertainty explicitly as a function of which modalities are absent for a given subject, rather than relying on raw model confidence alone.

In clinical decision-support systems, reliable estimates of prediction trustworthiness are as important as predictive accuracy, particularly in multimodal settings where data completeness varies across subjects. Shannon entropy computed from predicted class probabilities provides a simple sample-level measure of uncertainty. When critical diagnostic modalities are unavailable, a model should express higher uncertainty even if its raw output probabilities appear confident. As such, we inflated these sample entropy values based on the amount of important missing features and an inflation factor $\beta$. Inflated uncertainty values can then be thresholded; the logic is that if uncertainty is too high, it is better to withhold a prediction until more assessments are made. We fix the uncertainty threshold at $\tau$=0.900 for selective prediction (reject option). The scaling factor $\beta$ is selected by maximising the Spearman rank correlation between adjusted uncertainty and prediction errors, weighted by the retention gap (the difference in accuracy between retained and rejected samples). Retained samples are those whose adjusted uncertainty falls below $\tau$ and are therefore used for prediction, while rejected samples exceed the threshold and are abstained from (i.e., prediction is not trustworthy). We present the results in Fig.~\ref{fig:uncertainty}.

On the validation set ($N = 882$), base Shannon entropy averaged $0.428 \pm 0.309$. After adjustment ($\beta = 0.400$), entropy increased to $0.446 \pm 0.323$. Without selective abstention, predictions on the full dataset achieved accuracy 0.862 and 
F1 0.826. Applying uncertainty thresholding at $\tau = 0.900$, the adjusted entropy outperformed raw entropy-based thresholding: adjusted uncertainty retained 85.7\% of samples (756/882) with accuracy 0.900 and F1 0.869, compared to raw entropy which retained 87.8\% (774/882) with accuracy 0.893 and F1 0.861. Beyond this comparison, the adjusted uncertainty scores correlated significantly with prediction errors (Spearman $\rho$ = 0.348, $p< 10^{-26}$), and rejected samples showed markedly lower accuracy (0.635) than retained samples, indicating that the adjustment has identified a genuinely less reliable subset of predictions rather than discarding samples arbitrarily.

The missingness-aware adjustment offers measurable, if modest, improvement over raw entropy filtering on the validation set (+0.007 in accuracy) at the cost of a 2.0\% reduction in coverage. This increment is small in absolute terms but consistent in direction across the metrics examined, and it is achieved on top of a selective-prediction mechanism that is already effective at isolating unreliable predictions, as shown by the retained-versus-rejected accuracy gap above. The marginal benefit is expected given the limited missingness patterns in ADNI, where only CSF biomarkers are systematically absent (70.1\% missing) while other modalities remain near-complete. The adjustment may become more impactful in cohorts with heterogeneous or multi-modality missingness patterns, where importance-weighted scaling can more effectively capture the relative contribution of unavailable features to diagnostic uncertainty.

\section{Discussion}\label{sec:discussion}

In this paper, we presented NITROGEN, an attention-based architecture, for the joint tasks of performing AD diagnosis and cognitive score prediction from incomplete multimodal clinical data. To evaluate its generalizability and reproducibility, we applied NITROGEN across data from three independent cohorts with heterogeneous characteristics.

Computationally, our results demonstrated that NITROGEN conferred meaningful advantages in reliability and cross-cohort robustness over standard tree ensemble methods while achieving competitive classification and prediction performance. Additionally, our results highlighted important differences in model reliability that are not captured by AUC alone. Tree-based ensemble methods, particularly LightGBM, achieved the highest discriminative performance on ADNI but exhibited systematic overconfidence and poor probability calibration, most pronounced under domain shift,  a setting where inflated confidence on incorrect predictions is particularly consequential for clinical decision-making. In contrast, NITROGEN maintained well-calibrated predictions across cohorts, as reflected in the lowest ECE on OASIS-3 and a larger confidence gap between correct and incorrect predictions, indicating meaningful uncertainty awareness rather than uniform overconfidence. These findings reinforce the argument that clinical machine learning benchmarks should systematically evaluate models along multiple dimensions,  discrimination, calibration, and cross-cohort generalisation, particularly for decision support applications where reliability is no less critical than predictive accuracy~\cite{vavekanandMultimodalMachineLearning2026}.

Recent advances in artificial intelligence have substantially improved the performance of automated AD classification using MRI data. Specialised three-dimensional convolutional architectures trained on large imaging cohorts have reported near-ceiling performance for binary CN versus AD discrimination under controlled experimental settings, with accuracies approaching 99\% in some studies~\cite{shuklaAlzConvNetsClassificationAlzheimer2023}. More recently, MRI foundation models pre-trained on large collections of brain scans have demonstrated improved data efficiency and transferability, achieving strong performance even with limited downstream training data~\cite{takGeneralizableFoundationModel2026}. These advances highlight the substantial diagnostic information contained within structural MRI alone and demonstrate the capabilities of modern imaging-based deep learning models. However, such approaches address a different problem setting from the one considered here. MRI foundation models and specialised three-dimensional convolutional architectures typically operate on a single, highly informative modality and assume the availability of complete imaging data. By contrast, our objective was to model heterogeneous clinical datasets composed of demographic, genetic, biomarker, and neuroimaging variables that are frequently incomplete and unevenly sampled across individuals. Although this tabular multimodal setting may not achieve the near-perfect classification performance reported for dedicated MRI models, it addresses a broader class of prediction problems encountered in clinical research, where multiple data sources must be integrated despite substantial missingness. Rather than competing directly with imaging foundation models, our results suggest that these approaches may ultimately be complementary. Future work may leverage powerful modality-specific representations, such as pre-trained MRI encoders, while preserving downstream missingness-aware integration and uncertainty modelling. Such a modular strategy may combine the representational strength of foundation models with the flexibility required for heterogeneous clinical data.

Moreover, predicting continuous cognitive outcomes proved more challenging than binary classification across all models. This reduced external generalisation likely reflects the inherent difficulty of cross-cohort score prediction in the absence of harmonised assessment protocols, compounded by a scale mismatch between CDR-SB, the training target, and the global CDR used as observed scores in OASIS-3. Furthermore, predicted values exhibited a compressed dynamic range relative to the clinician rated scores, with a systematic tendency toward regression to the population mean, a well-characterised phenomenon in neural network regression. Taken together, these observations suggest that targeted methodological refinements, such as \textit{post-hoc} output calibration, for example, that of temperature scaling, distributional loss functions, or architectural modifications, may be useful to better capture the full variability of cognitive outcomes.

Scientifically, cross-cohort and cross-method validation revealed key brain regions consistently identified across datasets and machine learning methods. Notably, cortical thickness in the bilateral temporal pole emerged as the most consistently implicated MRI region across models and cohorts. While hippocampal atrophy is classically considered the hallmark structural signature of AD, the prominence of temporal pole thinning is consistent with evidence that cortical thickness loss in this area correlates with cognitive decline and plasma tau levels in both MCI and AD populations~\cite{frisoniClinicalUseStructural2010a, 
advancingresearchandtreatmentforfrontotemporallobardegenerationartflinvestigatorsDiagnosticValuePlasma2020}. As a core node of the limbic network, the temporal pole subserves semantic memory and social cognition and undergoes early atrophy in both AD and frontotemporal dementia; its predictive relevance may therefore extend across the full diagnostic spectrum. However, sufficiency analysis indicated that robust AD discrimination arises not from temporal pole atrophy in isolation, but from its integration with complementary clinical, genetic, biomarker, and imaging information.

Beyond convergent attribution evidence, our results revealed a clear dissociation in generalisation performance between AIBL and OASIS-3. Generalisation to AIBL was consistent with the shared recruitment strategies and clinical protocols between ADNI and AIBL. By contrast, the marked performance degradation on OASIS-3 when applying ADNI-trained models is consistent with known differences between the two cohorts. Specifically, unlike ADNI and AIBL, OASIS-3 enrols preclinical populations, comprising a higher proportion of cognitively normal participants, fewer AD patients, and no MCI subjects~\cite{lamontagneOASIS3LongitudinalNeuroimaging2019a}. Nevertheless, the sensitivity improvements observed for NITROGEN on both external cohorts suggest that cross-sample attention may confer a form of implicit regularisation that enhances robustness to distributional shift, even in the absence of explicit domain adaptation.

Finally, we introduced a missingness-aware adjustment of Shannon entropy by scaling prediction uncertainty according to modality-level importance, in order to better reflect confidence under incomplete multimodal inputs and  while this approach yielded only marginal improvements in accuracy over standard uncertainty filtering, it provides a simple mechanism for more reliable uncertainty estimation in decision-support settings where robustness to missing modalities is essential.

We acknowledge several limitations of our study. 
First, all models were trained exclusively on ADNI. We do so because ADNI is a well-characterised research cohort with relatively uniform acquisition protocols; training the model only on one cohort, despite evaluating externally on two other datasets, however, may limit generalizability of the trained model to clinical populations with greater demographic and technical heterogeneity. Consistent with this, we observe substantial performance degradation on OASIS-3, likely attributable to distributional shift between cohorts. Future work may consider transfer learning or domain adaptation strategies to improve robustness across datasets, sites and scanners. 
Second, prediction of cognitive scores remained moderate across all architectures, potentially reflecting the inherent variability of psychometric instruments and the complex, non-linear relationship between neuroimaging features and cognitive outcomes~\cite{mcdonoughLinearNonlinearRelationships2020}. In parallel, the training instability of NITROGEN on MMSE and CDR-SB prediction suggests that task-specific modifications, such as auxiliary losses or score-stratified sampling, may be useful  to improve continuous outcome prediction. 
Third, the sufficiency analysis revealed limitations of conventional performance metrics for diagnosing feature importance under extreme feature ablation. Although the identified temporal pole region was consistently selected across datasets and methods, restricting the models to this feature frequently produced degenerate behaviours, including constant predictions or uniformly uncertain outputs. In several cases, standard metrics alone failed to adequately characterise these failure modes, requiring direct inspection of prediction distributions and probability outputs. More informative evaluation strategies may therefore be required to distinguish genuine feature sufficiency from prediction collapse under severe information restriction. 
Fourth, the proposed missingness-aware uncertainty adjustment may partially capture information already encoded in the predictive distribution itself. Missing features can reduce confidence both directly, through the absence of informative inputs, and indirectly, because incomplete feature combinations deviate from patterns observed during training. Consequently, the additional uncertainty induced by missing modalities may overlap with uncertainty already reflected in the predicted probabilities. More generally, disentangling epistemic uncertainty from uncertainty attributable to missing or incomplete observations remains an open challenge. Future work may therefore investigate alternative uncertainty estimation approaches, including deep ensembles, Monte Carlo dropout, or explicit aleatoric-epistemic decompositions.
Finally, while our study included individuals spanning the CN, MCI, and AD spectrum, the analyses remained cross-sectional and therefore cannot directly characterise within-subject disease trajectories, the proposed NITROGEN architecture, however is naturally suitable to modelling relationships across samples and disease stages. Future work may extend the framework to longitudinal settings using mixed-effects formulations, repeated-measures designs, neural dynamical systems, or physics-informed approaches to better capture the temporal progression of neurodegeneration.

\section{Methods}\label{sec:methods}

\subsection{Data}

Three cohorts were included in this study: ADNI~\cite{shawCerebrospinalFluidBiomarker2009b} as the primary dataset, and OASIS-3~\cite{lamontagneOASIS3LongitudinalNeuroimaging2019a} and AIBL~\cite{ellis_australian_2009} for external validation. Structural MRI scans were processed using our previously established preprocessing pipeline~\cite{diazOPTIMUSPredictingMultivariate2025}. The feature set used in this study comprises MRI-derived regional cortical thickness measures, demographic variables (age, sex, and years of education), \textit{APOE} $\varepsilon$4 allele counts, and cerebrospinal fluid (CSF) biomarkers, including amyloid-$\beta$ (A$\beta$), total tau (t-tau), and phosphorylated tau (p-tau). 

\paragraph{Alzheimer's Disease Neuroimaging Initiative (ADNI)}
Data used in the preparation of this article were obtained from the Alzheimer’s Disease Neuroimaging Initiative (ADNI) database (\url{adni.loni.usc.edu}). The ADNI was launched in 2003 as a public-private partnership, led by Principal Investigator Michael W. Weiner, MD. The primary goal of ADNI has been to test whether serial magnetic resonance imaging (MRI), positron emission tomography (PET), other 
biological markers, and clinical and neuropsychological assessment can be combined to measure the 
progression of mild cognitive impairment (MCI) and Alzheimer’s disease (AD). The ADNI dataset comprised 7,858 cross-sectional samples derived from repeated visits from 1,776 unique individuals, of whom 56.4\% were male. Diagnostic groups comprise 2,497 cognitively normal (CN), 3,583 mild cognitive impairment (MCI), and 1,778 Alzheimer's disease (AD) subjects. APOE$\varepsilon$4 genotyping was available for 1,769 unique subjects, of whom 53.3\% carried no risk allele, 36.8\% carried one allele, and 9.9\% carried two alleles. Participants had a mean age of 73.7 years (range: 54.4--91.4) and a mean educational attainment of 16.1 years (range: 4--20). CSF biomarker measurements were available for 2,353 samples or observations, with mean concentrations of 956.5~pg/mL for A$\beta$ (range: 200.0 -1,700.0~pg/mL), 293.2~pg/mL for total tau (range: 80.0--1,300.0~pg/mL), and 28.1~pg/mL for phosphorylated tau (range: 8.0--120.0~pg/mL), consistent with previously reported ADNI biomarker distributions~\cite{shawCerebrospinalFluidBiomarker2009b}.

\paragraph{Open Access Series of Imaging Studies (OASIS)-3}
The OASIS-3 dataset comprised 2,675 cross-sectional samples from 1,078 unique individuals, of whom 54.9\% were female. Diagnostic groups comprise 2,228 CN, 49 MCI, and 398 AD samples. Participants had a mean age of 71.4 years (range: 45--97) and a mean educational attainment of 15.8 years (range: 6--29). APOE$\varepsilon$4 genotyping was available for all 1,078 unique subjects, of whom 59.7\% carried no risk allele, 34.8\% carried one allele, and 5.5\% carried two alleles. CSF biomarker variables were included as fully missing columns to harmonise the feature space with ADNI.

\paragraph{Australian Imaging, Biomarkers and Lifestyle (AIBL)}
Data was collected by the AIBL study group. AIBL study methodology has been reported previously (Ellis et al. 2009). 
The AIBL dataset comprised 1,286 cross-sectional samples from 684 unique individuals, of whom 56.0\% were female. Diagnostic groups included 947 CN, 184 MCI, and 155 AD samples. Participants had a mean age of 73.7 years (range: 55--96). MRI-derived cortical thickness features were available for 1,282 of 1,286 samples. APOE$\varepsilon$4 genotyping was available for all 684 subjects, of whom 70.0\% carried no risk allele, 24.9\% carried one allele, and 5.1\% carried two alleles. Education and CSF biomarker variables were included as fully missing columns to harmonise the feature space with ADNI.

\paragraph{Missing Variables} Not all modalities were available across cohorts: CSF biomarkers were missing in OASIS-3, while years of education were unavailable in AIBL. To ensure a consistent feature space and avoid unnecessary sample exclusion, these variables were retained as missing features. Key cohort characteristics are summarised in Fig.~\ref{fig:dataset_characteristics}.

\subsection{Models}\label{sec:model}

All modalities in this study were represented as tabular data. Clinical tabular data in healthcare often suffer from missing values, high variability, and noise, which pose significant challenges for predictive modelling~\cite{vavekanandMultimodalMachineLearning2026}. In the context of AD, where integrating diverse data sources---including clinical assessments, imaging, and fluid biomarkers, is crucial, improving current architectures becomes a pressing need. As such, here we introduce an attention based architecture that operate directly on incomplete tabular representations, enabling robust integration of heterogeneous modalities without requiring explicit imputation. This approach supports predictive tasks of clinical relevance under realistic conditions of missingness and noise.

\begin{figure}[t!]
    \centering
    \begin{subfigure}[b]{1\textwidth}
        \centering
        \includegraphics[width=\textwidth]{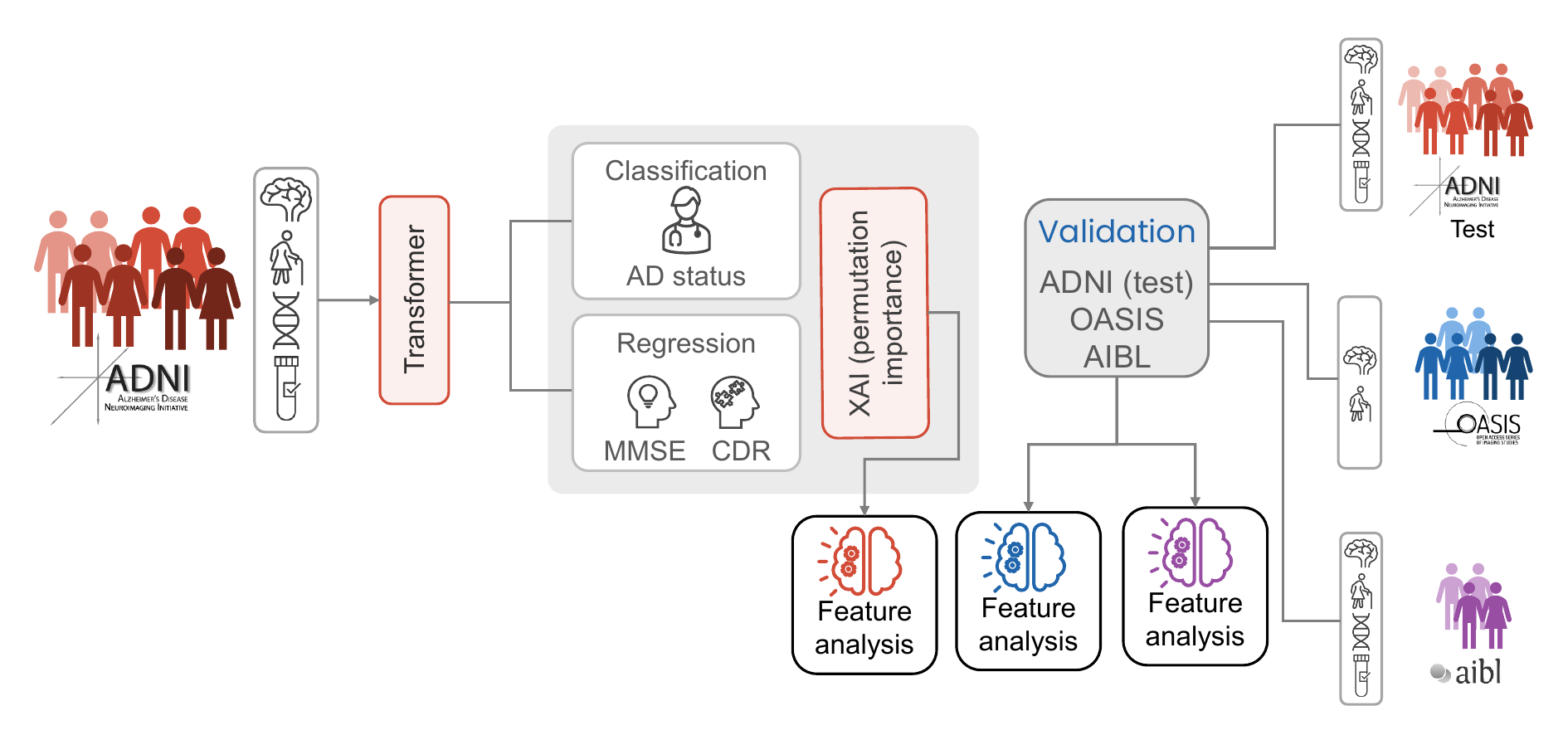}
    \end{subfigure}
    \hfill
     \begin{subfigure}[b]{\textwidth}
        \centering
        \includegraphics[width=\textwidth]{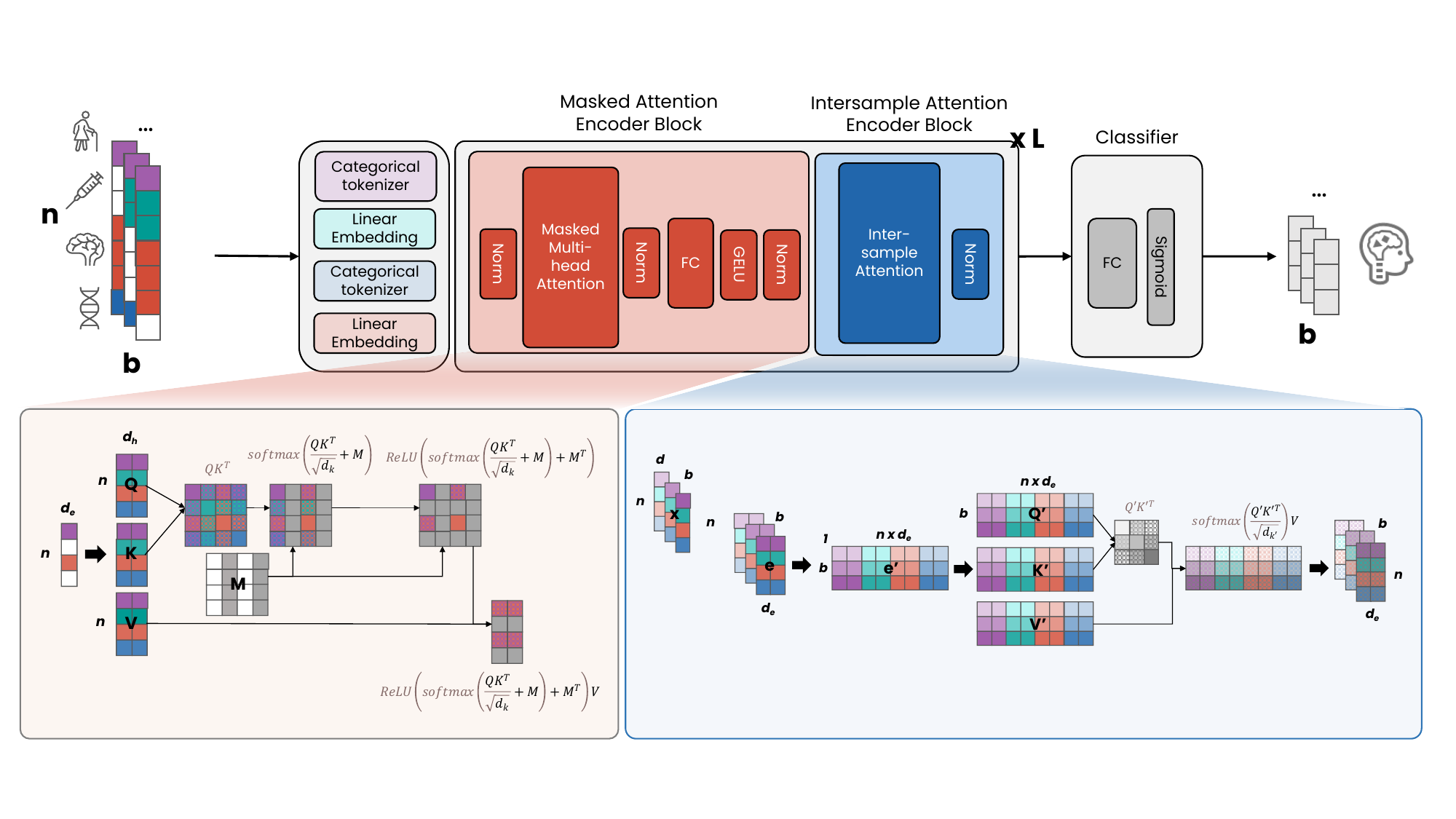}
    \end{subfigure}
    \caption{
        Top: Overview of the experimental design across datasets (ADNI, OASIS-3, AIBL), illustrating training and cross-cohort evaluation settings used to assess generalisation across independently acquired clinical populations. 
        Bottom : Overview of the NITROGEN architecture. 
        The model is composed of stacked \textit{Masked Attention Encoder Blocks}, which handle feature-wise masked attention as described in Eq.~\eqref{eq:naim_masked_attention}, 
        followed by \textit{Intersample Attention Blocks}, which allow the model to incorporate information across samples according to the mechanism detailed in Eq.~\eqref{eq:intersample_attention}. 
        This modular design enables modelling of both feature-level dependencies and cross-sample interactions, enhancing performance in noisy and heterogeneous tabular data settings.
    }
    \label{fig:model_architecture}
\end{figure}

\subsubsection{Architecture}

The main modules in our architecture combine NAIM~\cite{carusoNotAnotherImputation2025}, which introduces a reformulation of masked self-attention to handle missing values by ignoring their contributions during training, and SAINT~\cite{somepalliSAINTImprovedNeural2021b}, which leverages intersample attention to stabilise predictions in the presence of noisy or incomplete inputs. The backbone of the proposed architecture is illustrated in Figure~\ref{fig:model_architecture}.

The NAIM model introduces a masked self-attention mechanism specifically designed to handle missing values in tabular data~\cite{carusoNotAnotherImputation2025}. In NAIM, input features are first transformed into embeddings, with categorical features being transformed into learnable token embeddings that encode both feature identity and value, while numerical variables are projected into the same embedding space through learned linear fully connected layer, yielding a unified token representation suitable for attention-based modelling. 

Then, to explicitly exclude contributions from missing data, NAIM introduces feature-wise masking directly within the attention computation. The modified attention is defined in~\cite{carusoNotAnotherImputation2025} as:

\begin{equation}
    FeatureMaskedAttention(Q, K, V) = \mathrm{ReLU}\left(\mathrm{softmax}\left(\frac{QK^T}{\sqrt{d_h}} + M\right) + M' \right)V,
\end{equation}
\label{eq:naim_masked_attention}

\noindent where $Q, K, V \in \mathbb{R}^{n \times d_h}$ are query, key, and value projections, respectively, derived from input embeddings, and $M \in \mathbb{R}^{n \times n}$ is a masking matrix encoding missing features. The additional mask $M' = M^T$ further suppresses contributions from missing values after the attention weights are computed. 

This ``double masking`` ensures that missing features do not influence the learned dependencies, either directly or indirectly, allowing the model to operate robustly on incomplete inputs without requiring explicit imputation. A graphical representation is in Figure~\ref{fig:model_architecture}.

While the masked attention mechanism operates at the feature level within each sample, it does not explicitly leverage relationships between different samples. To address this, in our architecture, we extend NAIM with an intersample attention module inspired by SAINT~\cite{somepalliSAINTImprovedNeural2021b}. The key idea is to enable the model to learn similarities across samples (e.g., across subject), allowing it to exploit shared patterns within a mini-batch. This may be beneficial in clinical settings, where subjects may exhibit similar profiles despite missing or noisy measurements.

Given feature embeddings \( e \in \mathbb{R}^{b \times n \times d_e} \), where \( b \) is the batch size, \( n \) the number of features, and \( d_e \) the embedding dimension, each sample is reshaped into a single token of dimension \( d_h = n \cdot d_e \), yielding \( e' \in \mathbb{R}^{b \times d_h} \).
Query, key, and value projections are computed as:
\begin{equation}
    Q' = e'W_Q', K' = e'W_K', V' = e'W_V' \in \mathbb{R}^{b \times d_h},
\end{equation}
\label{eq:intersample_attention}
respectively, with $d_h = n \cdot d_e$, and $W_Q', W_K', W_V' \in  \mathbb{R}^{d \times d}$. 

Self-attention is then applied across the batch dimension in the following way:
\begin{equation}
    IntersampleAttention(Q', K', V') = softmax(\frac{Q'K'^T}{\sqrt{d_h}})V'.
\end{equation}

Finally, the resulting representations are reshaped back to the original feature-wise structure \( \mathbb{R}^{b \times n \times d_e} \).

This allows each sample to attend to other samples in the batch, capturing inter-patient relationships that can help stabilise predictions in the presence of missing or noisy data.

\subsubsection{Attention-Based Models}
We evaluated two attention-based models across all tasks: the proposed NITROGEN model, which extends NAIM by incorporating intersample attention inspired by SAINT~\cite{somepalliSAINTImprovedNeural2021b} and a baseline NAIM model having the same parameters as in the original paper~\cite{carusoNotAnotherImputation2025}. Both models share the same input processing pipeline: categorical variables are embedded using dataset-specific indices and cardinalities, continuous features are projected into a shared token space via learnable linear transformations, and missing values are handled natively through a dedicated $-\infty$ token that is excluded by the masked self-attention mechanism. 

Both NAIM and NITROGEN were trained using the AdamW optimiser with a batch size of 128 and early stopping based on validation loss, with a patience of 50 epochs for both models. Output layers and loss functions were adapted per task: a single sigmoid unit with binary cross-entropy loss for binary classification tasks, and two linear output units with mean squared error loss for simultaneous MMSE and CDR-SB prediction.

NAIM comprised six transformer encoder layers with token embedding dimension $d_{\text{token}} = 6$, three self-attention heads, a feed-forward dimension of 1,000, GELU activation, and no dropout. NAIM was trained with a learning rate of $1 \times 10^{-4}$, final model was saved after 31 epochs as it reached minimum validation loss.

NITROGEN extended this architecture with four transformer encoder layers, $d_{\text{token}} = 32$, four self-attention heads, a feed-forward dimension of 463, and a single intersample attention layer following the encoder layers, enabling the model to attend jointly across feature tokens and across samples within a batch. 
NITROGEN was trained with a learning rate of $1 \times 10^{-6}$. The final model was saved after 7 epochs, due to the very fast convergence of NITROGEN despite reduced learning rate. 

\subsection{Explainable AI}
To characterise feature relevance across models, we computed post-hoc perturbation-based feature attributions using the Captum interpretability framework~\cite{kokhlikyanCaptumUnifiedGeneric2020a}. Specifically, we employed feature permutation attribution, computed independently for each trained model on the held-out test sets from ADNI, OASIS-3, and AIBL. Attributions were computed with respect to the per-sample log-loss rather than raw predicted probability, providing a loss-sensitive measure of feature relevance that accounts for both the direction and confidence of predictions. For each feature, attribution was obtained by independently permuting that feature across all test samples while keeping all other features fixed, and measuring the resulting change in model log-loss. To obtain global importance profiles, sample-level attributions were aggregated using the mean absolute attribution score across the test set. This perturbation-based approach is model-agnostic and applicable across both transformer-based and tree-based architectures, enabling a consistent attribution procedure across all evaluated models~\cite{fisherAllModelsAre2019}.

We note that feature permutation shuffles all feature values uniformly across samples, including missing value tokens ($-\infty$), redistributing missingness patterns across samples. Attribution scores for partially observed features therefore reflect the joint contribution of feature magnitude and missingness structure to model predictions.

\subsection{Uncertainty quantification}

For clinical deployment, predictive performance alone is insufficient; models must also provide reliable estimates of predictive uncertainty. In healthcare applications, uncertainty estimation is essential for assessing model reliability, identifying potentially incorrect predictions, and supporting safe decision-making.

Uncertainty can arise from two primary sources: data uncertainty (aleatoric) and model uncertainty (epistemic)~\cite{gawlikowskiSurveyUncertaintyDeep2023, heSurveyUncertaintyQuantification2024}. In this work, we focus on predictive uncertainty derived from model outputs, with particular attention to how this uncertainty is affected by missing modalities in incomplete multimodal clinical data.

In standard neural network classifiers, predictive confidence is commonly approximated using the softmax probabilities produced by the final output layer. Although these probabilities are not guaranteed to represent well-calibrated posterior probabilities, prior work has shown that maximum softmax probability can serve as a simple yet effective baseline for detecting misclassified or out-of-distribution samples~\cite{hendrycks17baseline}.

Beyond mis-classification detection, predictive entropy and related measures derived from the softmax distribution have been proposed as indicators of model reliability and generalisation performance~\cite{tornettaEntropyMethodsConfidence2021}. However, softmax-based confidence estimates are often overconfident, particularly under dataset shift, motivating the need for calibration-aware uncertainty assessment. Alternative approaches have therefore explored transformations of the logits---such as kurtosis or normalised logit differences---as improved proxies for prediction correctness~\cite{tahaConfidenceEstimationClassification2022}.

More generally, predictive uncertainty can be modelled explicitly. For instance, discriminative models may be extended to output parameters of predictive distributions (e.g., mean and variance), while generative approaches estimate uncertainty by modelling the underlying data distribution~\cite{heSurveyUncertaintyQuantification2024}. Although these methods provide principled uncertainty estimates, they typically require architectural modifications and additional training constraints.

We start with the assumption the predicted class probabilities can serve as a preliminary reflection of predictive confidence~\cite{hendrycks17baseline}. Specifically, for a model output $h(x)$, the softmax probability for class $i$ is defined as follows:

\begin{equation}
\pi_i = \frac{\exp(h_i(x))}{\sum_{c=1}^{C} \exp(h_c(x))},
\end{equation}
\noindent where $\pi_i$ denotes the predicted probability of class $i$, for $i = 0, 1, \dots, C$, and the uncertainty is subsequently summarised using the entropy of the predictive distribution.

Further, to quantify uncertainty, we compute the Shannon entropy of the predictive distribution~\cite{pearceUnderstandingSoftmaxConfidence2021}:

\begin{equation}
H(\boldsymbol{\pi}(x)) = - \sum_{i=1}^{C} \pi_i(x) \log \pi_i(x).
\end{equation}

In the binary classification setting ($C=2$), this reduces to:
\begin{equation}
H(p) = - p \log p - (1 - p)\log(1 - p),
\end{equation}
where $p = \pi_{1}(x)$ denotes the predicted probability of the positive class. 
The entropy is maximal at $p=0.5$ (maximum uncertainty) and vanishes as $p \to 0$ or $p \to 1$, reflecting increasingly confident predictions. 

\subsection{Adjusting Prediction Confidence for Missing Modalities}

In multimodal clinical settings, entire diagnostic modalities are frequently unavailable due to cost, acquisition constraints, or clinical workflows. While many machine learning models are designed to handle missing data, they typically produce predictions without explicitly quantifying how the absence of specific modalities impacts predictive confidence. To the best of our knowledge, existing approaches either focus on handling missing inputs during training or on estimating predictive uncertainty, but do not connect these two aspects in a principled manner.

To address this gap, we propose a missingness-aware uncertainty framework that provides a simple framework for incorporating modality-level missingness into prediction confidence estimates. 

\paragraph{Modality importance estimation.}

To quantify the contribution of each modality, we used the feature attribution scores obtained with permutation importance in the explainable AI analysis. Let $p_{i,f}$ denote the permutation importance associated with feature $f$ for sample $i$, and let $\mathcal{M}_k$ be the set of features belonging to modality $k$.

For each sample, the importance of modality $k$ is computed by summing the absolute permutation importance values of all features belonging to that modality and averaging by the number of features:
\begin{equation}
m_{i,k} = \frac{1}{|\mathcal{M_{k}}|} \sum{f \in \mathcal{M_k}} |p_{i,f}|.
\end{equation}

The overall importance of modality $k$ is then obtained by averaging across all samples:
\begin{equation}
I_k = \frac{1}{N} \sum_{i=1}^{N} m_{i,k},
\end{equation}
where $N$ is the number of samples.

Finally, to enable comparison across modalities, the modality importance scores are normalised:
\begin{equation}
\tilde{I}_k = \frac{I_k}{\sum_j I_j},
\end{equation}
where the sum runs over all modalities $j$. The normalised importance $\tilde{I}_k \in [0,1]$ represents the relative contribution of modality $k$ to the model predictions.

\paragraph{Missingness-aware uncertainty adjustment.} For a subject $x$ with a set of missing modalities $M_{\text{miss}}$, we define a missingness-aware uncertainty inflation factor
\begin{equation}
    IF(x) = 1 + \beta \sum_{k \in M_{\text{miss}}} \tilde{I}_k,
\end{equation}
where $\beta \ge 0$ controls the sensitivity of the uncertainty adjustment. The adjusted uncertainty interval width is then given by
\begin{equation}
    PI_{\text{adj}}(x) = IF(x) \cdot PI_{\text{width}}.
\end{equation}

This formulation increases predictive uncertainty in proportion to the estimated importance of the missing modalities. As a result, the absence of highly informative modalities leads to wider uncertainty intervals, providing a method for incorporating missing modality information into confidence estimates.

\subsection*{Code availability}
Code to reproduce the experiments described in this study is publicly available at \url{https://github.com/cschneuw/nitrogen}.

\section{Experiments}

\subsection{Baseline Models}

In addition to NAIM and NITROGEN, five baseline models were evaluated for benchmarking: XGBoost, LightGBM, Random Forest, MA-LASSO, and MA-GBT. These baselines were selected to include both widely used tree-based methods and recently proposed missing-aware tabular learning approaches. XGBoost~\cite{chenXGBoostScalableTree2016a} was configured with 400 boosting rounds, unrestricted maximum depth, a learning rate of 0.15, subsampling and column sampling ratios of 0.6, and histogram-based tree construction. LightGBM~\cite{keLightGBMHighlyEfficient2017} was used with 400 boosting rounds, a maximum depth of 8, a learning rate of 0.1, and full row and column sampling. Both XGBoost and LightGBM handle missing values natively and were restricted to classification tasks. Random Forest~\cite{breimanRandomForests2001} comprised 200 trees with Gini impurity criterion, unrestricted depth, and a minimum of 4 samples per leaf, with missing values handled through mean imputation. Two missing-aware baselines---MA-LASSO and MA-GBT~\cite{samadImputationfreeLearningTabular2025}---were evaluated on classification tasks only, as their formulation does not extend to regression. MA-LASSO was trained with regularisation parameters $\alpha = 100$ and $\beta = 100$; MA-GBT used 200 estimators, a 
maximum depth of 5, a learning rate of 0.1, and a subsampling ratio of 0.6. Both missing-aware models explicitly penalise reliance on missing features through modified objective functions, with input features zero-imputed at training and inference time. Owing to its substantially longer training time, MA-GBT was not evaluated in all experimental settings. 

\subsection{Binary Classification Tasks}
\label{subsec:binary_classification}

Two binary classification tasks were defined and evaluated on stratified 60/20/20 train/validation/test splits derived from ADNI only, with OASIS-3 and AIBL reserved exclusively as held-out external cohorts. Splits were performed using group-stratified shuffling to ensure that repeated measurements from the same individual were always assigned to the same partition, preventing subject-level data leakage. Continuous features were 
standardised using a scaler fitted on the training set only and applied to all subsequent splits and external cohorts. 

Classification performance was assessed over ten independent training runs initialised with different random seeds (123, 42, 0, 7, 13, 21, 37, 55, 77, and 99). Metrics reported in Table~\ref{tab:cn_ad_results_seeds} and the ROC curves in Figure~\ref{fig:classification_performance} are presented as the mean and standard deviation across runs. All other analyses were performed using a fixed random seed (123).

\paragraph{Task 1: CN vs AD}
Retaining only CN and AD samples from ADNI, the training set comprised 2,560 samples (1,452 CN, 1,108 AD), the validation set 882 samples (524 CN, 358 AD; 59.4\% vs 40.6\%), and the test set 833 samples (521 CN, 312 AD; 62.5\% vs 37.5\%). External cohorts comprised 2,625 samples from OASIS-3 (2,228 CN, 398 AD) and 1,102 samples from AIBL (947 CN, 155 AD).

\paragraph{Task 2: CN vs MCI/AD}
Using all available ADNI samples, the training set comprised 4,704 samples (1,438 CN, 3,266 MCI/AD), the validation set 1,561 samples (538 CN, 1,023 MCI/AD), and the test set 1,593 samples (521 CN, 1,072 MCI/AD). The class imbalance reflects the natural diagnostic distribution in ADNI and was preserved across splits without resampling. External cohorts comprised 2,675 samples from OASIS-3 (2,228 CN, 447 demented) and 1,286 samples from AIBL (947 CN, 339 MCI/AD).

\subsection{Multi-class Diagnostic Classification}

All architectures were adapted for three-class classification (CN / MCI / AD) by changing the output to 3 classe while retaining the hyper-parameters established in the binary setting. For tree-based models, objectives were updated to their multi-class equivalents. For NAIM and NITROGEN, the output layer was adapted to produce three logits passed through a softmax activation, with the predicted class determined by argmax over the probability vector and optimised with cross-entropy loss. Probabilities were formatted as an $N \times C$ matrix to enable per-class and macro-averaged AUC computation under the one-vs-rest (OvR) strategy.

Following the same group-stratified 60/20/20 splitting strategy as in Section~\ref{subsec:binary_classification}, the ADNI training set comprised 4,704 samples (CN: 1,438, MCI: 2,196, AD: 1,070), the validation set 1,561 samples (CN: 538, MCI: 680, AD: 343), and the test set 1,593 samples (CN: 521, MCI: 707, AD: 365). OASIS-3 (2,675 samples; CN: 2,228, MCI: 49, AD: 398) and AIBL (1,286 samples; CN: 947, MCI: 184, AD: 155) were used as held-out external validation cohorts. The near-absence of MCI samples in OASIS-3 reflects differences in recruitment strategy relative to ADNI and AIBL, with OASIS-3 specifically targeting preclinical populations, resulting in a higher proportion of cognitively normal participants and fewer dementia cases. 

\subsection{Cognitive Score Prediction}

To extend the evaluation beyond diagnostic classification, models were trained on ADNI to jointly predict Mini-Mental State Examination (MMSE) and Clinical Dementia Rating Sum of Boxes (CDR-SB) scores from multimodal input features. For NAIM and NITROGEN, the output layer was replaced with a two-dimensional linear unit optimised with mean squared error loss, while all other architectural components and training settings remained identical to the classification setting; tree-based models were similarly adapted to predict both targets simultaneously.

The ADNI cohort (7,858 samples; CN: 2,497, MCI: 3,583, AD: 1,778) was partitioned into training, validation, and test sets using a 60/20/20 group-stratified split (seed = 123), yielding 4,704 training samples (CN: 1,256, MCI: 2,236, AD: 1,212), 1,561 validation samples (CN: 446, MCI: 723, AD: 392), and 1,593 test samples (CN: 521, MCI: 707, AD: 365). OASIS-3 (CN: 2,228, MCI: 49, AD: 398) and AIBL (CN: 947, MCI: 184, AD: 155) were reserved exclusively as external validation cohorts. As OASIS-3 and AIBL provide only the global Clinical Dementia Rating (CDR) rather than CDR-SB, CDR was used as a surrogate target for out-of-distribution evaluation; while CDR and CDR-SB differ in scale and granularity, they are expected to be strongly correlated, allowing cross-cohort regression performance to be approximately assessed.

\subsection{Test of Sufficiency} 

To verify that the identified MRI features are sufficient to sustain classification performance, we performed a sufficiency test on held-out subsets not used during feature attribution computation.

The training and validation splits were identical to those described in Section~\ref{subsec:binary_classification}. To prevent data leakage between attribution computation and sufficiency evaluation, each test cohort was further partitioned into a 70\% attribution subset and a 30\% held-out evaluation subset using stratified splits with subject 
identifiers as grouping variable. The attribution subsets comprised 580 ADNI samples (CN: 353, MCI/AD: 227), 1,833 OASIS-3 samples (CN: 1,554, MCI/AD: 279), and 762 AIBL samples (CN: 649, MCI/AD: 113). The corresponding held-out evaluation subsets comprised 253 ADNI samples (CN: 168, MCI/AD: 85), 793 OASIS-3 samples (CN: 674, MCI/AD: 119), 
and 340 AIBL samples (CN: 298, MCI/AD: 42). Permutation-based feature attributions were computed exclusively on the 70\% attribution subsets. Features were ranked by mean absolute attribution score for each model-dataset combination, and the most consistent MRI features were identified by retaining those appearing within the top-30 across all 
combinations. This procedure yielded one cortical thickness features from the temporal pole: L Limbic TempPole 3.

Sufficiency was evaluated on the remaining 30\% of the ADNI test set using models trained on the original ADNI training data described in Section~\ref{subsec:binary_classification}. The trained models were applied without retraining under three input configurations:\\ (i) the full multimodal feature set, used as the baseline;\\ (ii) the two selected temporal pole features together with all non-MRI features, with all other MRI cortical thickness features masked; and\\ (iii) the two selected temporal pole features only, with all remaining features masked.

Performance metrics and ROC curves are reported in Figure~\ref{fig:feature_attributions}f and Table~\ref{tab:sufficiency_analysis}.

\subsection{Hyper-parameter sweeps}

To ensure fair comparison across architectures, hyperparameter selection was performed separately for each model family using the Task 1 (CN vs AD) training and validation sets (splits were generated with fixed seed = 123). The resulting hyperparameter configurations were then kept fixed across all tasks and cohorts to maintain a consistent experimental framework and limit the influence of repeated task-specific optimisation.

All experiments were tracked and managed using Weights \& Biases (W\&B) sweeps~\cite{wandb}. For NITROGEN, Bayesian search with hyperband-based early termination (minimum 15 iterations, $\eta = 2$, $s = 5$) was performed to maximise balanced accuracy on the validation set, over the following hyperparameter space: number of encoder layers (2, 3, 4, 6), token embedding dimension $d_{\text{token}}$ (32, 48, 64), number of self-attention heads (4, 8), feed-forward hidden dimension (integer-uniform between 128 and 1,024), dropout rate (0, 0.1, 0.2), activation function (ReLU, GELU), presence of bias terms (learnable intercepts), mask type (0: missing-feature masking; 1: missing-feature masking with masking of interactions involving missing features; 2: both forward and reverse missing-feature masks), number of intersample attention layers (1, 2), and intersample layer placement (applied after the encoder, or interleaved with encoder layers); intersample attention was enabled in all sweep configurations. The optimiser (AdamW), embedder initialisation (uniform), and maximum number of training epochs (200, with early stopping) were fixed across runs, and the learning rate was sampled log-uniformly between $1 \times 10^{-5}$ and $1 \times 10^{-3}$.

During both hyperparameter search, the learning rate was adaptively reduced using a scheduler (factor = 0.5, patience = 5 epochs), and training was terminated early if the monitored validation metric failed to improve by at least $1 \times 10^{-4}$ over a fixed patience of 20 epochs, with the same configuration retained for all downstream training runs.

For tree-based models, grid search was conducted over commonly used ranges. XGBoost sweeps varied number of estimators (100, 200, 400), maximum tree depth (4, 6, 8 or None), learning rate (0.01--0.2), and row and column subsampling ratios (0.6, 0.8, 1.0). LightGBM sweeps additionally explored number of leaves and subsampling ratios. Random Forest sweeps varied number of estimators, maximum depth, maximum features per split, and minimum samples per leaf or split.

For missing-aware models, MA-LASSO sweeps explored combinations of regularisation parameters $\alpha$ and $\beta$ $\in \{0.1, 1, 10, 100\}$. MA-GBT sweeps varied number of estimators, maximum depth, learning rate, sub-sampling ratio, and missingness penalty $\alpha$, with optional node-wise regularisation enabled or disabled.

In all cases, the final hyperparameter configuration was selected based on validation balanced accuracy and held fixed for all subsequent tasks and cross-dataset evaluations, with only output layer and loss function adapted per task. 

\subsection{Selection of the Uncertainty Sensitivity Parameter}
\label{subsec:selection_beta}

The sensitivity parameter $\beta$ controls the influence of missing modalities on 
predictive uncertainty. To select an appropriate value, we performed a grid search over $\beta \in [0, 2]$ on the ADNI validation set ($N=882$) for the CN versus AD 
classification task, with a fixed uncertainty threshold $\tau = 0.900$, chosen to retain a meaningful proportion of samples while allowing selective abstention on uncertain 
predictions.

For each value of $\beta$, we computed the adjusted uncertainty $H_{\text{adj}}$ and 
retained samples satisfying:
\begin{equation}
    H_{\text{adj}}(x) \leq \tau.
\end{equation}
To identify the optimal $\beta$, we maximised the product of two complementary criteria: the Spearman rank correlation between adjusted uncertainty and prediction errors, $\rho(H_{\text{adj}}, \mathbf{1}[\hat{y} \neq y])$, which measures how well uncertainty ranks incorrect predictions; and the retention gap, defined as the difference in accuracy between retained and rejected samples. Let $\text{Acc}_{\text{retained}}$ denote the classification accuracy on samples with $H_{\text{adj}}(x) \leq \tau$ (retained for prediction) and $\text{Acc}_{\text{rejected}}$ denote the accuracy on samples with $H_{\text{adj}}(x) > \tau$ (abstained from):
\begin{equation}
    S(\beta) = 
    \max\!\left(0,\, \rho(H_{\text{adj}}, \mathbf{1}[\hat{y} \neq y])\right) 
    \times 
    \max\!\left(0,\, \text{Acc}_{\text{retained}} - \text{Acc}_{\text{rejected}}\right).
\end{equation}
This joint criterion rewards configurations where uncertainty is both well-calibrated with respect to errors and effectively separates reliable from unreliable predictions, while remaining insensitive to trivial solutions at $\beta = 0$ where no modality adjustment is applied.

The optimal parameter was identified at $\beta^* = 0.400$ with $\tau = 0.900$, achieving a Spearman correlation of $\rho = 0.348$ ($p < 10^{-26}$) between adjusted uncertainty and prediction errors (Fig.~\ref{fig:beta_grid_search}). On the validation set, uncertainty thresholding retained 85.7\% of samples (756/882), improving accuracy from 0.862 to 0.900 ($\Delta = +0.038$), with rejected samples exhibiting substantially lower accuracy (0.635), confirming that the adjusted uncertainty effectively identifies 
unreliable predictions.

\subsection{Uncertainty adjustment evaluation}
\label{sec:uncertainty_evaluation}
 
To assess whether the improvement in selective accuracy obtained through uncertainty thresholding is statistically significant, we conducted permutation tests on both the validation (see Fig.~\ref{fig:uncertainty}). For each permutation, true diagnostic labels were randomly shuffled and the same fraction of samples as retained by the uncertainty threshold was selected at random, without any uncertainty-based filtering. This procedure was repeated 1,000 times to construct a null distribution of accuracy improvements expected by chance. The observed accuracy improvement, defined as the difference between selective accuracy on uncertainty-retained samples and accuracy on the full dataset, was then compared against this null distribution. A one-sided empirical $p$-value was computed as the fraction of permuted improvements exceeding the observed value.

In addition to selective accuracy, we assessed the quality of the adjusted uncertainty estimates through two complementary analyses. First, we computed the Spearman rank correlation $\rho$ between adjusted uncertainty scores and binary prediction errors, measuring whether higher uncertainty reliably identifies misclassified samples. Second, we evaluated calibration by partitioning samples into decile bins of uncertainty and computing the Spearman correlation between mean bin uncertainty and mean bin error rate, 
providing a coarser but more stable measure of monotonic calibration across the uncertainty range. Both analyses were performed for raw and missingness-adjusted uncertainty to quantify the improvement attributable to the modality adjustment.

\bmhead{Acknowledgements}
C.S.D., N.B. and O.Y.C. conceptualised the study. C.S.D. developed the methods, wrote the codes, performed data processing and analysis. N.B. contributed to uncertainty quantification adjustment.  C.S.D. and O.Y.C. wrote the manuscript, with comments from all other authors. 

This work was supported by Swiss National Science Foundation under Grants \texttt{CR00I5-235987} and \texttt{3200-0-239967}.

Data collection and sharing for the Alzheimer's Disease Neuroimaging Initiative (ADNI) is funded by the National 
Institute on Aging (National Institutes of Health Grant U19 AG024904). The grantee organisation is the Northern 
California Institute for Research and Education. In the past, ADNI has also received funding from the National 
Institute of Biomedical Imaging and Bioengineering,  the Canadian Institutes of Health Research, and private 
sector contributions through the Foundation for the National Institutes of Health (FNIH) including generous 
contributions from the following: AbbVie, Alzheimer’s Association; Alzheimer’s Drug Discovery Foundation; 
Araclon Biotech; BioClinica, Inc.; Biogen; Bristol-Myers Squibb Company; CereSpir, Inc.; Cogstate; Eisai Inc.; 
Elan Pharmaceuticals, Inc.; Eli Lilly and Company; EuroImmun; F. Hoffmann-La Roche Ltd and its affiliated 
company Genentech, Inc.; Fujirebio; GE Healthcare; IXICO Ltd.; Janssen Alzheimer Immunotherapy Research \& 
Development, LLC.; Johnson \& Johnson Pharmaceutical Research \& Development LLC.; Lumosity; Lundbeck; 
Merck \& Co., Inc.; Meso Scale Diagnostics, LLC.; NeuroRx Research; Neurotrack Technologies; Novartis 
Pharmaceuticals Corporation; Pfizer Inc.; Piramal Imaging; Servier; Takeda Pharmaceutical Company; and 
Transition Therapeutics.  

Data used in the preparation of this article was obtained from the Australian Imaging Biomarkers and Lifestyle flagship study of ageing (AIBL) funded by the Commonwealth Scientific and Industrial Research Organisation (CSIRO) which was made available at the ADNI database (\url{www.loni.usc.edu/ADNI}). 

Data were provided in part by OASIS-3: Longitudinal Multimodal Neuroimaging: Principal Investigators: T. Benzinger, D. Marcus, J. Morris; NIH P30 AG066444, P50 AG00561, P30 NS09857781, P01 AG026276, P01 AG003991, R01 AG043434, UL1 TR000448, R01 EB009352. AV-45 doses were provided by Avid Radiopharmaceuticals, a wholly owned subsidiary of Eli Lilly.

\section*{Declarations}

\subsection*{Data availability}

Three cohorts were used in this study: ADNI (primary dataset), and OASIS-3 and AIBL (external validation). ADNI data are available at \url{https://adni.loni.usc.edu/}. Access can be requested for scientific investigation, teaching, or clinical research planning via the Image and Data Archive (IDA) platform (\url{https://ida.loni.usc.edu}). AIBL data are available to authorised users through a request process integrated with the ADNI access procedure via the LONI infrastructure, at \url{https://ida.loni.usc.edu/collaboration/access/appApply.jsp?project=AIBL}, or jointly with ADNI at \url{https://ida.loni.usc.edu/collaboration/access/appApply.jsp?project=AIBL&project=ADNI}, subject to agreement with the relevant project terms of use. OASIS-3 data are available at \url{https://sites.wustl.edu/oasisbrains/home/oasis-3/} and can also be accessed via the LONI/IDA platform subject to a Data Use Agreement (\url{https://ida.loni.usc.edu/collaboration/access/appLicense.jsp})

\bibliography{sn-bibliography}
\newpage


\begin{appendices}
\section{Supplementary information}

\begin{table}[b!]
\centering
\caption{Performance of models trained on ADNI for CN vs. AD classification and evaluated on the ADNI train, validation, and test splits, as well as on the external OASIS-3 and AIBL cohorts. Models are sorted by AUC within each split. Seed = 123. }
\label{tab:cn_ad_results}
\begin{tabular}{llcccccc}
\toprule
Data/Split & Model & Acc. & Bal. Acc. & Spec. & Sens. & F1 & AUC \\
\midrule
\multirow{7}{*}{ADNI (Train)}
 & XGBoost           & 1.000 & 1.000 & 1.000 & 1.000 & 1.000 & 1.000 \\
 & LightGBM          & 1.000 & 1.000 & 1.000 & 1.000 & 1.000 & 1.000 \\
 & MA-GBT            & 1.000 & 1.000 & 1.000 & 1.000 & 1.000 & 1.000 \\
 & Random Forest      & 0.986 & 0.985 & 0.994 & 0.975 & 0.984 & 0.999 \\
 & NITROGEN  & 0.914 & 0.908 & 0.952 & 0.863 & 0.896 & 0.972 \\
 & NAIM              & 0.917 & 0.912 & 0.951 & 0.873 & 0.901 & 0.969 \\
 & MA-Lasso          & 0.910 & 0.908 & 0.924 & 0.893 & 0.896 & 0.964 \\
\midrule
\multirow{7}{*}{ADNI (Val)}
 & NITROGEN  & 0.862 & 0.853 & 0.899 & 0.807 & 0.826 & 0.929 \\
 & XGBoost           & 0.856 & 0.853 & 0.870 & 0.835 & 0.825 & 0.925 \\
 & LightGBM          & 0.848 & 0.845 & 0.863 & 0.827 & 0.815 & 0.924 \\
 & NAIM              & 0.865 & 0.859 & 0.889 & 0.830 & 0.833 & 0.919 \\
 & MA-GBT            & 0.844 & 0.841 & 0.853 & 0.830 & 0.811 & 0.916 \\
 & MA-Lasso          & 0.854 & 0.850 & 0.868 & 0.832 & 0.822 & 0.910 \\
 & Random Forest      & 0.814 & 0.808 & 0.840 & 0.777 & 0.772 & 0.894 \\
\midrule
\multirow{7}{*}{ADNI (Test)}
 & LightGBM          & 0.854 & 0.846 & 0.875 & 0.817 & 0.807 & 0.925 \\
 & Random Forest      & 0.863 & 0.852 & 0.896 & 0.808 & 0.816 & 0.925 \\
 & MA-GBT            & 0.856 & 0.848 & 0.879 & 0.817 & 0.810 & 0.923 \\
 & XGBoost           & 0.861 & 0.854 & 0.881 & 0.827 & 0.816 & 0.922 \\
 & NITROGEN  & 0.846 & 0.841 & 0.862 & 0.821 & 0.800 & 0.915 \\
 & NAIM              & 0.826 & 0.824 & 0.831 & 0.817 & 0.779 & 0.912 \\
 & MA-Lasso          & 0.822 & 0.826 & 0.812 & 0.840 & 0.780 & 0.909 \\
\midrule
\multirow{7}{*}{AIBL}
 & NITROGEN  & 0.894 & 0.811 & 0.926 & 0.697 & 0.649 & 0.936 \\
 & MA-Lasso          & 0.902 & 0.840 & 0.926 & 0.755 & 0.684 & 0.931 \\
 & LightGBM          & 0.911 & 0.786 & 0.960 & 0.613 & 0.660 & 0.929 \\
 & XGBoost           & 0.917 & 0.782 & 0.970 & 0.594 & 0.669 & 0.923 \\
 & NAIM              & 0.905 & 0.812 & 0.941 & 0.684 & 0.669 & 0.919 \\
 & MA-GBT            & 0.913 & 0.806 & 0.955 & 0.658 & 0.680 & 0.917 \\
 & Random Forest      & 0.897 & 0.765 & 0.949 & 0.581 & 0.614 & 0.913 \\
\midrule
\multirow{7}{*}{OASIS-3}
 & MA-GBT            & 0.716 & 0.673 & 0.735 & 0.611 & 0.395 & 0.738 \\
 & NITROGEN  & 0.560 & 0.638 & 0.526 & 0.751 & 0.341 & 0.718 \\
 & NAIM              & 0.639 & 0.642 & 0.638 & 0.646 & 0.352 & 0.714 \\
 & XGBoost           & 0.542 & 0.630 & 0.504 & 0.756 & 0.334 & 0.714 \\
 & LightGBM          & 0.561 & 0.621 & 0.535 & 0.706 & 0.328 & 0.701 \\
 & MA-Lasso          & 0.572 & 0.622 & 0.550 & 0.693 & 0.329 & 0.697 \\
 & Random Forest      & 0.456 & 0.573 & 0.405 & 0.741 & 0.292 & 0.667 \\
\bottomrule
\end{tabular}
\end{table}

\begin{landscape}
\begin{table}[t!]
\centering
\caption{Performance of models trained on ADNI for CN vs. AD classification on ADNI test set, AIBL, and OASIS-3 cohorts, reported as mean $\pm$ standard deviation across 10 random seeds (\texttt{seeds} = [123, 42, 0, 7, 13, 21, 37, 55, 77, 99]).
Models are sorted by AUC within each split.}
\label{tab:cn_ad_results_seeds}
\begin{tabular}{llcccccc}
\toprule
Data/Split & Model & Acc. & Bal. Acc. & Spec. & Sens. & F1 & AUC \\
\midrule
\multirow{6}{*}{ADNI (Test)}
 & LightGBM          & $0.869 \pm 0.018$ & $0.861 \pm 0.020$ & $0.895 \pm 0.031$ & $0.827 \pm 0.041$ & $0.829 \pm 0.030$ & $0.935 \pm 0.014$ \\
 & XGBoost           & $0.864 \pm 0.019$ & $0.857 \pm 0.019$ & $0.887 \pm 0.032$ & $0.827 \pm 0.034$ & $0.824 \pm 0.030$ & $0.933 \pm 0.012$ \\
 & NITROGEN  & $0.848 \pm 0.023$ & $0.845 \pm 0.022$ & $0.858 \pm 0.038$ & $0.832 \pm 0.035$ & $0.809 \pm 0.032$ & $0.924 \pm 0.016$ \\
 & Random Forest      & $0.851 \pm 0.025$ & $0.841 \pm 0.025$ & $0.888 \pm 0.037$ & $0.794 \pm 0.039$ & $0.806 \pm 0.036$ & $0.916 \pm 0.020$ \\
 & NAIM              & $0.846 \pm 0.024$ & $0.839 \pm 0.024$ & $0.870 \pm 0.036$ & $0.808 \pm 0.037$ & $0.802 \pm 0.037$ & $0.915 \pm 0.019$ \\
 & MA-Lasso          & $0.839 \pm 0.024$ & $0.837 \pm 0.023$ & $0.841 \pm 0.038$ & $0.833 \pm 0.037$ & $0.800 \pm 0.035$ & $0.913 \pm 0.018$ \\
\midrule
\multirow{6}{*}{AIBL}
 & XGBoost           & $0.917 \pm 0.004$ & $0.789 \pm 0.012$ & $0.967 \pm 0.004$ & $0.610 \pm 0.025$ & $0.674 \pm 0.018$ & $0.918 \pm 0.005$ \\
 & MA-Lasso          & $0.894 \pm 0.011$ & $0.838 \pm 0.013$ & $0.916 \pm 0.014$ & $0.761 \pm 0.030$ & $0.669 \pm 0.024$ & $0.917 \pm 0.008$ \\
 & LightGBM          & $0.911 \pm 0.006$ & $0.800 \pm 0.014$ & $0.954 \pm 0.008$ & $0.645 \pm 0.031$ & $0.670 \pm 0.020$ & $0.916 \pm 0.008$ \\
 & Random Forest      & $0.907 \pm 0.002$ & $0.771 \pm 0.009$ & $0.960 \pm 0.005$ & $0.582 \pm 0.022$ & $0.637 \pm 0.009$ & $0.913 \pm 0.003$ \\
 & NAIM              & $0.890 \pm 0.010$ & $0.809 \pm 0.009$ & $0.922 \pm 0.017$ & $0.695 \pm 0.032$ & $0.641 \pm 0.014$ & $0.905 \pm 0.012$ \\
 & NITROGEN  & $0.886 \pm 0.020$ & $0.807 \pm 0.023$ & $0.916 \pm 0.033$ & $0.697 \pm 0.072$ & $0.633 \pm 0.026$ & $0.901 \pm 0.011$ \\
\midrule
\multirow{6}{*}{OASIS-3}
 & LightGBM          & $0.613 \pm 0.050$ & $0.660 \pm 0.032$ & $0.593 \pm 0.060$ & $0.727 \pm 0.036$ & $0.365 \pm 0.034$ & $0.719 \pm 0.020$ \\
 & XGBoost           & $0.551 \pm 0.074$ & $0.641 \pm 0.039$ & $0.512 \pm 0.090$ & $0.770 \pm 0.028$ & $0.345 \pm 0.036$ & $0.713 \pm 0.031$ \\
 & NITROGEN  & $0.556 \pm 0.060$ & $0.636 \pm 0.025$ & $0.521 \pm 0.078$ & $0.751 \pm 0.054$ & $0.341 \pm 0.023$ & $0.710 \pm 0.018$ \\
 & NAIM              & $0.490 \pm 0.056$ & $0.602 \pm 0.026$ & $0.442 \pm 0.071$ & $0.763 \pm 0.033$ & $0.314 \pm 0.021$ & $0.703 \pm 0.012$ \\
 & Random Forest      & $0.610 \pm 0.168$ & $0.621 \pm 0.048$ & $0.606 \pm 0.222$ & $0.636 \pm 0.140$ & $0.349 \pm 0.058$ & $0.700 \pm 0.031$ \\
 & MA-Lasso          & $0.593 \pm 0.021$ & $0.616 \pm 0.019$ & $0.583 \pm 0.029$ & $0.650 \pm 0.047$ & $0.326 \pm 0.016$ & $0.689 \pm 0.017$ \\
\bottomrule
\end{tabular}
\end{table}
\end{landscape}

\begin{table}[t!]
\centering
\caption{Performance of models trained on ADNI for cognitively normal (CN) vs. cognitively impaired (MCI/AD) classification and evaluated across datasets. Models are sorted by AUC within each split. Seed = 123.}
\label{tab:dementia_results}
\begin{tabular}{llcccccc}
\toprule
Data/Split & Model & Acc. & Bal. Acc. & Spec. & Sens. & F1 & AUC \\
\midrule
\multirow{7}{*}{ADNI (Train)}
 & XGBoost           & 1.000 & 1.000 & 1.000 & 1.000 & 1.000 & 1.000 \\
 & LightGBM          & 1.000 & 1.000 & 1.000 & 1.000 & 1.000 & 1.000 \\
 & MA-GBT            & 0.998 & 0.998 & 0.996 & 0.999 & 0.999 & 1.000 \\
 & Random Forest      & 0.992 & 0.987 & 0.974 & 1.000 & 0.994 & 1.000 \\
 & NAIM              & 0.825 & 0.773 & 0.640 & 0.906 & 0.878 & 0.880 \\
 & MA-Lasso          & 0.780 & 0.770 & 0.746 & 0.795 & 0.834 & 0.862 \\
 & NITROGEN  & 0.781 & 0.706 & 0.514 & 0.899 & 0.851 & 0.846 \\
\midrule
\multirow{7}{*}{ADNI (Val)}
 & LightGBM          & 0.744 & 0.677 & 0.459 & 0.894 & 0.821 & 0.810 \\
 & MA-GBT            & 0.721 & 0.655 & 0.442 & 0.868 & 0.803 & 0.795 \\
 & XGBoost           & 0.718 & 0.647 & 0.416 & 0.877 & 0.803 & 0.781 \\
 & Random Forest      & 0.700 & 0.584 & 0.214 & 0.955 & 0.806 & 0.769 \\
 & NITROGEN  & 0.703 & 0.633 & 0.409 & 0.857 & 0.791 & 0.761 \\
 & NAIM              & 0.702 & 0.646 & 0.467 & 0.826 & 0.784 & 0.749 \\
 & MA-Lasso          & 0.658 & 0.645 & 0.602 & 0.687 & 0.725 & 0.725 \\
\midrule
\multirow{7}{*}{ADNI (Test)}
 & MA-GBT            & 0.741 & 0.668 & 0.457 & 0.880 & 0.821 & 0.802 \\
 & LightGBM          & 0.744 & 0.672 & 0.463 & 0.881 & 0.822 & 0.799 \\
 & NITROGEN  & 0.727 & 0.648 & 0.418 & 0.877 & 0.812 & 0.794 \\
 & XGBoost           & 0.733 & 0.662 & 0.459 & 0.866 & 0.813 & 0.790 \\
 & Random Forest      & 0.706 & 0.581 & 0.221 & 0.942 & 0.812 & 0.785 \\
 & NAIM              & 0.731 & 0.668 & 0.488 & 0.849 & 0.809 & 0.781 \\
 & MA-Lasso          & 0.681 & 0.660 & 0.599 & 0.721 & 0.753 & 0.748 \\
\midrule
\multirow{7}{*}{AIBL}
 & Random Forest      & 0.400 & 0.575 & 0.206 & 0.944 & 0.454 & 0.766 \\
 & NITROGEN  & 0.453 & 0.591 & 0.300 & 0.882 & 0.460 & 0.757 \\
 & LightGBM          & 0.529 & 0.632 & 0.414 & 0.850 & 0.487 & 0.744 \\
 & MA-GBT            & 0.519 & 0.627 & 0.399 & 0.855 & 0.484 & 0.738 \\
 & XGBoost           & 0.570 & 0.636 & 0.496 & 0.776 & 0.487 & 0.733 \\
 & NAIM              & 0.510 & 0.622 & 0.385 & 0.858 & 0.480 & 0.732 \\
 & MA-Lasso          & 0.588 & 0.635 & 0.535 & 0.735 & 0.484 & 0.709 \\
\midrule
\multirow{7}{*}{OASIS-3}
 & LightGBM          & 0.288 & 0.545 & 0.159 & 0.931 & 0.304 & 0.700 \\
 & MA-Lasso          & 0.383 & 0.577 & 0.286 & 0.868 & 0.320 & 0.697 \\
 & NITROGEN  & 0.250 & 0.536 & 0.107 & 0.964 & 0.301 & 0.690 \\
 & Random Forest      & 0.215 & 0.516 & 0.064 & 0.969 & 0.292 & 0.668 \\
 & XGBoost           & 0.276 & 0.548 & 0.140 & 0.955 & 0.306 & 0.661 \\
 & MA-GBT            & 0.269 & 0.543 & 0.132 & 0.955 & 0.304 & 0.660 \\
 & NAIM              & 0.304 & 0.561 & 0.175 & 0.946 & 0.312 & 0.639 \\
\bottomrule
\end{tabular}
\end{table}

\begin{table}[t!]
\centering
\caption{
Global classification performance of models trained on ADNI for three-class diagnostic classification (CN / MCI / AD), evaluated across datasets. ADNI comprised 7,858 samples divided into stratified 60/20/20 train/validation/test splits (Train: 4,704; Val: 1,561; Test: 1,593). OASIS-3 ($n$=2,675) and AIBL ($n$=1,286) were used as held-out external validation cohorts. Metrics reported are balanced accuracy (Bal. Acc.), macro-averaged F1, macro-averaged AUC under the one-vs-rest (OvR) scheme, Matthews Correlation Coefficient (MCC), and Cohen's $\kappa$. Macro-averaged metrics weight each class equally regardless of class frequency, which is appropriate given the class imbalance present across cohorts. Models are sorted by Macro AUC within each split. Seed = 123. Full per-class metrics are reported in Table~\ref{tab:multiclass_perclass_metrics}. }
\label{tab:multiclass_global_metrics}
\begin{tabular}{llcccccc}
\toprule
Split & Model & Accuracy & Balanced Acc. & Macro F1 & Macro AUC (OvR) & MCC & Cohen \\
\midrule
\multirow{5}{*}{ADNI (Train)} & XGBoost & 1.000 & 1.000 & 1.000 & 1.000 & 1.000 & 1.000 \\
 & LightGBM & 1.000 & 1.000 & 1.000 & 1.000 & 1.000 & 1.000 \\
 & Random Forest & 0.982 & 0.976 & 0.981 & 0.999 & 0.972 & 0.972 \\
 & NITROGEN & 0.692 & 0.681 & 0.691 & 0.861 & 0.511 & 0.509 \\
 & NAIM & 0.626 & 0.601 & 0.615 & 0.801 & 0.399 & 0.393 \\
\midrule
\multirow{5}{*}{ADNI (Val)} & XGBoost & 0.582 & 0.580 & 0.589 & 0.782 & 0.342 & 0.339 \\
 & LightGBM & 0.596 & 0.594 & 0.602 & 0.778 & 0.366 & 0.362 \\
 & Random Forest & 0.570 & 0.550 & 0.557 & 0.765 & 0.326 & 0.304 \\
 & NITROGEN & 0.541 & 0.543 & 0.552 & 0.737 & 0.277 & 0.276 \\
 & NAIM & 0.523 & 0.519 & 0.526 & 0.719 & 0.247 & 0.242 \\
\midrule
\multirow{5}{*}{ADNI (Test)} & XGBoost & 0.591 & 0.573 & 0.588 & 0.771 & 0.351 & 0.345 \\
 & LightGBM & 0.588 & 0.579 & 0.590 & 0.762 & 0.350 & 0.347 \\
 & Random Forest & 0.560 & 0.524 & 0.534 & 0.762 & 0.302 & 0.278 \\
 & NITROGEN & 0.556 & 0.543 & 0.556 & 0.752 & 0.295 & 0.292 \\
 & NAIM & 0.504 & 0.478 & 0.492 & 0.707 & 0.203 & 0.196 \\
\midrule
\multirow{5}{*}{OASIS-3} & LightGBM & 0.265 & 0.383 & 0.247 & 0.596 & 0.117 & 0.062 \\
 & Random Forest & 0.116 & 0.340 & 0.150 & 0.595 & 0.081 & 0.024 \\
 & NAIM & 0.209 & 0.331 & 0.177 & 0.594 & 0.052 & 0.026 \\
 & NITROGEN & 0.233 & 0.369 & 0.247 & 0.583 & 0.123 & 0.056 \\
 & XGBoost & 0.206 & 0.339 & 0.209 & 0.574 & 0.091 & 0.044 \\
\midrule
\multirow{5}{*}{AIBL} & NAIM & 0.394 & 0.479 & 0.410 & 0.729 & 0.195 & 0.127 \\
 & Random Forest & 0.374 & 0.468 & 0.395 & 0.728 & 0.189 & 0.115 \\
 & LightGBM & 0.467 & 0.492 & 0.454 & 0.726 & 0.209 & 0.158 \\
 & XGBoost & 0.473 & 0.470 & 0.442 & 0.725 & 0.192 & 0.149 \\
 & NITROGEN & 0.363 & 0.463 & 0.395 & 0.719 & 0.169 & 0.104 \\
\bottomrule
\end{tabular}
\end{table}

\begin{table}[t!]
\centering
\caption{Per-class classification performance (Recall or sensitivity, specificity, and AUC) for three-class diagnostic classification (CN / MCI / AD), evaluated across datasets. Sensitivity (Sens.) corresponds to the true positive rate for each class; specificity (Spec.) corresponds to the true negative rate. AUC is reported per class under the one-vs-rest (OvR) scheme. Models are sorted by Macro AUC within each split (same order as Table~\ref{tab:multiclass_global_metrics}). Seed = 123. Global metrics are reported in Table~\ref{tab:multiclass_global_metrics}. }
\label{tab:multiclass_perclass_metrics}
\begin{tabular}{llccccccccc}
\toprule
& & \multicolumn{3}{c}{CN} & \multicolumn{3}{c}{MCI} & \multicolumn{3}{c}{AD} \\
\cmidrule(lr){3-5} \cmidrule(lr){6-8} \cmidrule(lr){9-11}
Split & Model & Rec. & Spec. & AUC & Rec. & Spec. & AUC & Rec. & Spec. & AUC \\
\midrule
\multirow{5}{*}{ADNI (Train)} & XGBoost & 1.000 & 1.000 & 1.000 & 1.000 & 1.000 & 1.000 & 1.000 & 1.000 & 1.000 \\
 & LightGBM & 1.000 & 1.000 & 1.000 & 1.000 & 1.000 & 1.000 & 1.000 & 1.000 & 1.000 \\
 & Random Forest & 0.989 & 0.996 & 1.000 & 0.998 & 0.973 & 1.000 & 0.941 & 0.999 & 0.998 \\
 & NITROGEN & 0.648 & 0.882 & 0.876 & 0.740 & 0.674 & 0.799 & 0.654 & 0.933 & 0.907 \\
 & NAIM & 0.510 & 0.870 & 0.803 & 0.736 & 0.569 & 0.713 & 0.555 & 0.931 & 0.886 \\
\midrule
\multirow{5}{*}{ADNI (Val)} & XGBoost & 0.496 & 0.828 & 0.798 & 0.635 & 0.572 & 0.670 & 0.609 & 0.918 & 0.877 \\
 & LightGBM & 0.498 & 0.848 & 0.802 & 0.660 & 0.578 & 0.667 & 0.624 & 0.915 & 0.866 \\
 & Random Forest & 0.342 & 0.910 & 0.781 & 0.765 & 0.439 & 0.646 & 0.542 & 0.930 & 0.869 \\
 & NITROGEN & 0.513 & 0.778 & 0.748 & 0.550 & 0.557 & 0.589 & 0.566 & 0.918 & 0.874 \\
 & NAIM & 0.390 & 0.814 & 0.735 & 0.615 & 0.496 & 0.565 & 0.551 & 0.910 & 0.856 \\
\midrule
\multirow{5}{*}{ADNI (Test)} & XGBoost & 0.516 & 0.854 & 0.799 & 0.680 & 0.541 & 0.654 & 0.523 & 0.928 & 0.859 \\
 & LightGBM & 0.532 & 0.833 & 0.781 & 0.644 & 0.573 & 0.647 & 0.562 & 0.919 & 0.859 \\
 & Random Forest & 0.332 & 0.918 & 0.788 & 0.779 & 0.404 & 0.643 & 0.460 & 0.931 & 0.856 \\
 & NITROGEN & 0.478 & 0.809 & 0.775 & 0.631 & 0.535 & 0.619 & 0.521 & 0.926 & 0.863 \\
 & NAIM & 0.382 & 0.825 & 0.744 & 0.642 & 0.423 & 0.547 & 0.411 & 0.926 & 0.830 \\
\midrule
\multirow{5}{*}{OASIS-3} & LightGBM & 0.223 & 0.895 & 0.677 & 0.449 & 0.454 & 0.429 & 0.477 & 0.787 & 0.684 \\
 & Random Forest & 0.092 & 0.951 & 0.684 & 0.755 & 0.139 & 0.361 & 0.173 & 0.965 & 0.740 \\
 & NAIM & 0.136 & 0.937 & 0.686 & 0.245 & 0.648 & 0.441 & 0.613 & 0.489 & 0.656 \\
 & NITROGEN & 0.201 & 0.899 & 0.680 & 0.531 & 0.323 & 0.378 & 0.374 & 0.899 & 0.691 \\
 & XGBoost & 0.154 & 0.911 & 0.653 & 0.388 & 0.406 & 0.380 & 0.475 & 0.771 & 0.690 \\
\midrule
\multirow{5}{*}{AIBL} & NAIM & 0.332 & 0.873 & 0.750 & 0.745 & 0.351 & 0.555 & 0.361 & 0.981 & 0.881 \\
 & Random Forest & 0.304 & 0.888 & 0.762 & 0.777 & 0.311 & 0.537 & 0.323 & 0.993 & 0.886 \\
 & LightGBM & 0.447 & 0.805 & 0.724 & 0.630 & 0.460 & 0.557 & 0.400 & 0.979 & 0.896 \\
 & XGBoost & 0.470 & 0.785 & 0.728 & 0.598 & 0.467 & 0.543 & 0.342 & 0.984 & 0.905 \\
 & NITROGEN & 0.291 & 0.873 & 0.745 & 0.723 & 0.312 & 0.505 & 0.374 & 0.984 & 0.907 \\
\bottomrule
\end{tabular}
\end{table}

\begin{figure}[t!]
\centering
\begin{subfigure}[t]{0.9\textwidth}
    \centering
    \includegraphics[width=\linewidth]{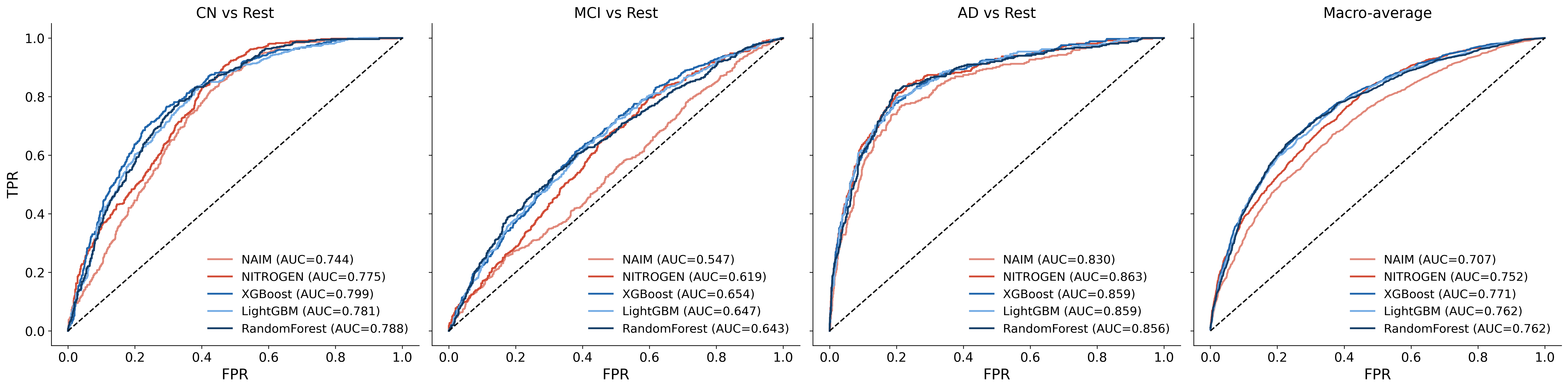}
    \caption{ADNI test subset}
    \label{fig:multiclass_val}
\end{subfigure}
\begin{subfigure}[t]{0.9\textwidth}
    \centering
    \includegraphics[width=\linewidth]{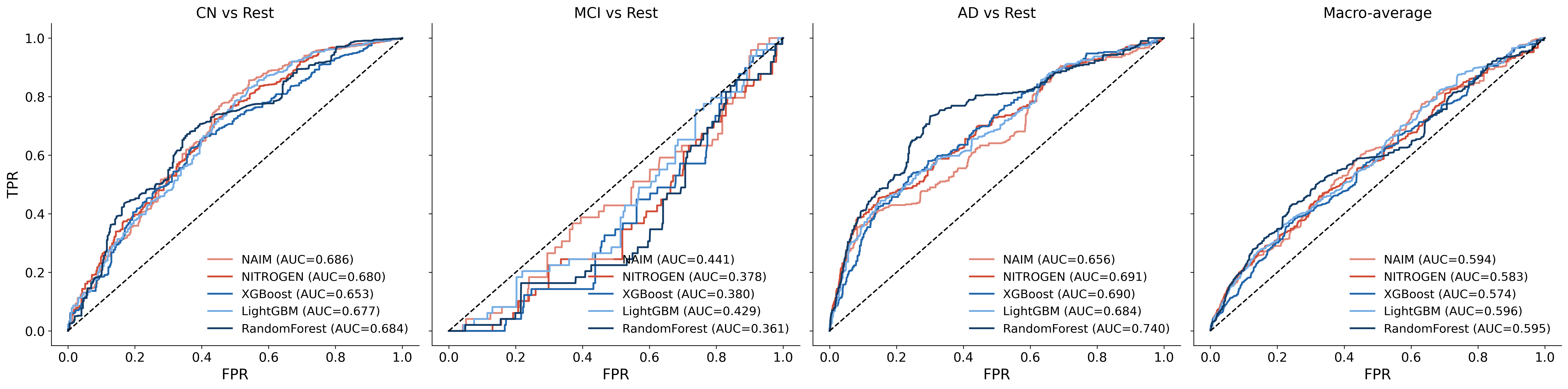}
    \caption{OASIS-3 test set}
    \label{fig:multiclass_oasis}
\end{subfigure}
\begin{subfigure}[t]{0.9\textwidth}
    \centering
    \includegraphics[width=\linewidth]{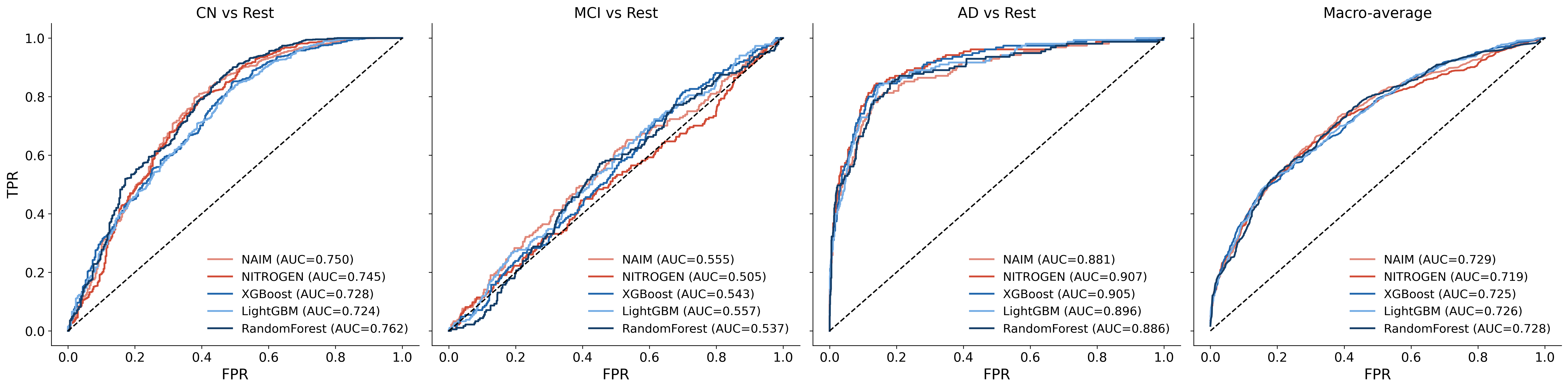}
    \caption{AIBL test set}
    \label{fig:multiclass_aibl}
\end{subfigure}
\caption{
\textbf{Classification performance in multi-class cognitive group classification.} ROC curves (One vs. Rest) for multi-class classification on (a) ADNI test set, (b) OASIS-3 and (c) AIBL. For each test set, the plots represent the ROC curves for the CN, MCI, AD vs. the rest and macro average from left to right. AUC values are reported in the legends. All models trained on ADNI. }
\label{fig:classification_multiclass}
\end{figure}

\begin{landscape}
\begin{table}[t!]
\centering
\caption{Performance of models trained on ADNI for regression of MMSE and CDR-SB scores, evaluated across datasets. ADNI comprised 7,858 samples (CN: 2,497, MCI: 3,583, AD: 1,778) divided into 60/20/20 train/validation/test splits. OASIS-3 (CN: 2,228, MCI: 49, AD: 398) and AIBL (CN: 947, MCI: 184, AD: 155) were used as held-out external validation cohorts. Within each split, models are sorted by MMSE R$^2$. Seed = 123. }
\label{tab:regression_results_supp}
\begin{tabular}{llcccccccccc}
\toprule
 & & \multicolumn{5}{c}{MMSE} & \multicolumn{5}{c}{CDR-SB} \\
\cmidrule(lr){3-7} \cmidrule(lr){8-12}
Data/Split & Model & MSE & MAE & R$^2$ & $r$ & $p$ & MSE & MAE & R$^2$ & $r$ & $p$ \\
\midrule
\multirow{4}{*}{ADNI (Train)}
 & XGBoost           & 0.000 & 0.000 & 1.000 & 1.000 & $0$                     & 0.000 & 0.000 & 1.000 & 1.000 & $0$ \\
 & Random Forest     & 3.075 & 1.159 & 0.788 & 0.918 & $0$                     & 1.626 & 0.828 & 0.736 & 0.892 & $0$ \\
 & NITROGEN          & 6.721 & 1.843 & 0.536 & 0.733 & $0$                     & 3.079 & 1.185 & 0.500 & 0.713 & $0$ \\
 & NAIM              & 7.151 & 1.876 & 0.507 & 0.714 & $0$                     & 3.121 & 1.223 & 0.493 & 0.703 & $0$ \\
\midrule
\multirow{4}{*}{ADNI (Val)}
 & NITROGEN          & 7.745 & 2.027 & 0.446 & 0.670 & $1.18 \mathrm{e}{-203}$ & 3.586 & 1.309 & 0.415 & 0.648 & $1.10 \mathrm{e}{-186}$ \\
 & Random Forest     & 8.015 & 2.076 & 0.427 & 0.661 & $6.46 \mathrm{e}{-197}$ & 3.633 & 1.363 & 0.407 & 0.650 & $4.09 \mathrm{e}{-188}$ \\
 & NAIM              & 8.389 & 2.081 & 0.400 & 0.636 & $5.69 \mathrm{e}{-178}$ & 3.815 & 1.384 & 0.377 & 0.616 & $1.41 \mathrm{e}{-163}$ \\
 & XGBoost           & 9.049 & 2.239 & 0.353 & 0.617 & $2.68 \mathrm{e}{-164}$ & 3.811 & 1.384 & 0.378 & 0.632 & $3.58 \mathrm{e}{-175}$ \\
\midrule
\multirow{4}{*}{ADNI (Test)}
 & NAIM              & 8.206 & 2.007 & 0.383 & 0.627 & $1.58 \mathrm{e}{-174}$ & 3.486 & 1.337 & 0.402 & 0.635 & $9.47 \mathrm{e}{-181}$ \\
 & NITROGEN          & 8.216 & 1.999 & 0.382 & 0.621 & $1.67 \mathrm{e}{-170}$ & 3.517 & 1.302 & 0.397 & 0.641 & $3.19 \mathrm{e}{-185}$ \\
 & Random Forest     & 8.417 & 2.069 & 0.367 & 0.610 & $8.87 \mathrm{e}{-163}$ & 3.578 & 1.352 & 0.386 & 0.633 & $6.63 \mathrm{e}{-179}$ \\
 & XGBoost           & 9.450 & 2.205 & 0.289 & 0.569 & $2.41 \mathrm{e}{-137}$ & 3.853 & 1.409 & 0.339 & 0.604 & $1.10 \mathrm{e}{-158}$ \\
\midrule
\multirow{4}{*}{AIBL}
 & NITROGEN          & 8.255 & 1.727 & 0.384 & 0.664 & $2.67 \mathrm{e}{-164}$ & 1.551 & 0.901 & -8.771 & 0.603 & $2.87 \mathrm{e}{-128}$ \\
 & NAIM              & 8.571 & 1.775 & 0.360 & 0.635 & $4.06 \mathrm{e}{-146}$ & 2.402 & 1.246 & -14.136 & 0.554 & $1.87 \mathrm{e}{-104}$ \\
 & Random Forest     & 8.744 & 1.767 & 0.347 & 0.666 & $2.45 \mathrm{e}{-165}$ & 1.642 & 1.095 & -9.344 & 0.656 & $2.91 \mathrm{e}{-159}$ \\
 & XGBoost           & 9.196 & 1.871 & 0.313 & 0.579 & $3.39 \mathrm{e}{-116}$ & 2.127 & 1.076 & -12.405 & 0.568 & $8.88 \mathrm{e}{-111}$ \\
\midrule
\multirow{4}{*}{OASIS-3}
 & Random Forest     & 9.738 & 2.426 & -0.637 & 0.285 & $3.02 \mathrm{e}{-51}$ & 4.797 & 1.879 & -62.886 & 0.284 & $6.31 \mathrm{e}{-51}$ \\
 & NAIM              & 11.928 & 2.770 & -1.005 & 0.303 & $5.02 \mathrm{e}{-58}$ & 7.896 & 2.456 & -104.163 & 0.247 & $2.24 \mathrm{e}{-38}$ \\
 & XGBoost           & 13.555 & 2.270 & -1.278 & 0.278 & $8.68 \mathrm{e}{-49}$ & 7.253 & 1.692 & -95.592 & 0.263 & $1.09 \mathrm{e}{-43}$ \\
 & NITROGEN          & 16.326 & 3.004 & -1.744 & 0.245 & $5.36 \mathrm{e}{-38}$ & 5.725 & 2.079 & -75.252 & 0.255 & $4.77 \mathrm{e}{-41}$ \\
\bottomrule
\end{tabular}
\end{table}
\end{landscape}

\begin{figure}[!t]
\centering
\begin{subfigure}[t]{0.32\textwidth}
\centering
\textbf{ADNI (Test)}\\[2pt]
\includegraphics[width=\textwidth]{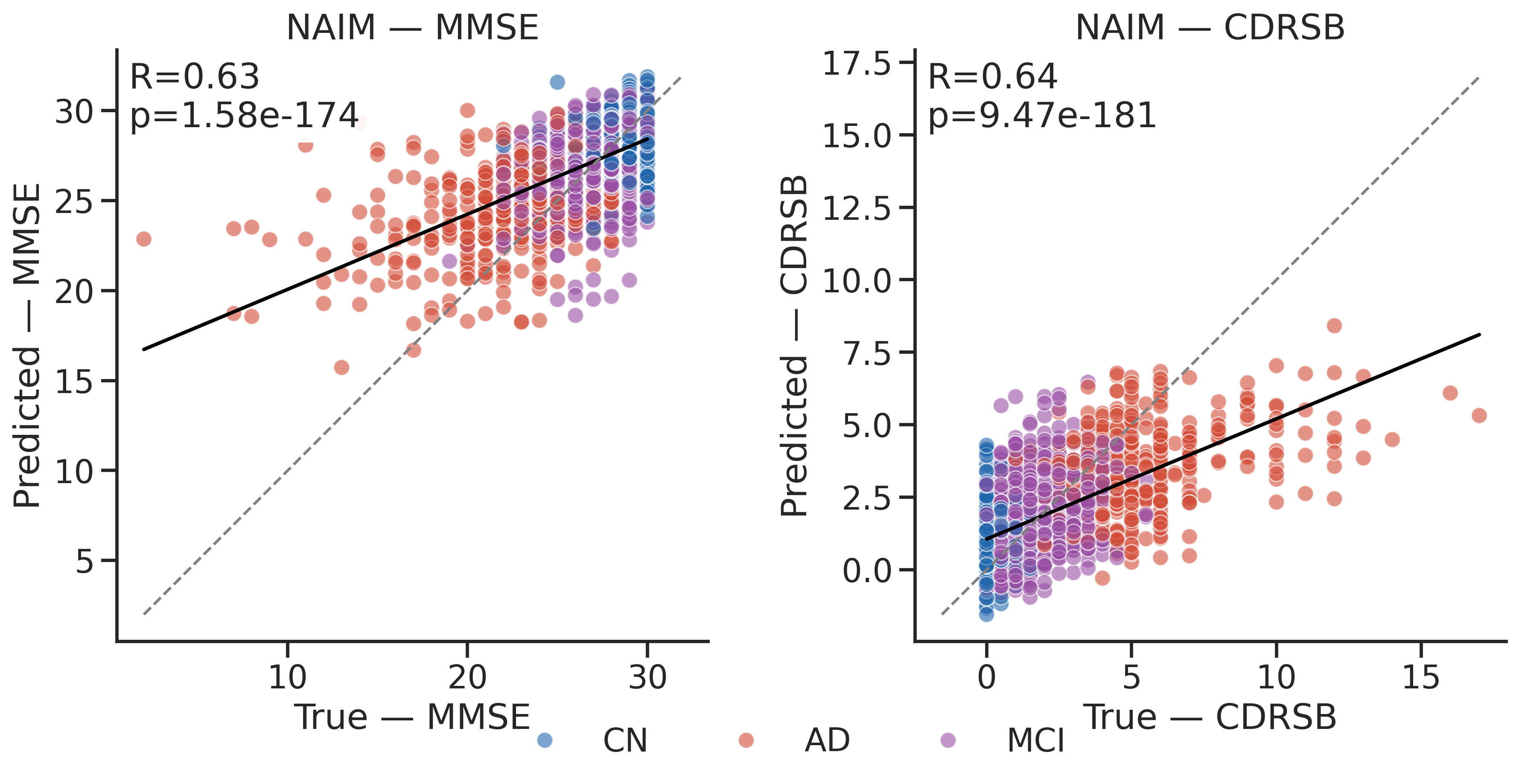}\\
\includegraphics[width=\textwidth]{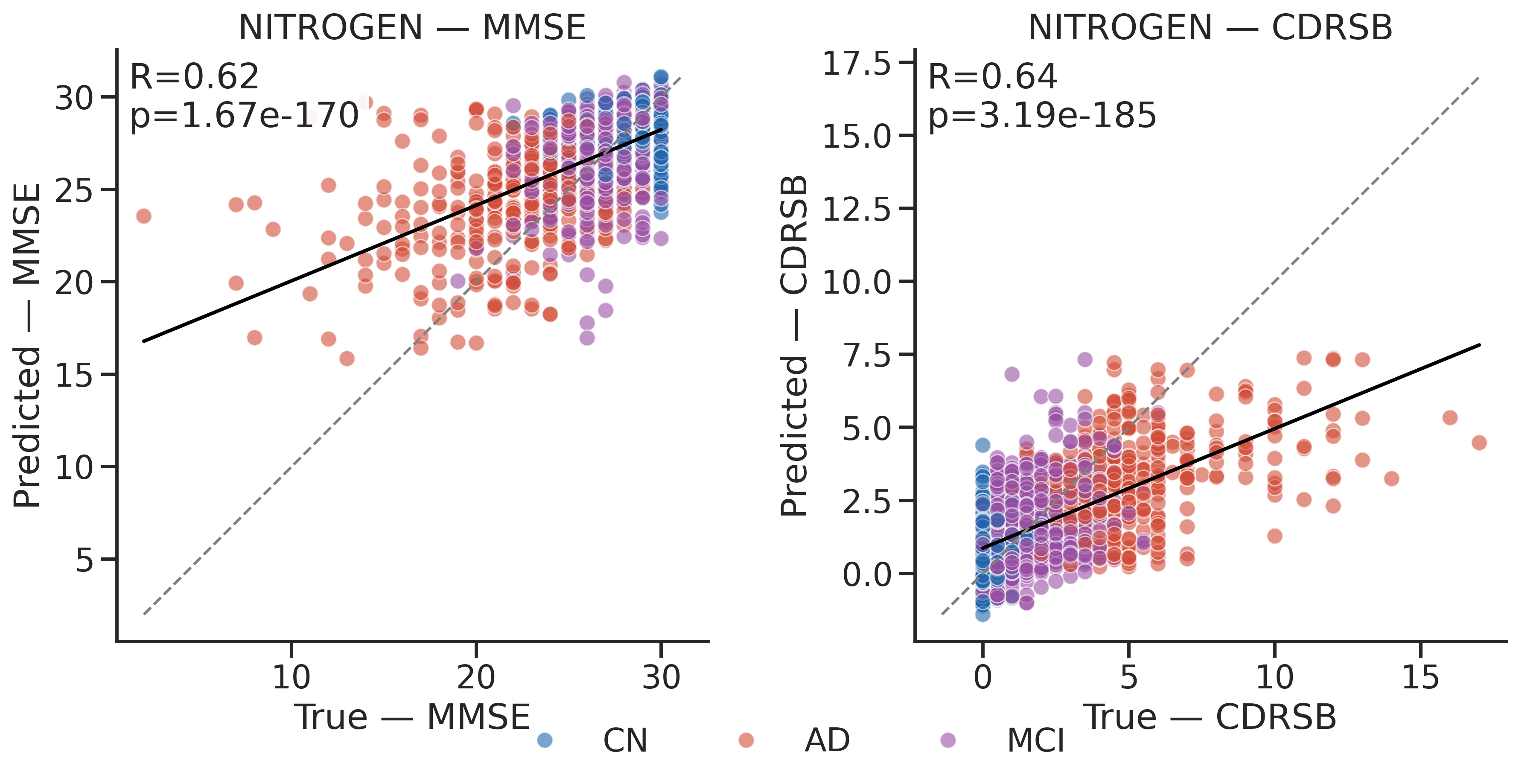}\\
\includegraphics[width=\textwidth]{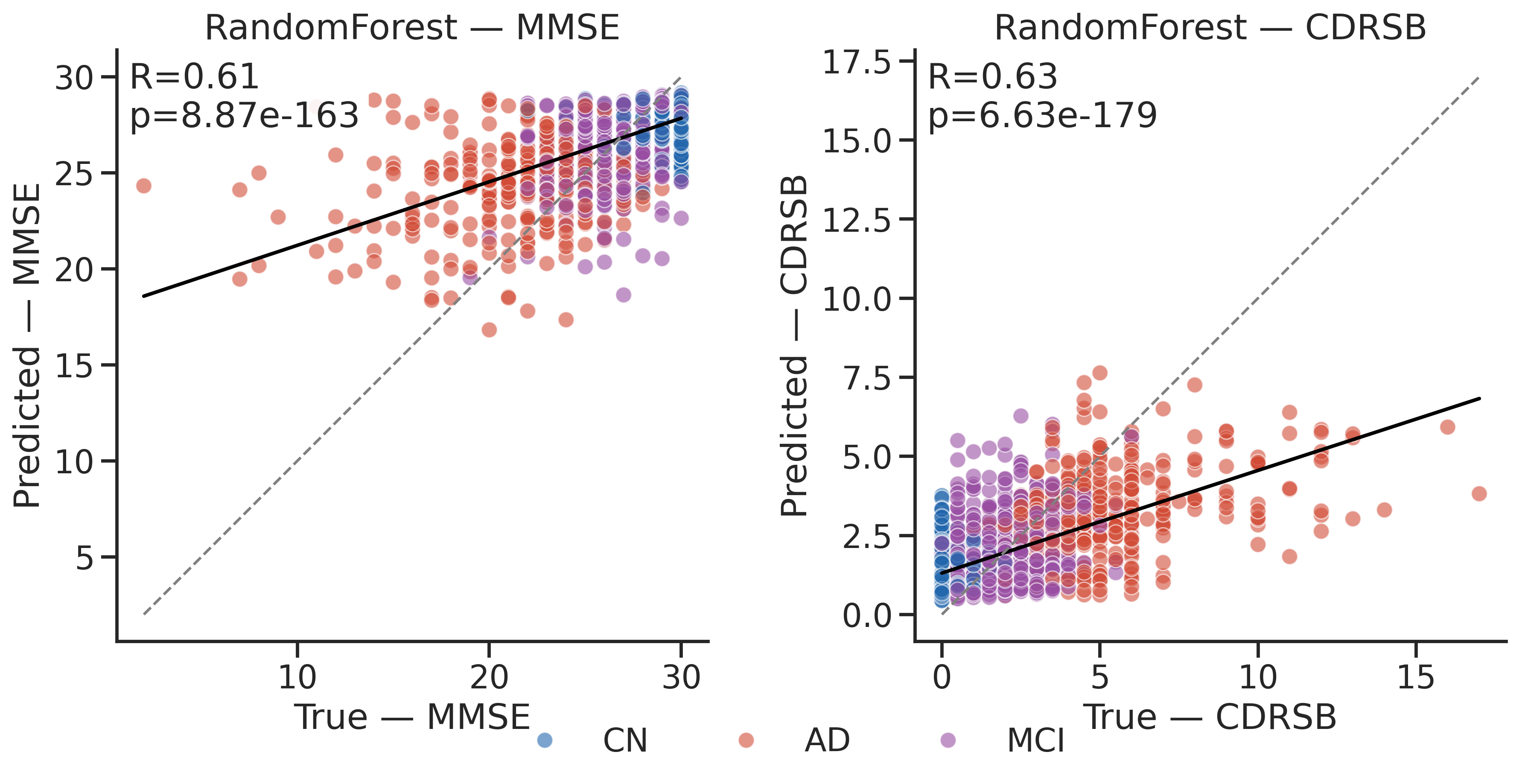}\\
\includegraphics[width=\textwidth]{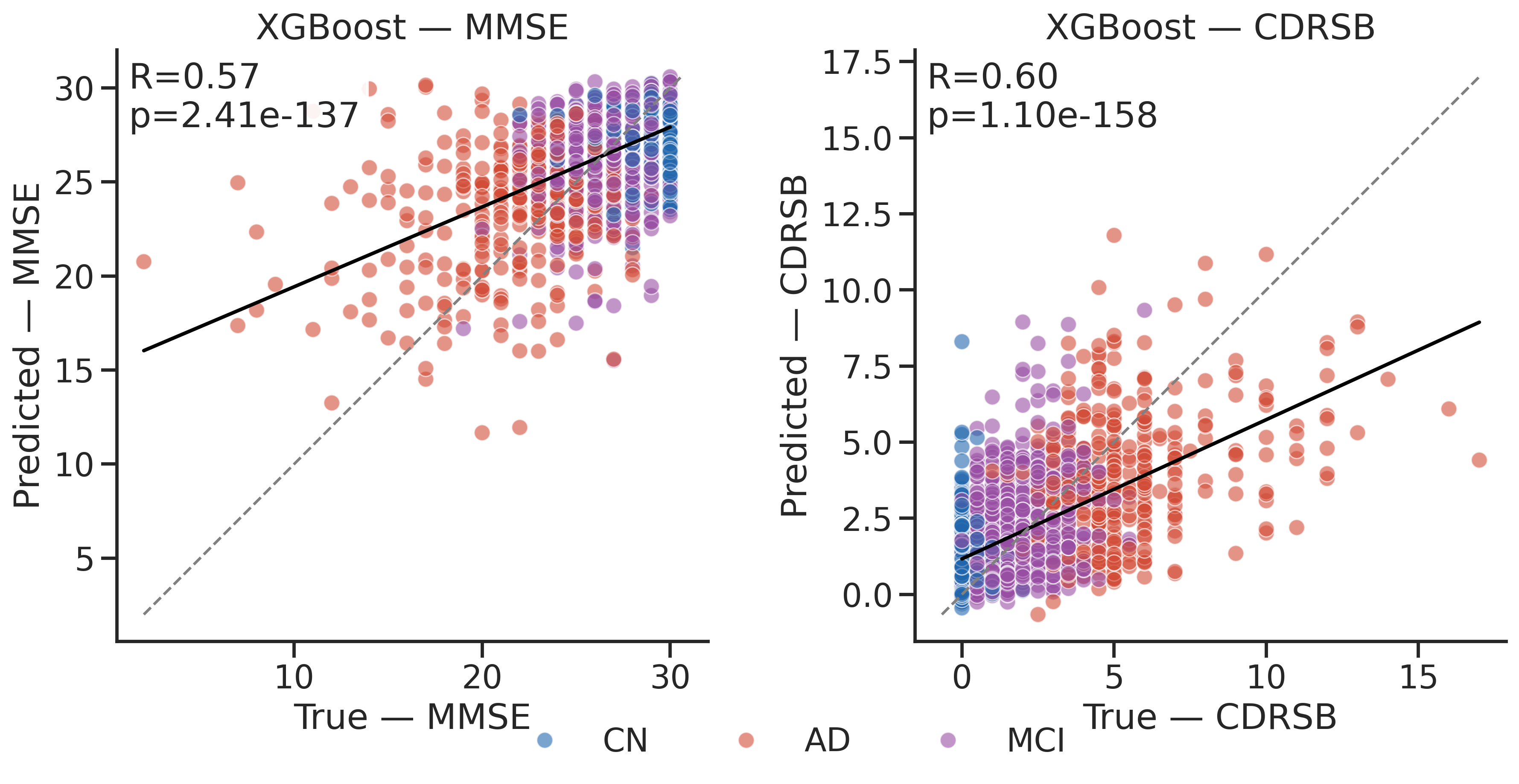}
\end{subfigure}
\begin{subfigure}[t]{0.32\textwidth}
\centering
\textbf{AIBL}\\[2pt]
\includegraphics[width=\textwidth]{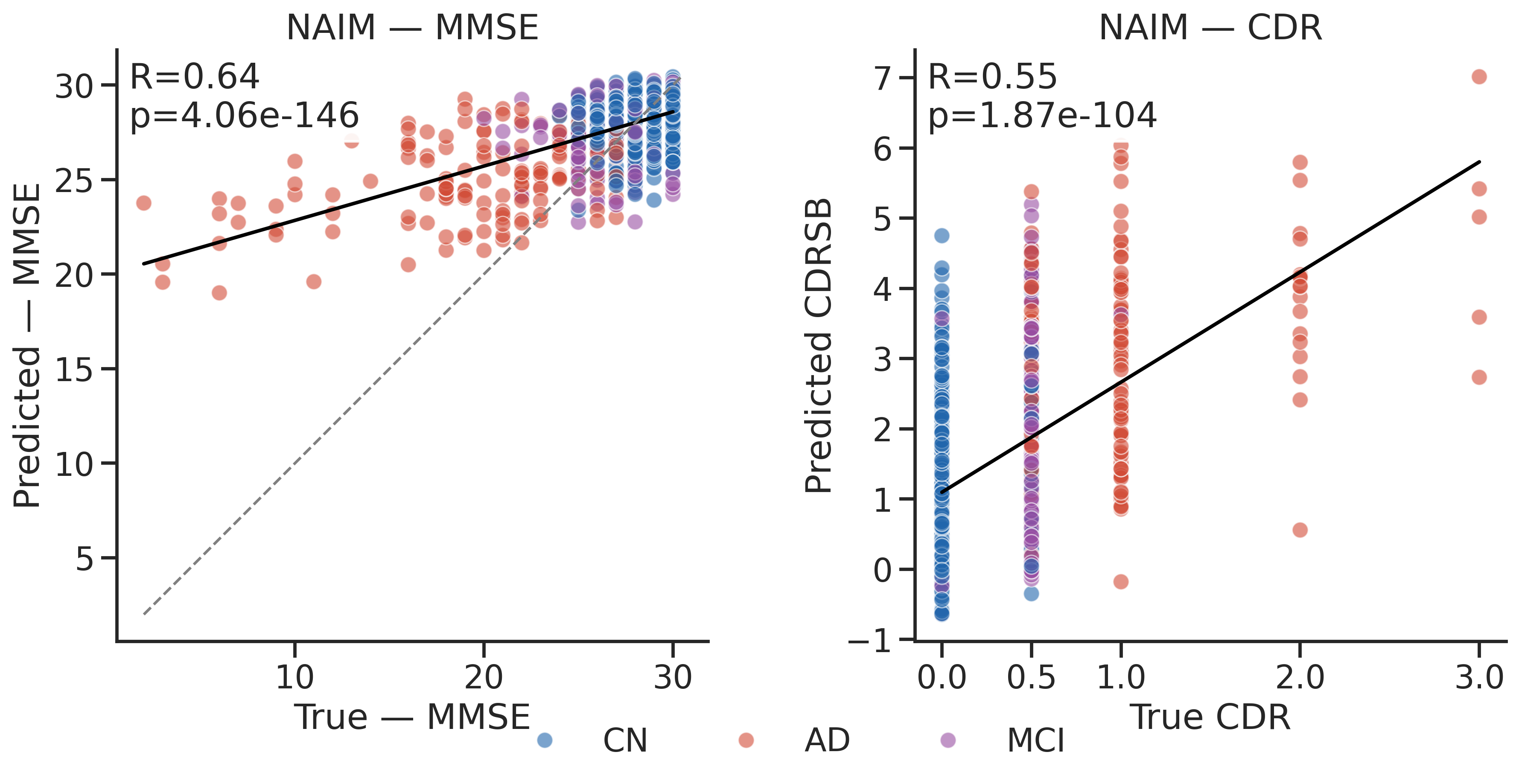}\\
\includegraphics[width=\textwidth]{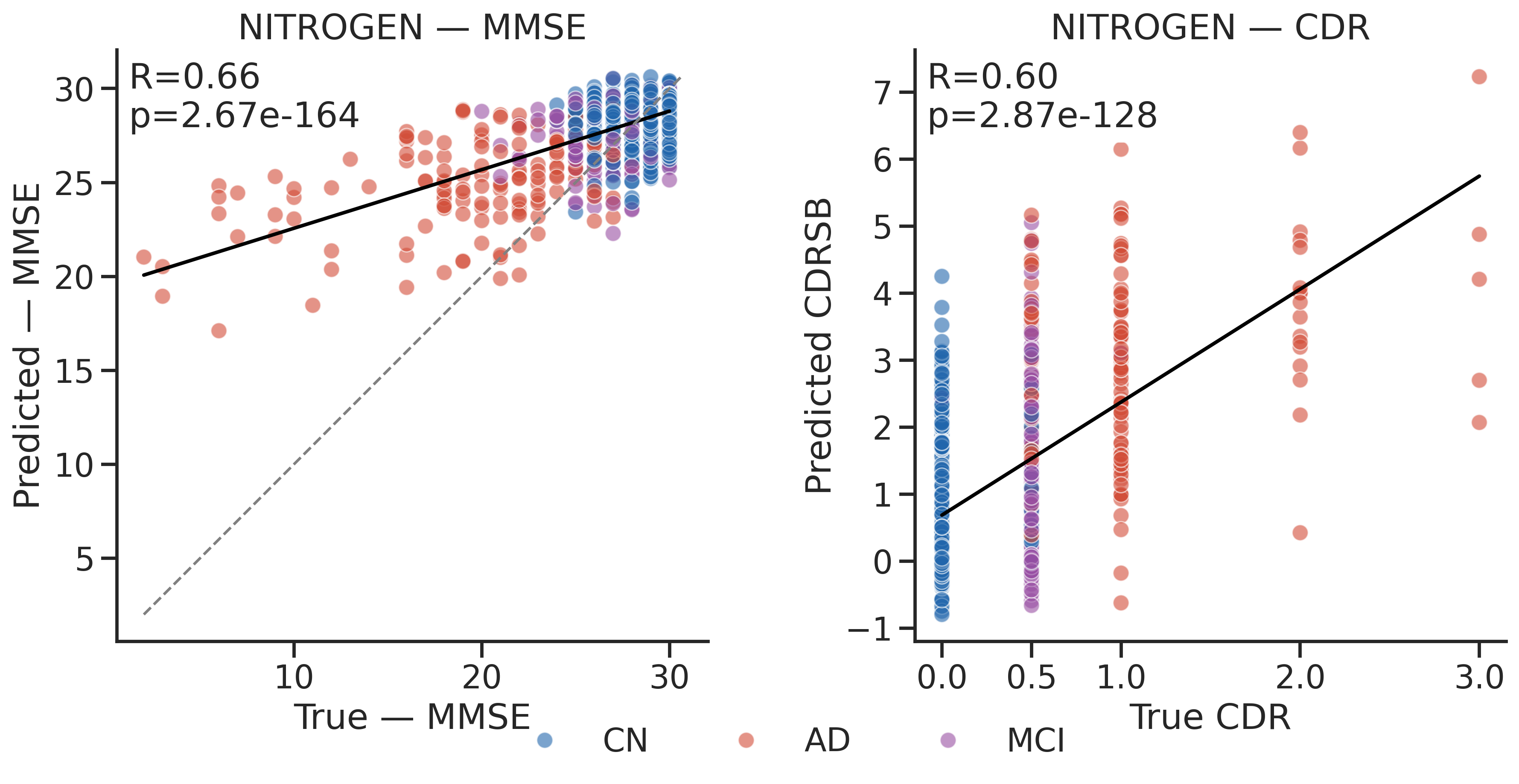}\\
\includegraphics[width=\textwidth]{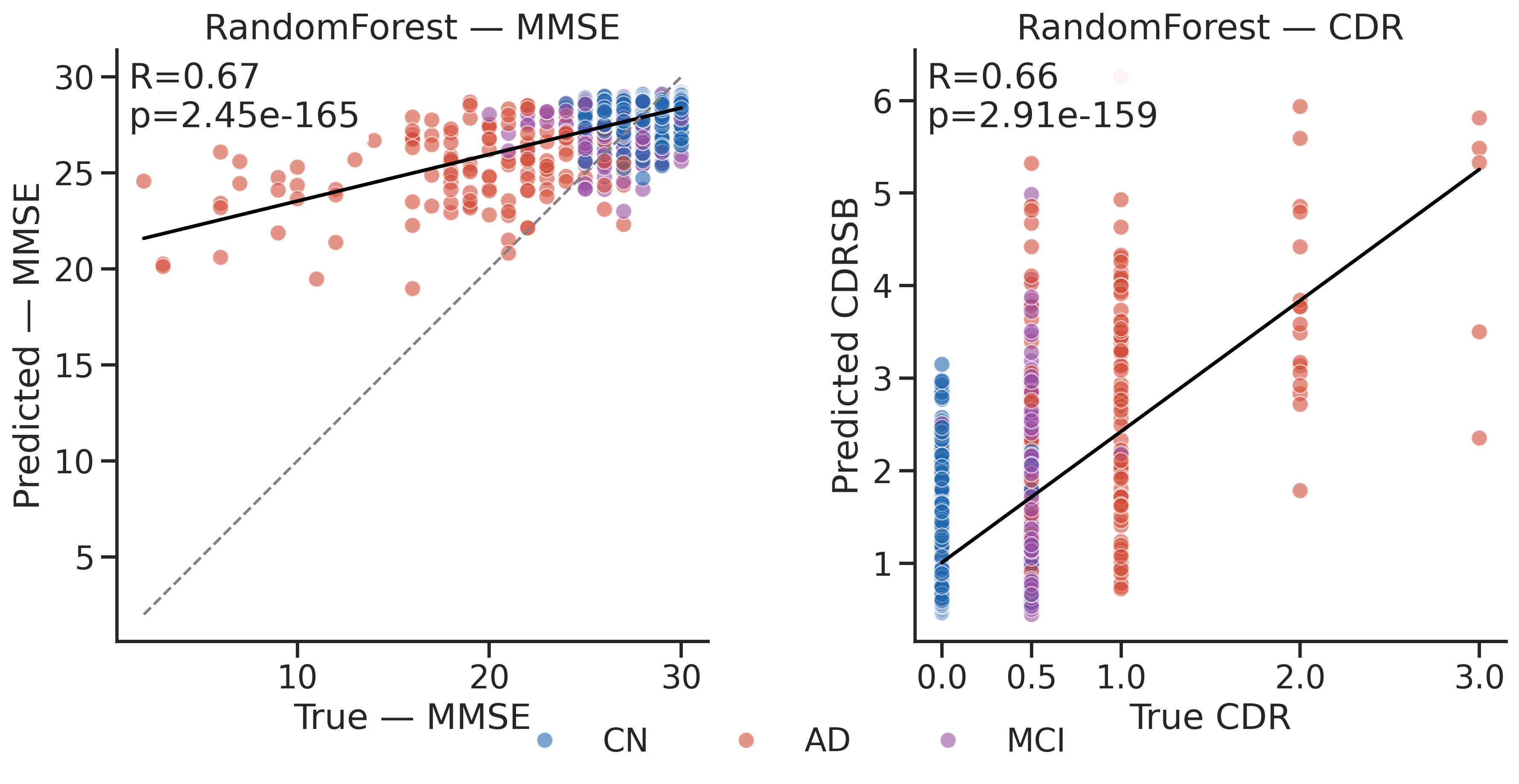}\\
\includegraphics[width=\textwidth]{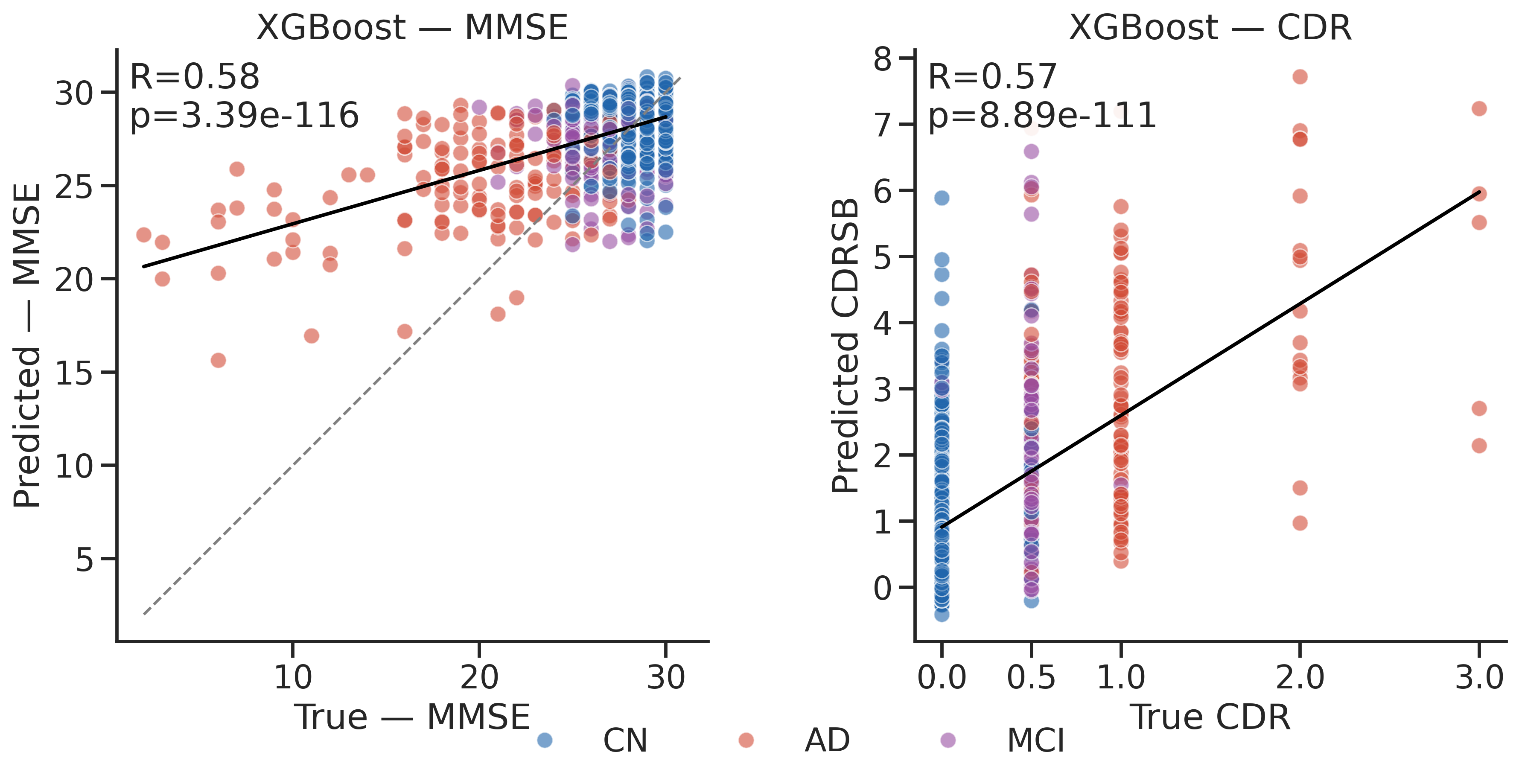}
\end{subfigure}
\begin{subfigure}[t]{0.32\textwidth}
\centering
\textbf{OASIS-3}\\[2pt]
\includegraphics[width=\textwidth]{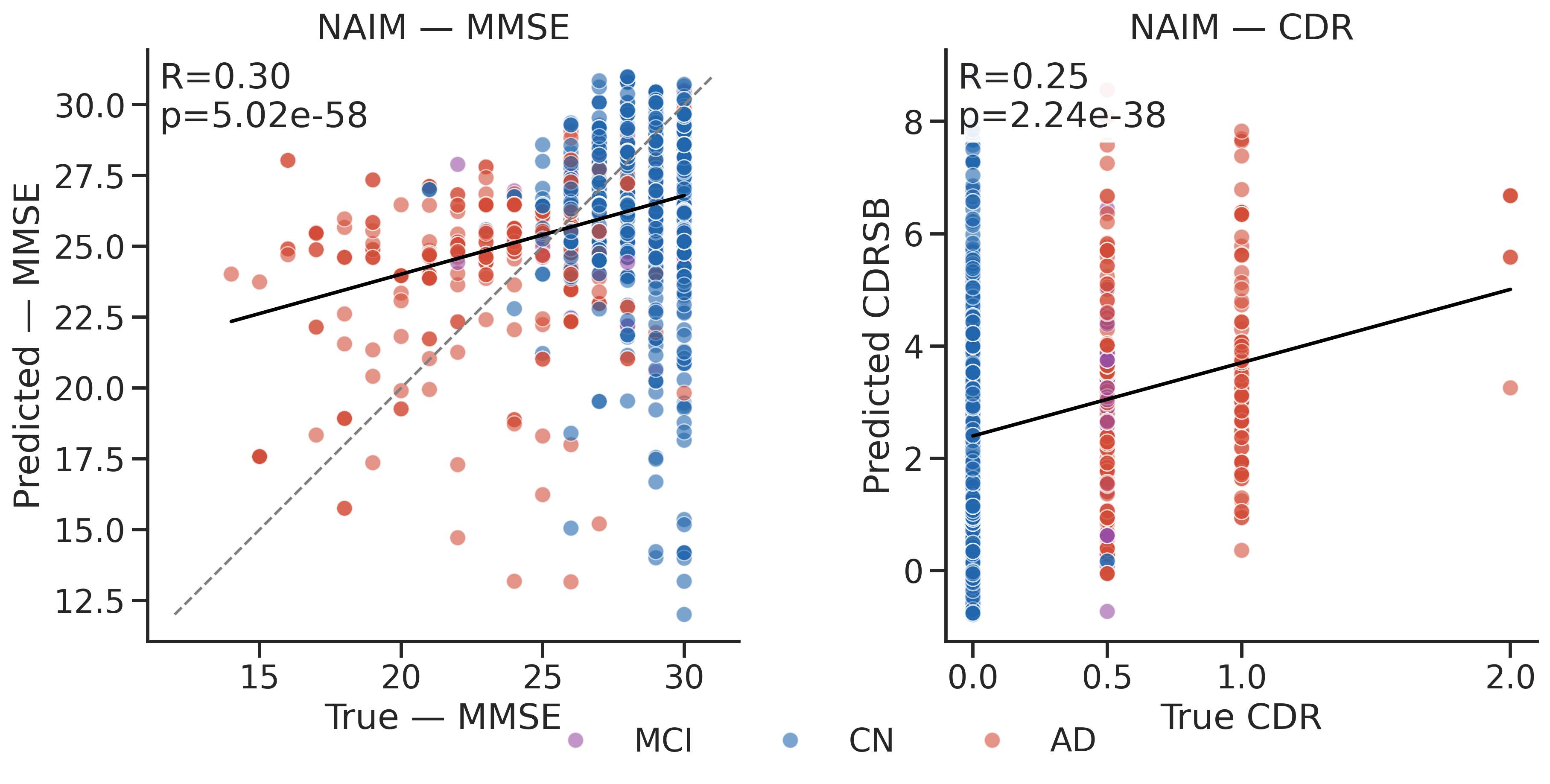}\\
\includegraphics[width=\textwidth]{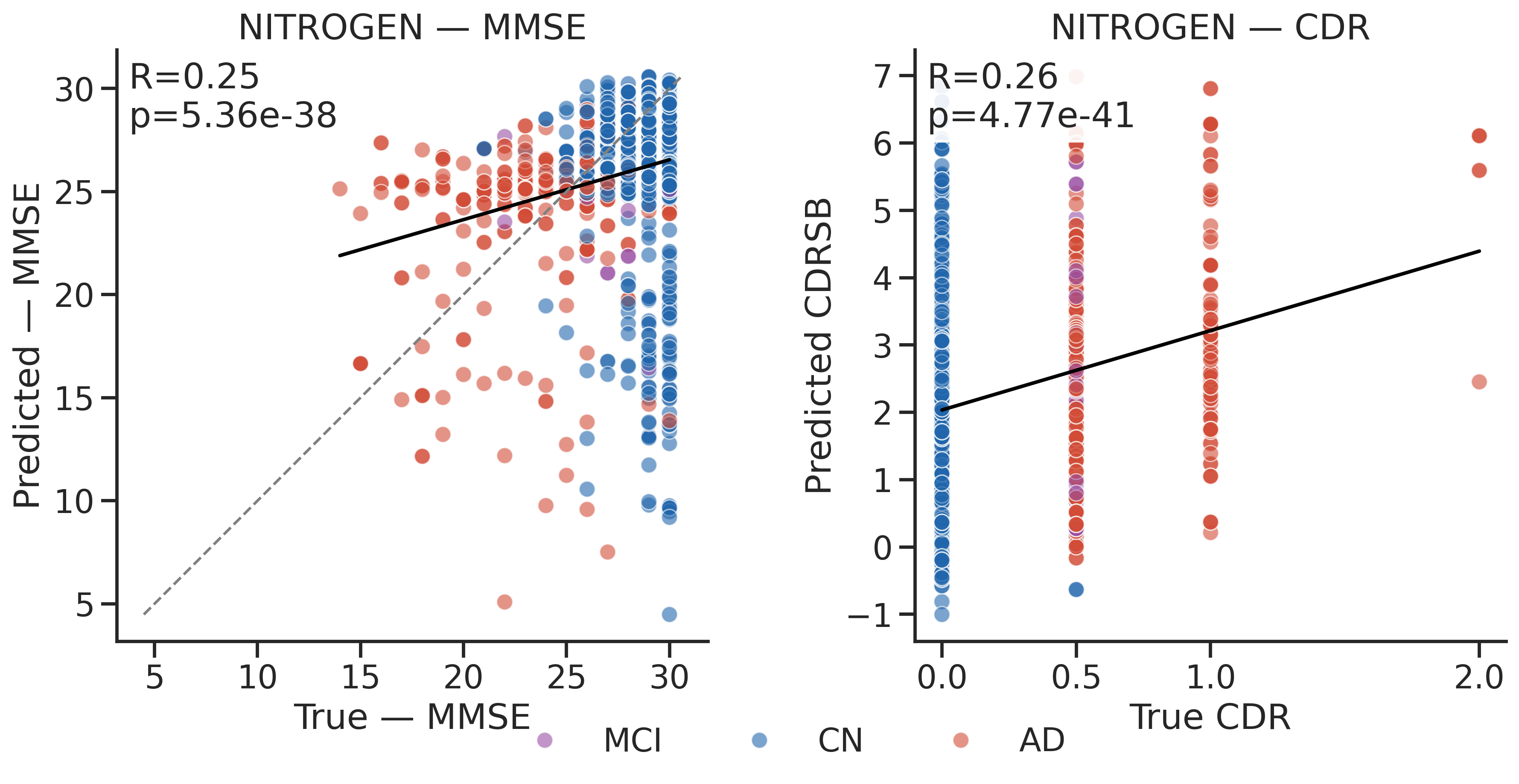}\\
\includegraphics[width=\textwidth]{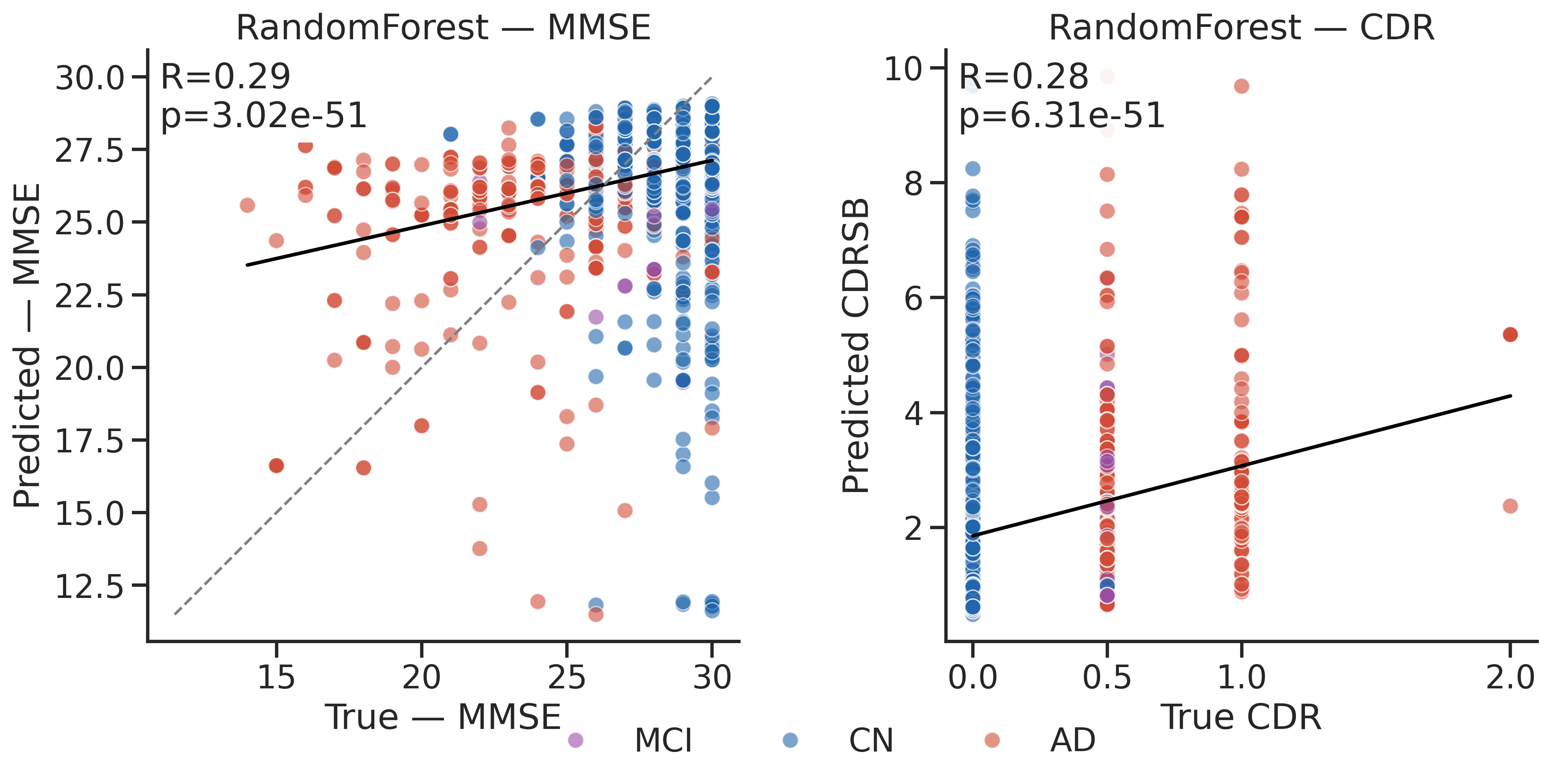}\\
\includegraphics[width=\textwidth]{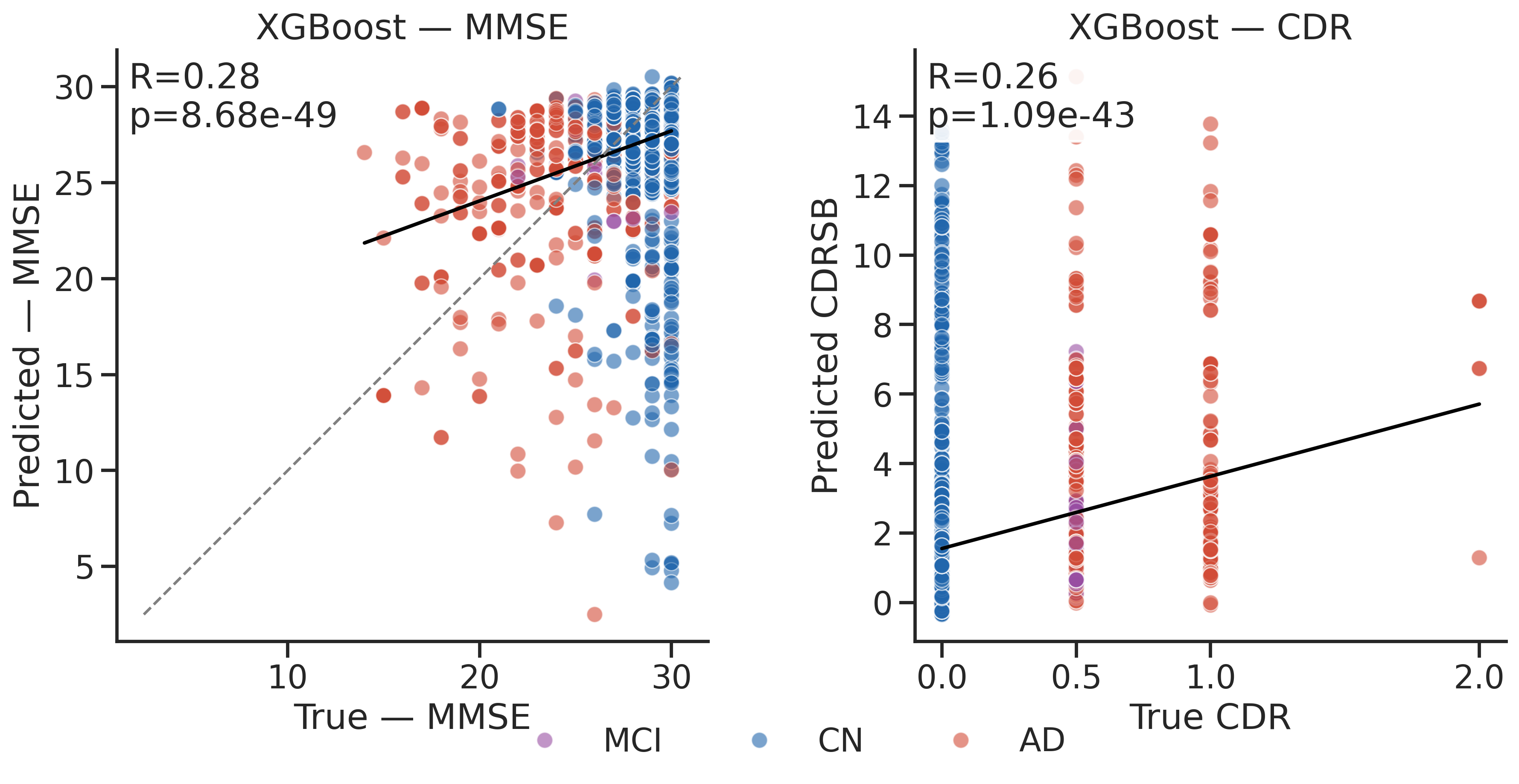}
\end{subfigure}
\caption{
\textbf{Regression performance across models and datasets.}
Predicted versus observed values for MMSE and CDR-SB scores across test datasets (columns: ADNI, AIBL, OASIS) and models (rows). 
Each point represents an individual, coloured by diagnostic group (CN: green, MCI: purple, AD: red). 
The dashed line indicates perfect regression, and the solid line shows the least-squares fit. 
Pearson correlation coefficients ($r$) are reported in each panel. 
Models were trained on ADNI and evaluated on internal (ADNI) and external (AIBL, OASIS) cohorts.
}
\label{fig:regression_all_models}
\end{figure}

\begin{figure}[t!]
\centering
\begin{subfigure}[t]{0.48\textwidth}
    \centering
    \includegraphics[width=\linewidth]{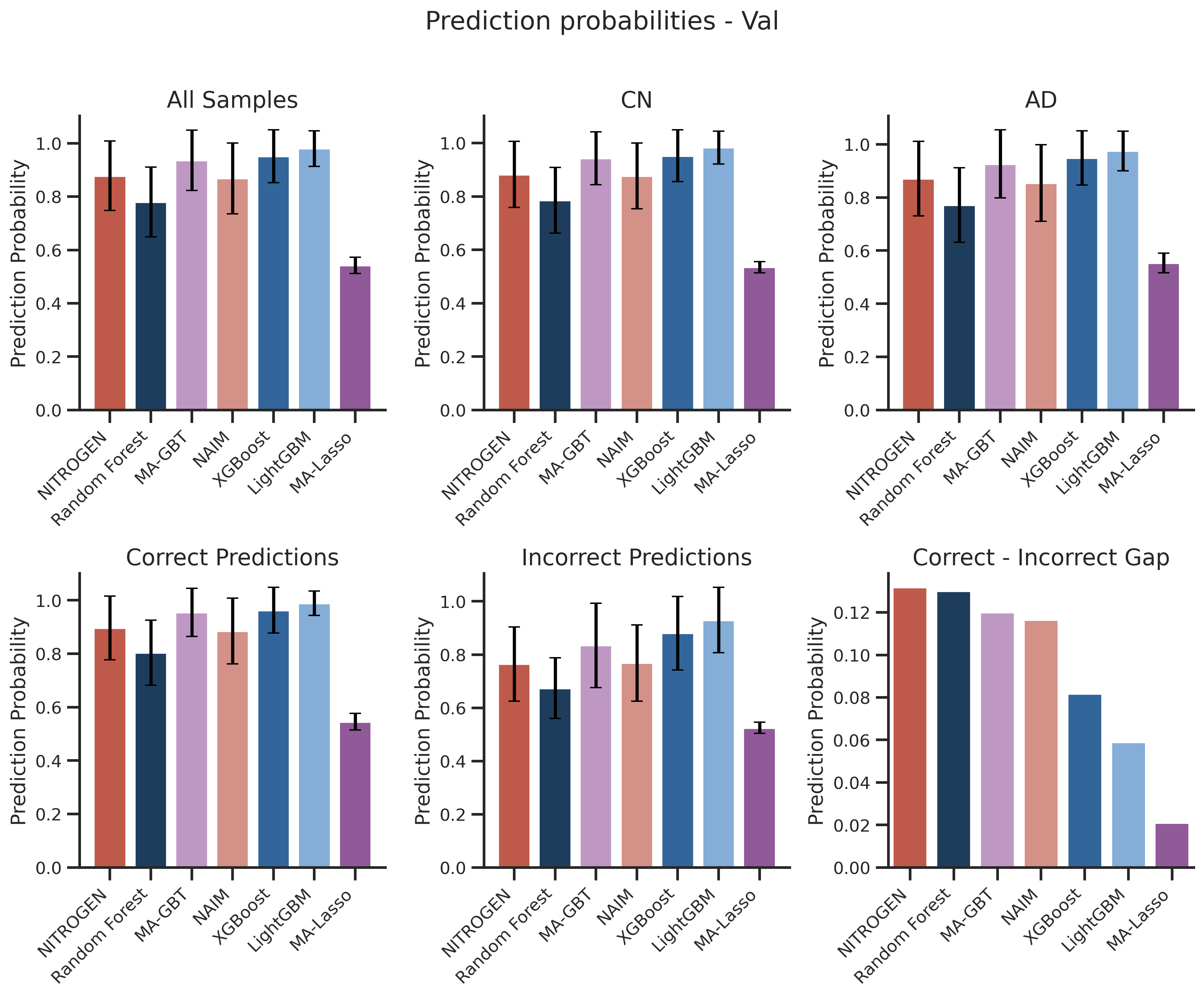}
    \caption{ADNI validation subset}
    \label{fig:predict_prob_val}
\end{subfigure}
\begin{subfigure}[t]{0.48\textwidth}
    \centering
    \includegraphics[width=\linewidth]{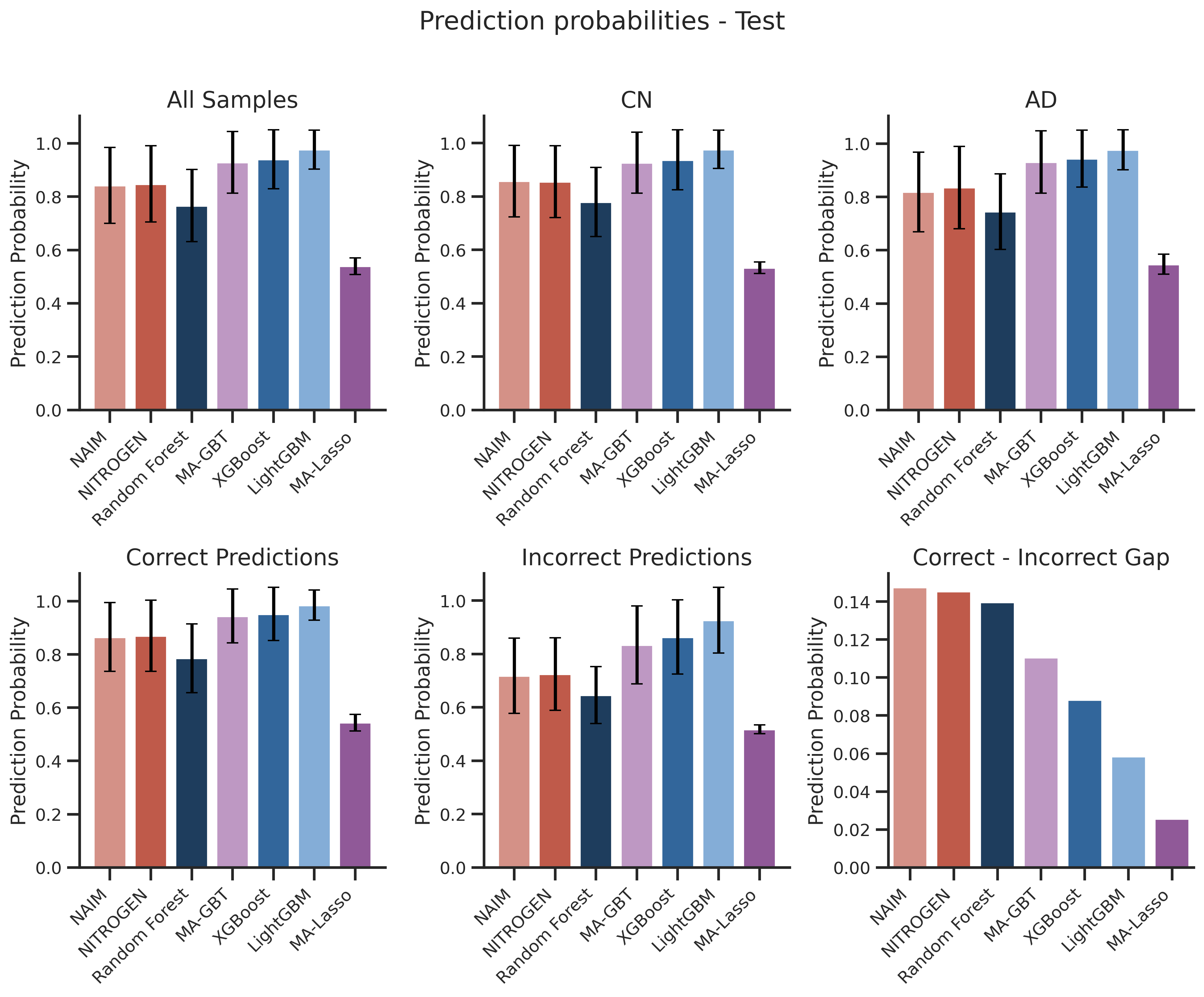}
    \caption{ADNI test subset}
    \label{fig:predict_prob_test}
\end{subfigure}
\begin{subfigure}[t]{0.48\textwidth}
    \centering
    \includegraphics[width=\linewidth]{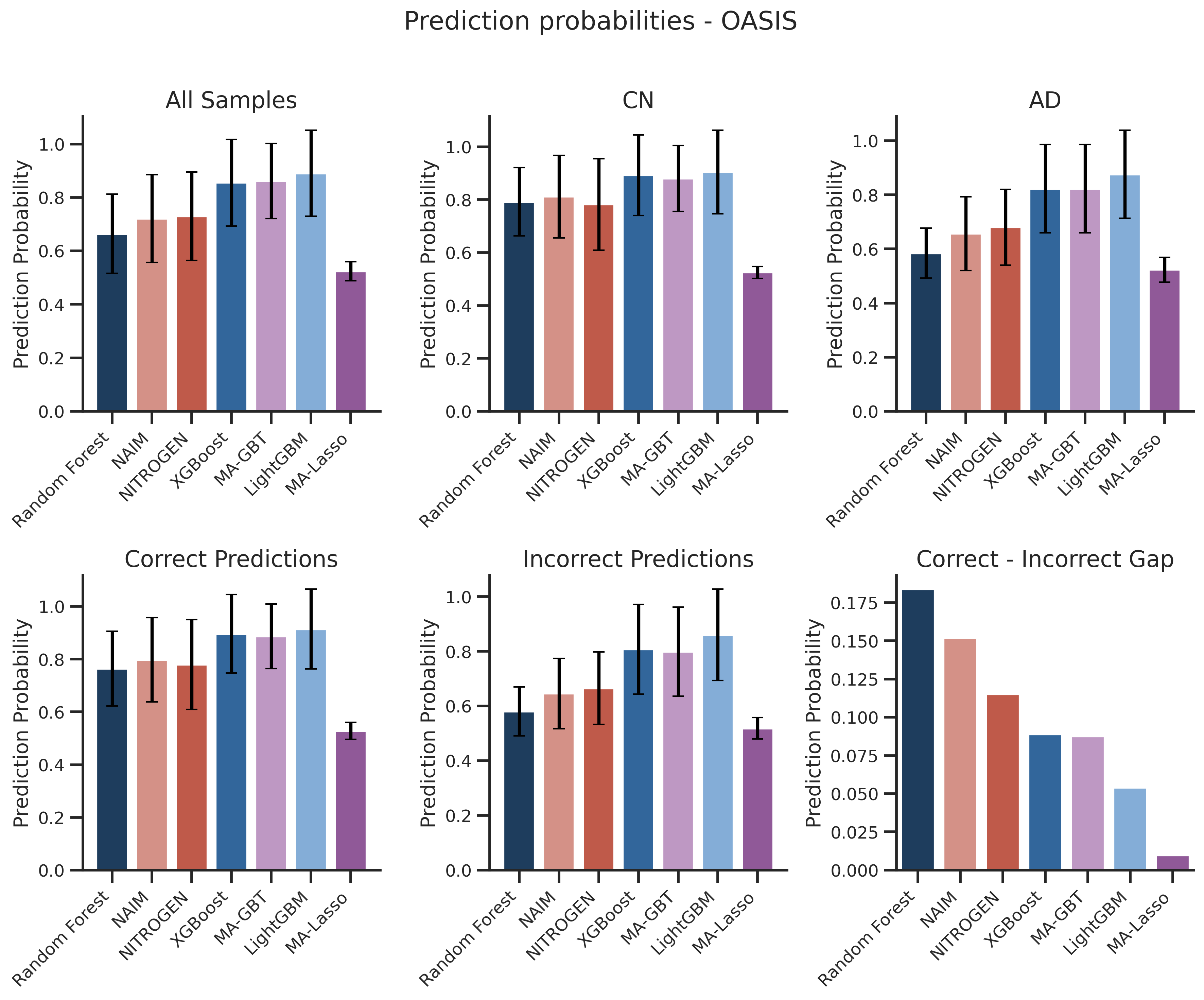}
    \caption{OASIS-3 test set}
    \label{fig:predict_prob_oasis}
\end{subfigure}
\begin{subfigure}[t]{0.48\textwidth}
    \centering
    \includegraphics[width=\linewidth]{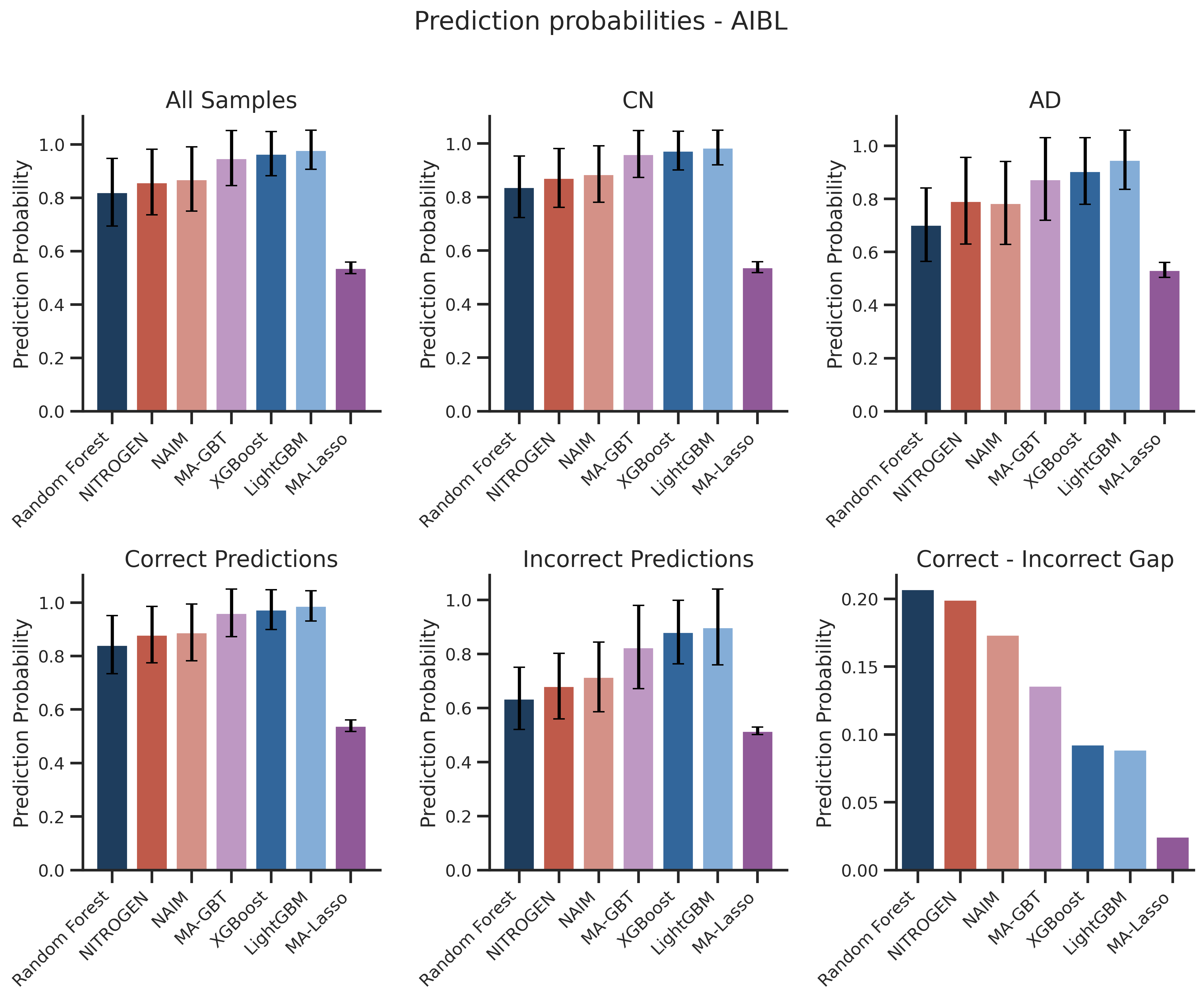}
    \caption{AIBL test set}
    \label{fig:predict_prob_aibl}
\end{subfigure}
\caption{
\textbf{Predicted class probabilities across internal and external datasets.}
Bar plots show mean predicted class probabilities across models, with error bars representing one standard deviation across samples (not shown for the correct–incorrect confidence gap, which reflects a derived difference between two conditions). Within each dataset, results are displayed for all samples, stratified by predicted class (CN and AD), by correctness of prediction (correct vs. incorrect), and by the correct–incorrect confidence gap. The confidence gap is defined as the difference between the mean predicted probability of correctly classified samples and that of incorrectly classified samples.
}
\label{fig:predict_probabilities}
\end{figure}

\begin{figure}[t!]
    \centering
    \includegraphics[width=0.75\textwidth]{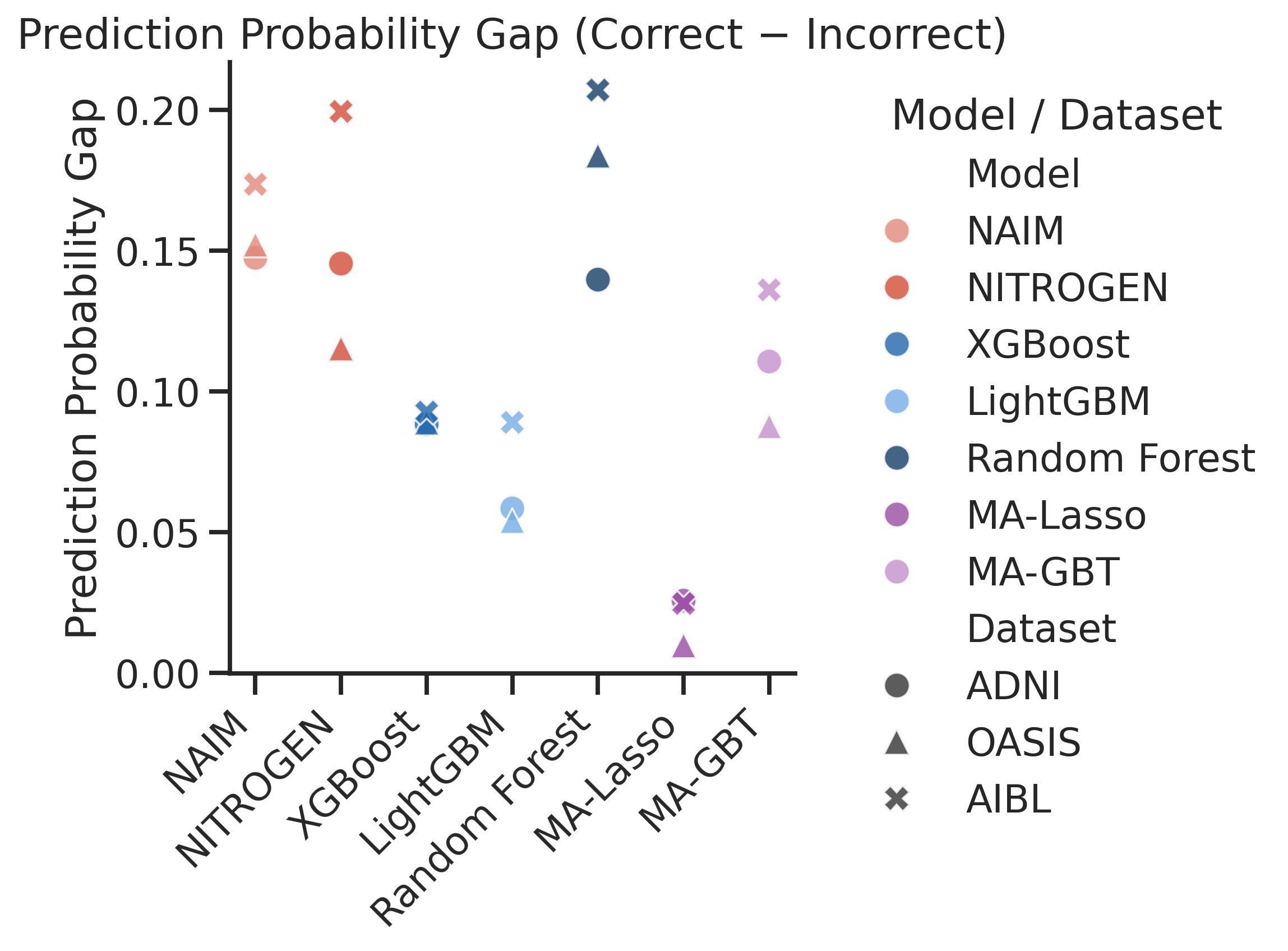}
    \caption{
    \textbf{Prediction probabilities gap across models and test datasets.}
    Scatter plot summarising the prediction confidence gap, defined as the difference between the mean predicted probability of correctly classified samples and that of incorrectly classified samples (correct $-$ incorrect). Each point represents a model–dataset pair. Colours indicate the classification model, and marker shapes denote the evaluation dataset (ADNI, OASIS-3, AIBL). Higher values indicate stronger separation between correct and incorrect predictions, reflecting improved confidence discrimination.
    }
    \label{fig:confidence_gap_summary}
\end{figure}

 \begin{figure}[t!]
    \centering
    \begin{subfigure}[t]{0.48\textwidth}
        \centering
        \includegraphics[width=\linewidth]{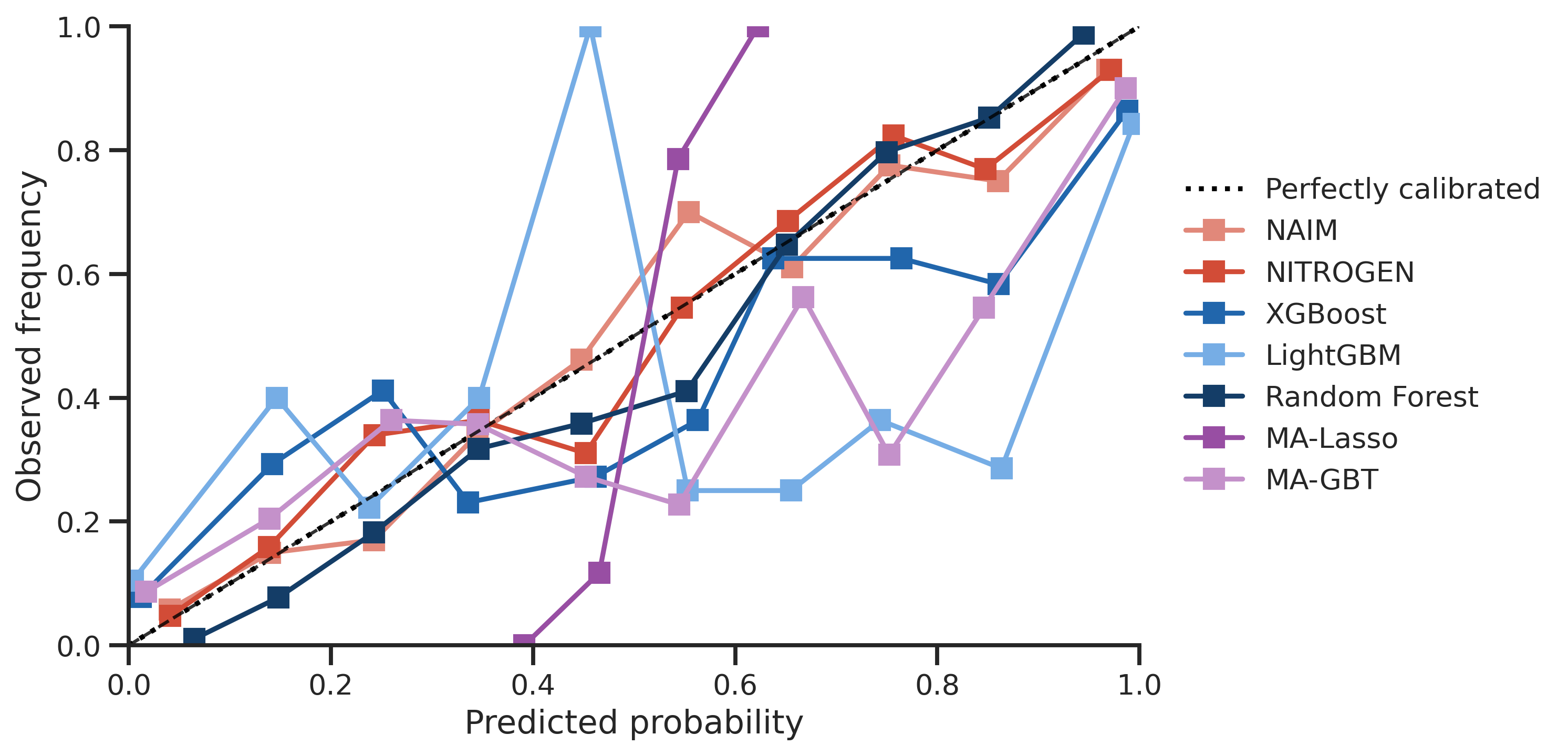}
        \caption{ADNI validation set}
        \label{fig:calibration_val}
    \end{subfigure}
    \hfill
    \begin{subfigure}[t]{0.48\textwidth}
        \centering
        \includegraphics[width=\linewidth]{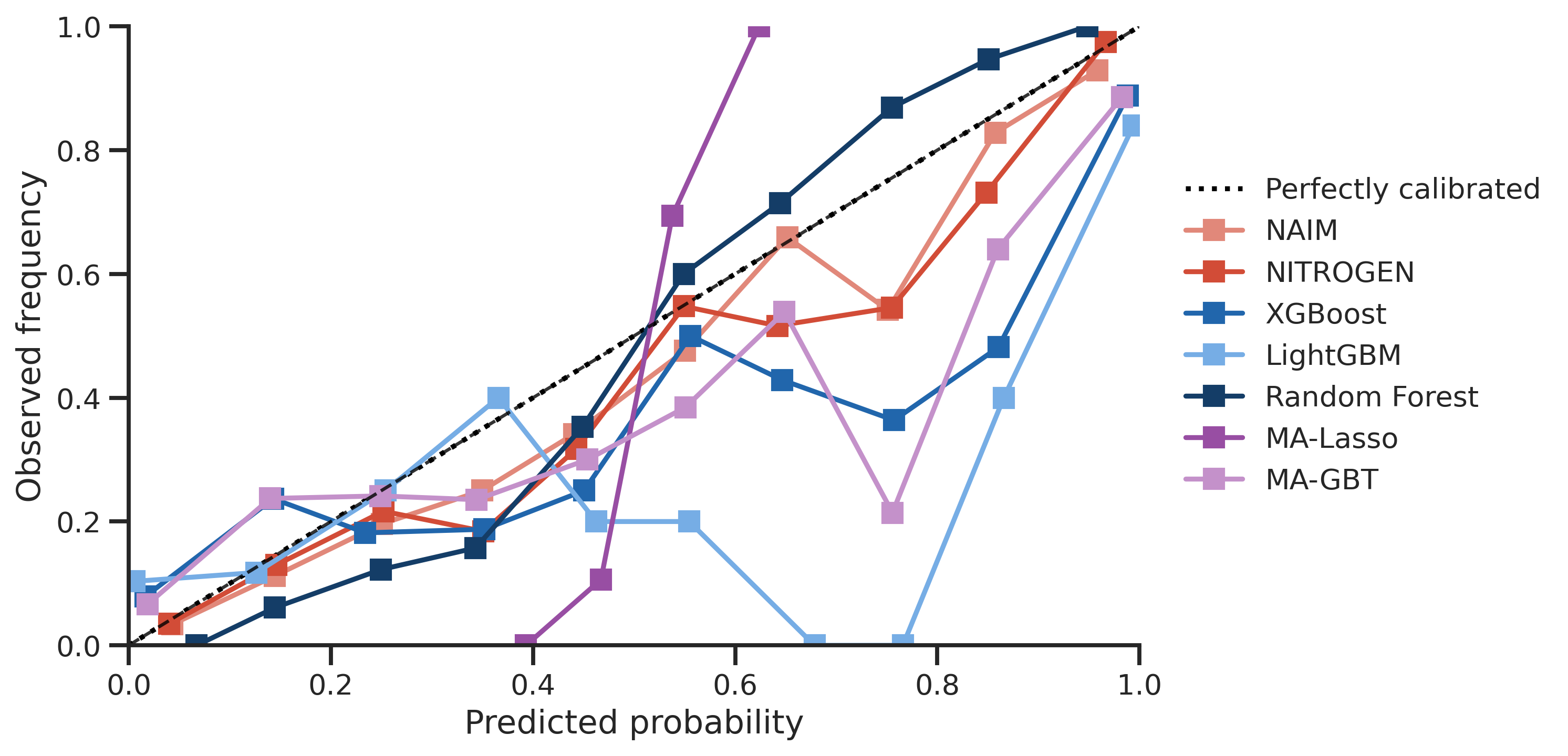}
        \caption{ADNI test set}
        \label{fig:calibration_test}
    \end{subfigure}
    \vspace{0.3cm}
    \begin{subfigure}[t]{0.48\textwidth}
        \centering
        \includegraphics[width=\linewidth]{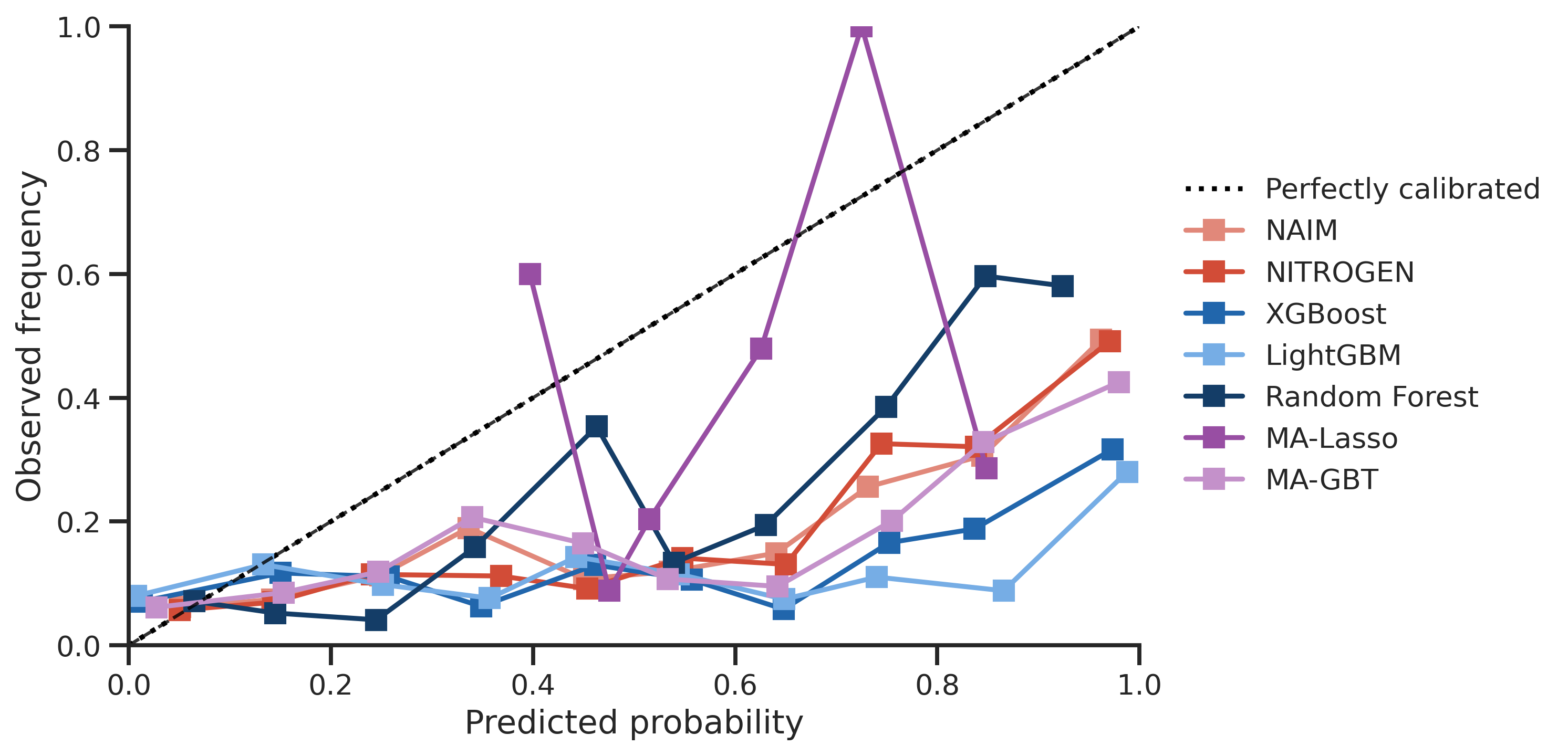}
        \caption{External OASIS-3 cohort}
        \label{fig:calibration_oasis}
    \end{subfigure}
    \hfill
    \begin{subfigure}[t]{0.48\textwidth}
        \centering
        \includegraphics[width=\linewidth]{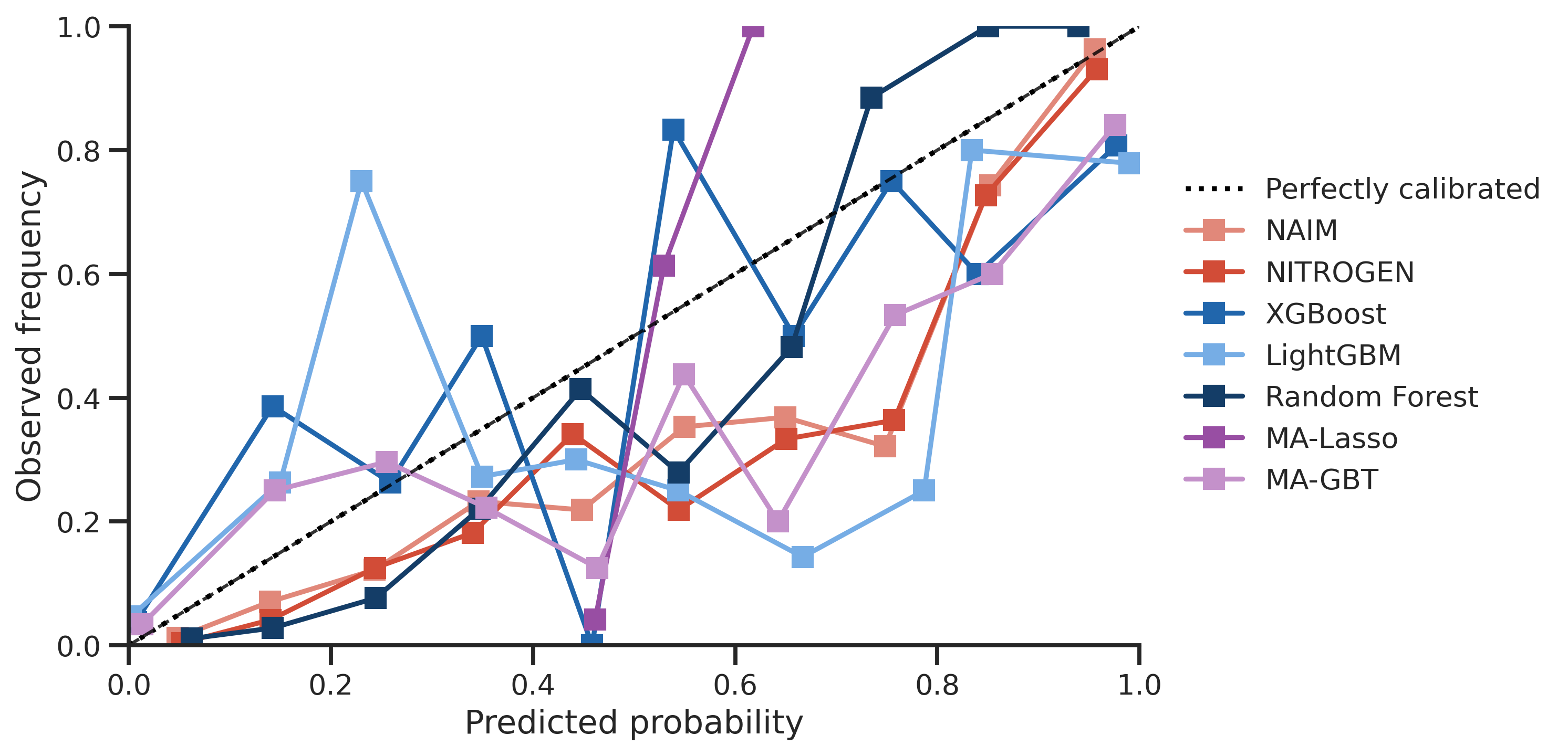}
        \caption{External AIBL cohort}
        \label{fig:calibration_aibl}
    \end{subfigure}
    \caption{
    \textbf{Calibration performance across internal and external datasets.}
    Reliability diagrams showing model calibration for the positive class (AD). 
    The mean predicted probability within each bin (x-axis) is plotted against the observed fraction of positives (y-axis). 
    Colours correspond to different classification models. 
    The dotted black diagonal represents perfect calibration. 
    Models were trained on ADNI and evaluated on the validation set (a), ADNI test set (b), OASIS-3 (c), and AIBL (d).
    }
    \label{fig:calibration_all}
\end{figure}

\begin{table*}[t!]
\centering
\scriptsize
\caption{Sufficiency analysis using the identified features (APOE$\varepsilon$4 and left temporal pole cortical thickness). Performance is reported for three input configurations: (i) full multimodal input (Base), (ii) only the identified MRI feature (LH\_Limbic\_TempPole\_3) with all other MRI modalities removed (set to NA) (MRI), and (iii) only the MRI feature with all others masked (ALL). Metrics include accuracy (Acc.), balanced accuracy (Bal. Acc.), specificity (Spec.), sensitivity (Sens.), and area under the ROC curve (AUC). Seed = 123. }
\label{tab:sufficiency_analysis}
\setlength{\tabcolsep}{4pt}
\begin{tabular}{lllccccc}
\toprule
Dataset & Modality & Model & Acc. & Bal. Acc. & Spec. & Sens. & AUC \\
\midrule
\multirow{21}{*}{ADNI (Test)}
& \multirow{7}{*}{Base}
& NAIM      & 0.87 & 0.86 & 0.88 & 0.85 & 0.91 \\
& & NITROGEN         & 0.85 & 0.84 & 0.86 & 0.82 & 0.91 \\
& & XGBoost          & 0.87 & 0.86 & 0.88 & 0.84 & 0.91 \\
& & LightGBM         & 0.85 & 0.84 & 0.86 & 0.82 & 0.92 \\
& & Random Forest     & 0.85 & 0.84 & 0.88 & 0.81 & 0.92 \\
& & MA-Lasso         & 0.82 & 0.83 & 0.80 & 0.86 & 0.91 \\
& & MA-GBT           & 0.83 & 0.82 & 0.85 & 0.79 & 0.90 \\
\cmidrule(lr){2-8}
& \multirow{7}{*}{MRI}
& NAIM      & 0.53 & 0.63 & 0.30 & 0.96 & 0.76 \\
& & NITROGEN         & 0.58 & 0.66 & 0.41 & 0.91 & 0.77 \\
& & XGBoost          & 0.65 & 0.71 & 0.54 & 0.87 & 0.79 \\
& & LightGBM         & 0.70 & 0.72 & 0.65 & 0.80 & 0.78 \\
& & Random Forest     & 0.51 & 0.61 & 0.30 & 0.92 & 0.77 \\
& & MA-Lasso         & 0.57 & 0.63 & 0.45 & 0.81 & 0.77 \\
& & MA-GBT           & 0.75 & 0.71 & 0.84 & 0.59 & 0.80 \\
\cmidrule(lr){2-8}
& \multirow{7}{*}{ALL}
& NAIM      & 0.34 & 0.50 & 0.00 & 1.00 & 0.77 \\
& & NITROGEN         & 0.34 & 0.50 & 0.00 & 1.00 & 0.77 \\
& & XGBoost          & 0.53 & 0.62 & 0.33 & 0.92 & 0.76 \\
& & LightGBM         & 0.66 & 0.50 & 1.00 & 0.00 & 0.76 \\
& & Random Forest     & 0.34 & 0.50 & 0.00 & 1.00 & 0.57 \\
& & MA-Lasso         & 0.72 & 0.71 & 0.76 & 0.66 & 0.77 \\
& & MA-GBT           & 0.66 & 0.50 & 1.00 & 0.00 & 0.76 \\
\midrule
\multirow{21}{*}{AIBL}
& \multirow{7}{*}{Base}
& NAIM      & 0.90 & 0.78 & 0.94 & 0.62 & 0.92 \\
& & NITROGEN         & 0.90 & 0.81 & 0.93 & 0.69 & 0.96 \\
& & XGBoost          & 0.93 & 0.79 & 0.98 & 0.60 & 0.94 \\
& & LightGBM         & 0.93 & 0.80 & 0.98 & 0.62 & 0.94 \\
& & Random Forest     & 0.90 & 0.75 & 0.95 & 0.55 & 0.92 \\
& & MA-Lasso         & 0.92 & 0.88 & 0.94 & 0.83 & 0.97 \\
& & MA-GBT           & 0.93 & 0.81 & 0.97 & 0.64 & 0.93 \\
\cmidrule(lr){2-8}
& \multirow{7}{*}{MRI}
& NAIM      & 0.38 & 0.63 & 0.30 & 0.95 & 0.81 \\
& & NITROGEN         & 0.57 & 0.71 & 0.53 & 0.88 & 0.79 \\
& & XGBoost          & 0.81 & 0.60 & 0.87 & 0.33 & 0.73 \\
& & LightGBM         & 0.71 & 0.66 & 0.72 & 0.60 & 0.70 \\
& & Random Forest     & 0.15 & 0.47 & 0.05 & 0.88 & 0.57 \\
& & MA-Lasso         & 0.74 & 0.75 & 0.74 & 0.76 & 0.81 \\
& & MA-GBT           & 0.87 & 0.64 & 0.95 & 0.33 & 0.79 \\
\cmidrule(lr){2-8}
& \multirow{7}{*}{ALL}
& NAIM      & 0.12 & 0.50 & 0.00 & 1.00 & 0.83 \\
& & NITROGEN         & 0.12 & 0.50 & 0.00 & 1.00 & 0.86 \\
& & XGBoost          & 0.40 & 0.65 & 0.32 & 0.98 & 0.86 \\
& & LightGBM         & 0.88 & 0.50 & 1.00 & 0.00 & 0.86 \\
& & Random Forest     & 0.12 & 0.50 & 0.00 & 1.00 & 0.41 \\
& & MA-Lasso         & 0.82 & 0.81 & 0.83 & 0.79 & 0.86 \\
& & MA-GBT           & 0.88 & 0.50 & 1.00 & 0.00 & 0.87 \\
\midrule
\multirow{21}{*}{OASIS-3}
& \multirow{7}{*}{Base}
& NAIM      & 0.49 & 0.60 & 0.44 & 0.76 & 0.68 \\
& & NITROGEN         & 0.54 & 0.65 & 0.49 & 0.81 & 0.72 \\
& & XGBoost          & 0.51 & 0.60 & 0.47 & 0.73 & 0.69 \\
& & LightGBM         & 0.52 & 0.57 & 0.49 & 0.66 & 0.67 \\
& & Random Forest     & 0.43 & 0.56 & 0.38 & 0.74 & 0.64 \\
& & MA-Lasso         & 0.55 & 0.61 & 0.52 & 0.70 & 0.70 \\
& & MA-GBT           & 0.68 & 0.65 & 0.69 & 0.60 & 0.72 \\
\cmidrule(lr){2-8}
& \multirow{7}{*}{MRI}
& NAIM      & 0.28 & 0.53 & 0.18 & 0.88 & 0.65 \\
& & NITROGEN         & 0.39 & 0.59 & 0.30 & 0.87 & 0.65 \\
& & XGBoost          & 0.33 & 0.58 & 0.23 & 0.93 & 0.65 \\
& & LightGBM         & 0.38 & 0.54 & 0.31 & 0.77 & 0.62 \\
& & Random Forest     & 0.19 & 0.51 & 0.05 & 0.97 & 0.59 \\
& & MA-Lasso         & 0.46 & 0.57 & 0.42 & 0.72 & 0.60 \\
& & MA-GBT           & 0.69 & 0.64 & 0.71 & 0.56 & 0.69 \\
\cmidrule(lr){2-8}
& \multirow{7}{*}{ALL}
& NAIM      & 0.15 & 0.50 & 0.00 & 1.00 & 0.62 \\
& & NITROGEN         & 0.15 & 0.50 & 0.00 & 1.00 & 0.55 \\
& & XGBoost          & 0.28 & 0.56 & 0.16 & 0.96 & 0.60 \\
& & LightGBM         & 0.85 & 0.50 & 1.00 & 0.00 & 0.56 \\
& & Random Forest     & 0.15 & 0.50 & 0.00 & 1.00 & 0.51 \\
& & MA-Lasso         & 0.76 & 0.57 & 0.84 & 0.30 & 0.57 \\
& & MA-GBT           & 0.85 & 0.50 & 1.00 & 0.00 & 0.54 \\
\bottomrule
\end{tabular}
\end{table*}

\begin{figure}[htbp]
    \centering
    \includegraphics[width=\textwidth]{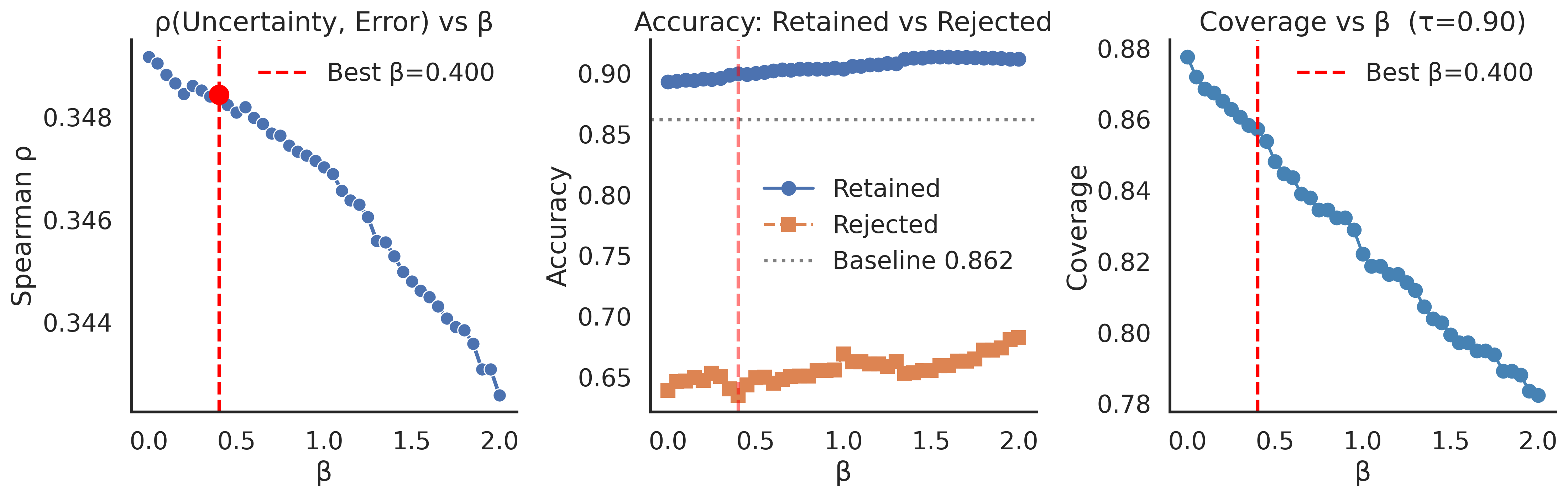}
    \caption{Grid search results for the sensitivity parameter $\beta$ at 
    fixed uncertainty threshold $\tau = 0.900$. 
    \textbf{(Left)} Spearman rank correlation $\rho$ between adjusted 
    uncertainty and prediction errors as a function of $\beta$, used as 
    the primary optimisation criterion. The optimal value $\beta^* = 0.400$ 
    is indicated by the vertical red dashed line. 
    \textbf{(Centre)} Accuracy on retained and rejected samples as a 
    function of $\beta$, with the baseline accuracy on the full dataset 
    shown as a horizontal dotted line. A well-calibrated $\beta$ should 
    increase the gap between retained and rejected accuracy. 
    \textbf{(Right)} Coverage (fraction of samples retained) as a function 
    of $\beta$, showing how the missingness adjustment progressively 
    shifts samples above the fixed threshold $\tau$ as $\beta$ increases. 
    The optimal configuration ($\beta^* = 0.400$, $\tau = 0.900$) retains 
    85.7\% of validation samples (756/882), improving accuracy from 0.8617 
    to 0.8995 ($\Delta = +0.0378$, $p < 0.001$ by permutation test), with 
    rejected samples exhibiting substantially lower accuracy (0.6349).}
    \label{fig:beta_grid_search}
\end{figure}



\end{appendices}



\end{document}